\include{macpap2}

\documentclass[12pt]{ociamthesis}

\usepackage{amssymb,epsfig}

\usepackage{axodraw}
\usepackage{color}

 \usepackage{setspace}

\newcommand{\SM}{Standard Model}
\newcommand{\TBM}{tri-bi-maximal mixing}
\newcommand{\SD}{sequential domination}
\newcommand{\GUT}{GUT}
\newcommand{\GUTs}{GUTs}
\newcommand{\GST}{Gatto-Sartori-Tonin}
\newcommand{\FN}{Froggatt-Nielsen}
\newcommand{\GJ}{Georgi-Jarlskog}
\newcommand{\FS}{family symmetry}
\newcommand{\FSs}{family symmetries}
\newcommand{\VEV}{vacuum expectation value}
\newcommand{\FCNC}{flavour changing neutral current}
\newcommand{\RH}{right-handed}
\newcommand{\LH}{left-handed}
\newcommand{\SUGRA}{supergravity}
\newcommand{\FP}{flavour problem}
\newcommand{\PS}{Pati-Salam}
\newcommand{\FMM}{fermion masses and mixings}

\graphicspath{{./}{Figs/}{Plots/}}

\title{Family symmetries \\[1ex]
and the origin of fermion masses and mixings}
\author{Ivo de Medeiros Varzielas}
\college{St Hugh's College}
\degree{Doctor of Philosophy}
\degreedate{Trinity 2007}
\renewcommand{\crest}{\beltcrest}

\begin{document}

\baselineskip=18pt plus1pt

\setcounter{secnumdepth}{3}
\setcounter{tocdepth}{3}

\maketitle

\begin{dedication}
This thesis is dedicated to my wife\\
Ana\\
for being herself\\
\end{dedication}

\begin{acknowledgements}
First of all, I would like to thank my wife and best friend Ana for her
love and support.\\
In particular I thank my parents Albano and Maria for always being
available for whatever I needed.\\
In general I would also like to thank my family and close
friends. I consider that my family members are also my friends and my friends are also my family.
\\
Thank you all for always being with me, regardless of physical distances!
\\
\\
Finally, thanks to Graham, whose guidance allowed me to have reached
this stage.
\end{acknowledgements}

\begin{abstract}

Family symmetries are possibly the most conservative extension of the
\SM \ that attempt explanations of the pattern of \FMM . The observed large mixing angles in the lepton sector may be the first
signal for the presence of a non-Abelian family symmetry.
We investigate the possibilities of simultaneously explaining
the observed pattern of masses of the quarks
(hierarchical masses and small mixing angles) and of the leptons
(near \TBM , thus large mixing angles). We show that such contrasting observations can be
achieved naturally via the seesaw mechanism, whether in models with continuous or discrete \FSs .

We consider also in some detail the constraints on \FCNC s arising from introducing a continuous \FS .
We show that, for a restricted choice of the
flavon sector, continuous \FSs \ are consistent with even the
most conservative limits both for the case of gauge mediated supersymmetry
breaking and the case of gravity mediated supersymmetry breaking.

\end{abstract}

\begin{romanpages}
\tableofcontents
\listoffigures
\listoftables
\end{romanpages}

\doublespacing

\chapter{Introduction \label{ch:intro}}

\section{Motivation and outline \label{sec:mot}}

While the \SM \ extended to include right-handed
neutrinos continues to successfully describe all existing data, there
are sound theoretical reasons to believe that there is physics beyond
the \SM. Appealing extensions of the \SM \ often include supersymmetry (SUSY), as well as Grand (\GUTs) . 
Very brief reviews of both can be found in subsections \ref{sub:SUSY} and \ref{sub:GUTs}.

The question why we have three generations of each type of fundamental
fermion remains without a convincing answer. In the \SM \ the masses and mixings of all
these fermions are simply parameters (the Yukawa couplings) that need
to be measured. When going beyond the \SM , those \FMM \ can arise through underlying mechanisms - examples of such being \FN , or the
seesaw mechanism, two
mechanisms that generate fermion masses through 
higher dimension operators involving heavier particles (both reviewed in section \ref{sec:mass mechs}).

The data from neutrino oscillations (reviewed in subsection
\ref{sub:summary}) compounds the puzzle of the \FMM . The data indicates that the leptonic 
mixing angles are large - in stark contrast with the small mixing
angles of the quark sector.

In the rest of chapter \ref{ch:intro} we briefly review the current status of \FMM , giving a very brief summary of the \SM , SUSY, \GUTs \ and of neutrino oscillations. We then present a \FS \ review, briefly discussing recent models in the literature and illustrating the main points by using a very simple $U(1)_{f}$ toy model, motivating us to conclude the chapter with a review of the \FN \ mechanism and of the seesaw mechanism.

Chapter \ref{ch:fs} presents original work on explaining the observed \FMM \ data with continuous \FSs , namely an $SU(3)_{f}$ \FS \ model. The \VEV \ alignment is then discussed (with more details in appendix \ref{app_A}). The messenger sector involved in the respective \FN \ mechanism is also presented. The chapter concludes with the phenomenological implications of the model and a brief discussion regarding the cancellation of anomalies.

In chapter \ref{ch:fsd} we present original work on approaching the \FMM \ problem with discrete \FSs. We start by presenting a simple but incomplete model based on the group $\Delta (12)$. Then we continue by studying a model based on the group $\Delta (27)$. The novel \VEV \ alignment mechanism used in the $\Delta(27)$ model is discussed in detail and then the chapter is concluded with the phenomenological predictions of the model.

Chapter \ref{ch:fcnc} contains original work on solving the \FS \ \FP \ (associated with continuous \FSs ). The topic is introduced by first re-examining the SUSY \FP \ and then identifying the contributions related to the continuous \FSs. A simple $U(1)_{f}$ model serves to illustrate the problem. We conclude the chapter using the same model to also illustrate possible solutions that apply in the general case. The solutions are discussed in relation to the respective SUSY breaking mechanism (gravity or gauge mediated).

Finally, in chapter \ref{ch:conc} we present a global conclusion and summary of the thesis.

\section{Fermion masses and mixings \label{sec:FMM}}

\subsection{Standard Model \label{sub:SM}}

The \SM \ is a quantum field theory describing the fundamental particles and their interactions, based in the local gauge group $SU(3)_{C} \times SU(2)_L
\times U(1)_{Y}$. The $SU(3)_{C}$ is the group of quantum chromodynamics (QCD), the theory of strong interactions of the coloured particles such as gluons (the gauge bosons of QCD) and quarks. The $SU(2)_L
\times U(1)_{Y}$ is the group of the electroweak interaction, whose gauge bosons include the weak gauge bosons $W^{+}$, $W^{-}$ and $Z^{0}$, as well as the photon $\gamma$ of electromagnetic interactions;
the electric charge is given by $Q_{em} = T_{3_{L}}+Y/2$ (the isospin $T_{3_{L}}$ is associated with $SU(2)_{L}$ and the hypercharge $Y$ is the $U(1)_{Y}$ quantum number).

The matter content of the \SM \ consists of three families of quarks and leptons. They transform as spinors under the Lorentz group, and as the \LH \ and \RH \ parts are treated differently under the gauge group, it is often more convenient to use 2 component Weyl spinors (dotted and undotted) rather than the Dirac spinor representation, although they are equivalent: the 4 component Dirac spinor $\Psi$ is composed of one undotted, or \LH , 2 component spinor $\zeta_{\alpha}$ and one dotted, or \RH , 2 component spinor $\eta^{\dagger \dot{\alpha}}$.
\begin{equation}
\Psi = 
\left(
\begin{array}{c}
\zeta_{\alpha}\\ 
\eta^{\dagger \dot{\alpha}}
\end{array}
\right)
\label{eq:petazeta}
\end{equation}
If the 4 component Dirac spinor has the same dotted and undotted Weyl spinors ($\eta \equiv \zeta$), it is called a Majorana spinor (the \LH \ and \RH \ spinors are equivalent). The hermitian conjugate of a \LH \ spinor is a \RH \ spinor and vice-versa:
\begin{equation}
\left( \eta^{\dagger \dot{\alpha}} \right)^{\dagger} = \eta^{\alpha}
\label{eq:RH into LH}
\end{equation}
Finally, the case (upper or lower) of the indices is relevant, and they can be raised or lowered with the appropriate anti-symmetric Levi-Civita tensors $\epsilon^{\alpha \beta}$ or $\epsilon_{\alpha \beta}$ (or the dotted versions): $\zeta^{\alpha} \equiv \epsilon ^{\alpha \beta} \zeta_{\beta}$ and so on.
In general we will omit the spinor indices for simplicity,
with the convention that two \LH \ Weyl spinors contract as $\zeta \eta \equiv \zeta^{\alpha} \eta_{\alpha}$ and two \RH \ spinors contract as $\zeta^{\dagger} \eta^{\dagger}= \zeta^{\dagger}_{\dot{\alpha}} \eta^{\dagger \dot{\alpha}}$.

We will use two distinct notations for the fermions, and it is useful to keep in mind that the charge conjugate $f^{c}$ of a \RH \ fermion $f_{R}$ transforms in the same way as a \LH \ fermion (see eq.(\ref{eq:RH into LH})). We denote the true \LH \ fermions either as $f$ or explicitly as $f_{L}$, so although similar it is important not to confuse $f^{c}$ with the charge conjugate of $f$, $f^{C}$. We won't mix notations, so we either use $f$ together with $f^{c}$, or particularly in chapter \ref{ch:fcnc}, $f_{L}$ together with $f_{R}$ (usually we use only $f$ and $f^{c}$, as in that case we deal solely with fields that transform as \LH \ fermions - very useful when discussing \GUTs ).

Each \SM \ family contains one \LH \ quark doublet $Q=(u,d)$, one \RH \ up type quark $u^{c}$, one \RH \ down type quark $d^{c}$, one \LH \ lepton doublet $L=(\nu, e)$ and one \RH \ charged lepton $e^{c}$. We will also include in each family one \RH \ neutrino $\nu^{c}$, even though it is a singlet of the \SM \ ($\nu^{c}$ is a natural feature of \GUTs \ - see subsection \ref{sub:GUTs}). The symmetry assignments of one such family under the \SM \ gauge group are displayed in table \ref{ta:SM}, along with the symmetry assignments of the Higgs boson $H$. A more detailed review can be found in \cite{PDG}.

\begin{table}[htb] \centering
\begin{tabular}{|c||c|c|c|}
\hline
Field & $SU(3)_{C}$ & $SU(2)_{L}$ & $U(1)_{Y}$ \\ \hline \hline
$H$ & $\mathbf{1}$ & $\mathbf{2}$ & $-\mathbf{1}$ \\ \hline
$Q$ & $\mathbf{3}$ & $\mathbf{2}$ & $\mathbf{1/3}$ \\
$u^{c}$ & $\mathbf{\bar{3}}$ & $\mathbf{1}$ & $-\mathbf{4/3}$\\
$d^{c}$ & $\mathbf{\bar{3}}$ & $\mathbf{1}$ & $\mathbf{2/3}$ \\ \hline
$L$ & $\mathbf{1}$ & $\mathbf{2}$ & $-\mathbf{1}$ \\
$\nu^{c}$ & $\mathbf{1}$ & $\mathbf{1}$ & $\mathbf{0}$ \\
$e^{c}$ & $\mathbf{1}$ & $\mathbf{1}$ & $\mathbf{2}$ \\ \hline
\end{tabular}
\caption{Assignment of quarks, leptons and Higgs under the \SM .}
\label{ta:SM}
\end{table}

The \SM \ gauge group prevents fermion mass terms from simply arising in the Lagrangian. For example, the Dirac mass term $m L e^{c}$ is not invariant under $SU(2)_{L}$. However, as $H$ is an $SU(2)_{L}$ doublet, one can form invariants by having $H$ together with $Q$ or $L$. The Yukawa Lagrangian $L_{Y}$ contains such terms and is invariant under the \SM \ gauge group:
\begin{equation}
L_{Y} =  Y^{u^{ij}} Q_{i} u^{c}_{j} H^{\dagger} + Y^{d^{ij}} L_{i} d^{c}_{j} H + Y^{e^{ij}} L_{i} e^{c}_{j} H + Y^{\nu^{ij}} L_{i} \nu^{c}_{j} H^{\dagger}
\label{eq:L_Y}
\end{equation}
$i$ and $j$ are family indices and we have suppressed the $SU(3)_{C}$ and $SU(2)_{L}$ indices.

It is through the Yukawa couplings of fermions to the Higgs boson $Y^{f^{ij}}$ that the fermions get their masses in the \SM \ (after the spontaneous breaking of $SU(2)_{L}$ when $H$ acquires a non-vanishing \VEV \ $\langle H \rangle = h$).

For completeness we must note that the Yukawa interactions (eq. (\ref{eq:L_Y})) only give rise to Dirac masses, mass terms where a \LH \ fermion and a \RH \ fermion are involved (as displayed, the charge conjugate of a \RH \ fermion). It is also possible to form Majorana mass terms - a term with two \LH \ fermions or a term with two \RH \ fermions. With the fields introduced so far, this is the case with $\nu^{c}$ only: being completely neutral under the \SM , a Majorana mass term $M \nu^{c} \nu^{c}$ is allowed.

The parameters $Y^{f^{ij}}$ constitute the majority of unknown
parameters of the \SM , corresponding to 6 quark masses, 3 mixing angles and 1 complex phase
for the quark sector; 6 lepton masses, 3 mixing angles and 1 complex phase (if
the light neutrinos have just Dirac masses) for the lepton sector.
Naturally, by including the possible Majorana masses of the $\nu^{c}$ there will be even more free parameters (see the seesaw mechanism review in subsection \ref{sub:ssm}).
The high number of parameters related to \FMM \ is further increased in the Higgs sector, where we have not just its \VEV \ $h$, but also its quartic coupling coefficient $\lambda$.
In the gauge sector, the $SU(3)_{C}$ gauge coupling $g_{3}$, the gauge coupling of
$SU(2)_{L}$, $g$, and the coupling of $U(1)_{Y}$, $g'$, are 3 more free
parameters of the theory. There is additionally $\theta_{QCD}$, parametrising CP violation in the strong interactions, although it isn't relevant for the remaining discussion.

\subsection{Supersymmetry \label{sub:SUSY}}

In the \SM , the Higgs mass parameter $m_H$ appearing in the Lagrangian is quadratically dependent on the cutoff scale at which new physics is introduced - this leads to the well known hierarchy problem of particle physics. Although $m_H$ hasn't been experimentally measured, it must be of order $10^{2}$ GeV as it sets the scale of electroweak breaking ($h$ depends on $m_H$ and $\lambda$, only two of the three parameters are independent). If the cutoff scale is taken to be the Planck scale as given by the Planck mass $M_{P}$, to keep $m_H$ relatively so tiny ($m_H / M_P$ is of order $10^{-17}$) requires a very high degree of fine tuning between the bare mass and the radiative corrections.

The most popular solution to the hierarchy problem is low energy $N=1$ SUSY. Under very general conditions, SUSY is the only possible extension of space-time symmetry beyond the Poincar\'{e} group (Lorentz group plus translations). On top of the Poincar\'{e} transformations, it adds peculiar fermionic transformations that happen to change the spin of fields (heuristically, the SUSY transformations are ``square roots'' of translations: the anti-commutator of two SUSY transformations is proportional to one translation operator). In $N=1$ SUSY only one distinct set of SUSY generators is introduced. We consider solely $N=1$, as with higher $N$ SUSY one can't have chiral fermions and parity violation as observed in the \SM \ without introducing extra states that conflict with precision tests.

In SUSY, states are assigned into superfields $\Phi$ (so each known \SM \ field is considered to be part of a superfield), and it is very useful to use the superpotential $P$ (to differentiate from the real potential, $V$). $P$ is an analytic function of the superfields $\Phi$, and as such is holomorphic ($P$ is a function only of the $\Phi$, and not of $\Phi^{\dagger}$). The terms in $P$ must be gauge invariant, and the renormalisable terms in the superpotential have dimension $3$ or less (in contrast with the Lagrangian, in which renormalisable terms have dimension $4$ or less).

In the minimal supersymmetric \SM \ (usually denoted as MSSM) the field content of the \SM \ is only increased by an extra Higgs $SU(2)_{L}$ doublet. We rename the \SM \ $H$ to $H_{d}$, responsible for giving mass to the down quarks and charged leptons. The extra Higgs, $H_{u}$, is required to generate the Dirac mass of the up quarks  (and of neutrinos if $\nu^{c}$ is included) as the holomorphicity requirement of $P$ prevents the charge conjugate of $H_{d}$ from playing that role (as opposed to what happened with $H$ in the \SM ).

The \SM \ fermions are placed in chiral superfields that contain also their respective superpartners, bosons with spin 0 usually denoted as sfermions (the squarks and sleptons). The \SM \ gauge bosons are instead part of vector superfields with their own superpartners, spin $1/2$ fermions, usually denoted as gauginos $\lambda$ (e.g. the gluinos, $\tilde{g}$, or the photino $\tilde{\gamma}$). The two Higgs belong to chiral superfields, although in this case obviously the spin $1/2$ fermions of the chiral superfields are actually the superpartners, the Higgsinos.

Each chiral superfield $\chi$ is composed of one complex scalar sfermion $\tilde{f}$ and one complex Weyl fermion\footnote{Another good reason to use Weyl spinors instead of Dirac spinors is that each chiral superfield includes one single, 2-component Weyl fermion.} $f$.
In turn, each vector superfield $W$ has one Weyl fermion $\lambda$, and the vector $A_{\mu}$.

The real potential $V$ is composed of two contributions. One is usually called the $F$-term, obtained from the superpotential: $F_{i} \equiv dP/d\tilde{f}_{i}$, where $i$ is an index labelling the components of whatever representation the field has under the gauge group (for example, three components if the chiral superfield containing $\tilde{f}_{i}$ is a triplet of $SU(3)$). The other contribution is usually called the $D$-term, and is associated with the gauge group: $D^{a} \equiv -g \tilde{f}^{\dagger} T^{a} \tilde{f} - g \xi^{a}$, where we now have added the Fayet-Iliopoulos term $\xi^{a}$ which may be non-zero only for Abelian $U(1)$ factors of the group (for example, if the group is $SU(3)$, $a$ labels the 8 generators $T^{a}$, and $\xi^{a}=0$). Excluding the Fayet-Iliopoulos term, we have then:

\begin{equation}
V = F^{\dagger} F + \frac{1}{2} D^{2} = \sum_{i} \left| \frac{dP}{d\tilde{f}_{i}} \right|^{2} +   \sum_{a} \frac{g_{a}^{2}}{2} \left( \tilde{f}^{\dagger} T^{a} \tilde{f}  \right)^{2}
\label{eq:FD}
\end{equation}

In terms of the hierarchy problem, SUSY relates the superpartners interactions with the interactions of their \SM \ counterparts in such a way that the loop diagrams contributing to the quadratic divergence of $m_H$ with superpartners in the loop give the exact same contribution as \SM \ contributions, but with opposite sign (due to the minus sign coming from the fermion loop): SUSY enables the exact cancellation of the quadratic divergence through the superpartners (for example, the stop squark cancels the leading top quark contribution), leaving only milder logarithmic divergences.

As superpartners haven't yet been observed, SUSY must be broken at some scale higher than the electroweak scale. If one wants to rely on SUSY to solve the hierarchy problem, this breaking scale $M_{SUSY}$ has to be relatively low, not much higher than $1$ TeV as $M_{SUSY}$ now serves as the cutoff scale of the \SM \ after which new physics is introduced (in this case, SUSY).

\begin{figure}[htb]
\centering
\resizebox{12 cm}{!}{
 \includegraphics{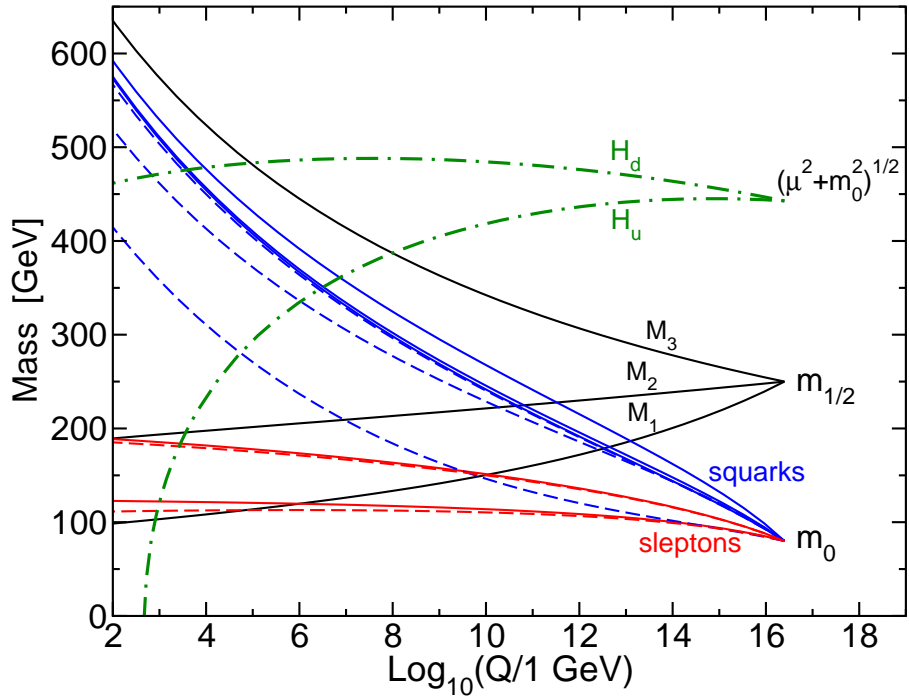}}
\caption[Running superpartner masses.]{Running superpartner masses (from \protect\cite{Martin:1997ns}).}
\label{fig:susy_mass}
\end{figure}

\begin{figure}[htb]
\centering
\resizebox{12 cm}{!}{
 \includegraphics{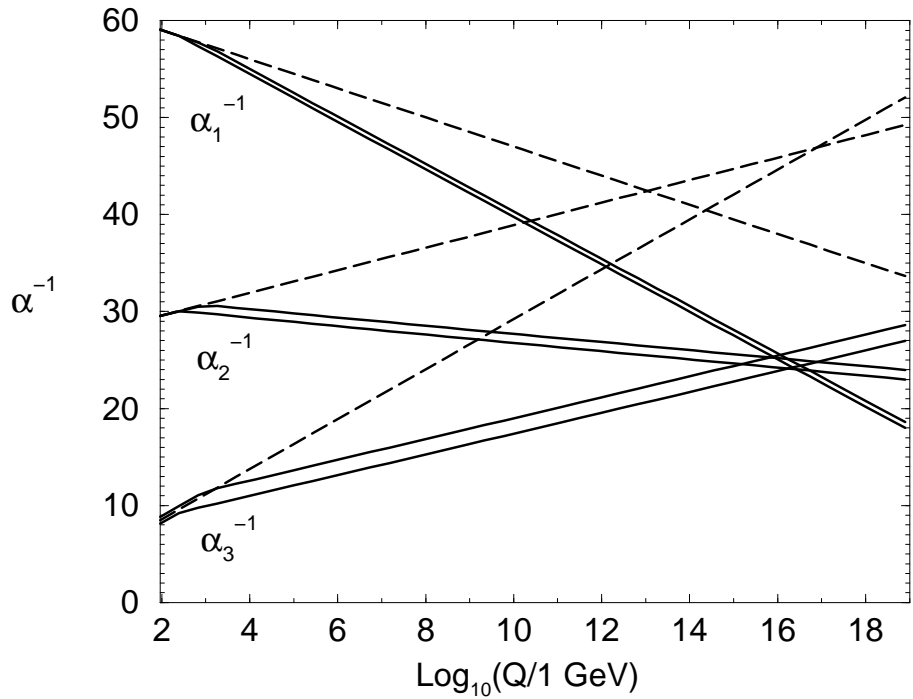}}
\caption[Running coupling constants.]{Running coupling constants (from \protect\cite{Martin:1997ns}).}
\label{fig:susy_gauge}
\end{figure}

The superpartners mass spectrum depends strongly on the SUSY breaking mechanism (see \cite{Martin:1997ns}). Figure \ref{fig:susy_mass} shows an illustrative example of the evolution of superpartner masses with energy scale $Q$, driven by radiative corrections of gauge (positive) and Yukawa (negative) contributions. The graph features the spectrum of the MSSM, with supergravity inspired boundary conditions (common masses $m_{0}$ for the scalar partners and $m_{1/2}$ for the fermion partners) imposed approximately at a unification scale $M_{GUT}$ of around $10^{16}$ GeV \cite{Martin:1997ns} (see also figure \ref{fig:susy_gauge}).

In figure \ref{fig:susy_mass}, $\mu$ is the coefficient of the $\mu$-term in the superpotential, coupling the two Higgs $\mu H_{u} H_{d}$. $M_{3}$, $M_{2}$ and $M_{1}$ are the gaugino masses (corresponding to the $SU(3)_{C}$, $SU(2)_{L}$ and $U(1)_{Y}$ gauge groups respectively) running from the common fermion mass $m_{1/2}$. The dashed lines correspond to the third generation sfermions, and the solid lines to the other sfermions, all running from the common scalar mass $m_{0}$.
Note that the strong interaction radiative corrections dominate, driving the gluinos and the squarks considerably heavier than the other gauginos and sleptons, and note also that the third generation sfermions are respectively lighter (particularly the stop and the sbottom) as they receive stronger Yukawa (negative) contributions.

Figure \ref{fig:susy_mass} displays another desirable feature of SUSY models - the Higgs mass (of $H_{u}$) can be driven negative at low $Q$, with the negative Yukawa contributions (largely due to coupling to the top quark) dominating over the gauge contributions and triggering a non-vanishing \VEV \ for $H_{u}$ that breaks electroweak symmetry. This mechanism was proposed originally in \cite{Ibanez:1982fr, Inoue:1982pi, AlvarezGaume:1983gj} and a recent review can be found in \cite{Ibanez:2007pf}.

Another very compelling reason for low energy SUSY to exist in nature is the apparent unification of coupling strengths. In the \SM \ the couplings don't quite unify (dashed lines in figure \ref{fig:susy_gauge}). However, with the introduction of the superpartners at the previously discussed SUSY scale $M_{SUSY}$ of around $1000$ GeV, the evolution is changed and the three couplings run together, as shown by the solid lines of figure \ref{fig:susy_gauge} (note that the unification scale in figure \ref{fig:susy_gauge} motivates the boundary conditions in figure \ref{fig:susy_mass}). In figure \ref{fig:susy_gauge}, the strong coupling represented by $\alpha_{3}(m_{Z})$ is varied from $0.113$ to $0.123$ and the mass thresholds are varied between $250$ and $1000$ GeV. Clearly, if $M_{SUSY}$ had been of a different order of magnitude, the couplings wouldn't run together: enticingly, introducing SUSY at the scale that leads to gauge coupling unification also solves the hierarchy problem.
$\alpha_{3}$, $\alpha_{2}$ and $\alpha_{1}$ are the hyperfine constants ($\alpha_{a}=g_{a}^{2}/4 \pi$) associated with $SU(3)_{C}$, $SU(2)_{L}$ ($g_{2}=g$ of subsection \ref{sub:SM}) and $U(1)_{Y}$ - although note $g_{1}=\sqrt{5/3} g'$ (the $g'$ of subsection \ref{sub:SM}) is normalised in order to be the coefficient in the canonical covariant derivative of $SU(5)$ or $SO(10)$ \GUT \ embeddings of the \SM \ gauge group.

A much more detailed review of SUSY can be found in \cite{Martin:1997ns}.

\subsection{Grand Unified Theories \label{sub:GUTs}}

Quarks and leptons share several properties, pointing towards the interesting
hypothesis that there might be some underlying fermion unification at some high energy scale $M_{GUT}$ (the unification scale).
Just as the \SM \ $SU(2)_{L}$ relates electrons and neutrinos, larger
symmetries can relate quarks and leptons. Extending the
\SM \ group to the \PS \ group $SU(4)_{C} \times SU(2)_{L} \times
SU(2)_{R}$  \cite{Pati:1974yy} is one example of a \GUT \ that ties quarks and leptons together: the leptons are seen as the extra
``colour'' of $SU(4)_{C}$. The $SU(2)_{R}$ factor makes the \PS \ \GUT \
left-right symmetric. 
Each of the three families now has one \LH \ multiplet including the respective quark and lepton doublets $(Q,L)$, and one \RH \ multiplet $(Q^{c},L^{c})$ including the charge conjugates of the \RH \ states that now belong to their own doublets ($Q^{c}$ and $L^{c}$). The full symmetry assignments
of fermions under the \PS \ group are in table \ref{ta:PS}, where the quantum number simplification is readily apparent (particularly when compared with the relatively strange hypercharge assignments of table \ref{ta:SM}).

\begin{table} [htb] \centering
\begin{tabular}{|c||c|c|c|}\hline
Field & $SU(4)_{C}$ & $SU(2)_{L}$ & $SU(2)_{R}$ \\ \hline \hline
$(Q, L)$ & $\mathbf{4}$ & $\mathbf{2}$ & $\mathbf{1}$ \\
$(Q^{c}, L^{c})$ & $\mathbf{\bar{4}}$ & $\mathbf{1}$ & $\mathbf{\bar{2}}$ \\ \hline
\end{tabular}
\caption{Assignment of fermions under \PS \ \GUT .}
\label{ta:PS}
\end{table}
From the multiplet structure in table \ref{ta:PS} one may see that $\nu^{c}$ is now naturally introduced together with the charge conjugates of the right-handed charged leptons, $e^{c}$ (unlike in the \SM ).

Although with \PS \ there was some extent of fermion unification, the gauge couplings remain independent parameters. To obtain gauge coupling unification at the unification scale $M_{GUT}$ (see figure \ref{fig:susy_gauge}), one can instead extend the  \SM \ gauge group into a single, simple group (simple in the group theory sense). This is possible as long as the rank of the group is larger or equal than the (combined) rank of the \SM \ gauge group. A commonly utilised example of such a group is $SU(5)$ \cite{Georgi:1974sy}. Although we won't go into the details, in $SU(5)$ \GUTs \ the fermions aren't fully unified, in the sense that they are introduced as two distinct irreducible representations like in \PS . Despite the appeal of gauge coupling unification, in terms of fermions $SU(5)$ is arguably less appealing than \PS \ is: the representations are a $\mathbf{10}$ containing $Q$, $u^{c}$ and $e^{c}$ and a $\bar{\mathbf{5}}$ containing $d^{c}$ and $L$ (note the absence of $\nu^{c}$, which can however be introduced as a singlet just like in the \SM ). 

If one is willing to go further one can extend the symmetry to $SO(10)$. With $SO(10)$ there is not only gauge coupling unification, but enticingly, every fermion of one family fits in one single fundamental representation: the $\mathbf{16}$ of $SO(10)$ (the charge conjugates of the \RH \ fermions fit together with the \LH \ fermions, including $\nu^{c}$ that nicely completes the multiplet). Another interesting point to note is that $SO(10)$ has as subgroups both \PS \ and $SU(5)$, and has inequivalent maximal breaking patterns into one or the other: $SO(10) \rightarrow SU(4)_{C} \times SU(2)_{L} \times
SU(2)_{R}$ or $SO(10) \rightarrow SU(5) \times U(1)_{X}$. See for example \cite{PDG} for a more complete review.

In any of the \GUTs \ discussed, however, the existence of three families remains unexplained. A possible explanation for the families arises from string theories, where the number of families can be related to the geometry of the extra dimensions in some way. In terms of quantum field theories, the more conservative explanation lies in extending the symmetry content with an additional family symmetry acting on the generations.

\subsection{Neutrino oscillations and data \label{sub:summary}}

Neutrino oscillation data implies the existence of neutrino masses. The associated parameters are an important part of the puzzle of \FMM , as the existence of neutrino masses leads to leptonic mixing. Neutrino oscillations arise from a straightforward quantum mechanical
phenomenon that occurs during the propagation of the neutrinos, causing them to change flavour. This is possible due to the existence of lepton mixing, which is entirely analogous to quark mixing (although the values of the mixing angles are quite different). Instead of the
Cabibbo-Kobayashi-Maskawa (CKM) matrix of the quark sector, the respective mixing matrix is sometimes denoted as the Pontecorvo-Maki-Nakagawa-Sakata (PMNS, or often only MNS) matrix. In the basis where the charged lepton
mass matrix is diagonal:

\begin{equation}
\nu_{i} = \sum_{\alpha} U_{\alpha i} \nu_{\alpha}
\label{eq:lc}
\end{equation}
$\nu_{i}$ are the mass eigenstates, $\nu_{\alpha}$ the flavour
eigenstates. The unitary matrix $U$ expressing the linear combination in eq.(\ref{eq:lc}) is the PMNS matrix (here we use Greek letters to clearly distinguish the flavour indices $\alpha$, $\beta$ from the mass indices $i$, $j$). With this it is easy to
understand how a specific flavour eigenstate can oscillate to a
different one as it propagates: it is composed of a linear combination
of mass eigenstates with masses $m_{i}$. The proportion of mass eigenstates will change during the propagation due to the phase factors $e^{-i m_{i} \tau}$ in the $\nu_{i}$ rest frame. In the laboratory frame, the phase factor becomes $e^{-i(E_{i}t-p_{i}L)}$ ($E_{i}$ and $p_{i}$ being the energy and momentum of $\nu_{i}$, $t$ and $L$ the time and position, all quantities in the laboratory frame). The neutrinos are very light, with $m_{i} \ll p_{i}$, so one can take $t \simeq L$ (natural units). Furthermore, considering that a specific neutrino flavour $\nu_{\alpha}$ is produced with definite momentum $p$, we have $E_{i} = \sqrt{p^2 + m_{i}^2} \simeq p + m_{i}^{2}/2p$ to good approximation. The phase factor becomes (approximately) $e^{-i(m_{i}^{2}/2p)L}$, and considering the average energy of the various mass eigenstates $E \simeq p$, we can obtain the formula for probability of flavour change from flavour state $\alpha$ into flavour state $\beta$ after propagation for a distance $L$ in the vacuum:

\begin{equation}
P_{\alpha \rightarrow \beta} = \left| \sum_{i} U_{\alpha i}^{\ast} U_{\beta i} e^{-i \frac{m_{i}^{2} L}{2 E}} \right|^{2}
\label{eq:probab}
\end{equation}
Eq.(\ref{eq:probab}) may be conveniently expressed as $P_{\alpha \rightarrow \beta}=\delta_{\alpha \beta}+Q_{\alpha \beta}$, with $Q_{\alpha \rightarrow \beta}$ being:
\begin{equation}
Q_{\alpha \rightarrow \beta} = - 4 \sum_{i>j} \mathrm{Re} \left( U_{\alpha i} ^{\ast} U_{\beta i}  U_{\alpha
  j} U_{\beta j}^{\ast}  \right) \sin ^{2} \left( \frac{\Delta
  m_{ij}^{2}L}{4 E} \right) +2 \mathrm{Im} \left( U_{\alpha i} ^{\ast} U_{\beta i}  U_{\alpha
  j} U_{\beta j}^{\ast}  \right) \sin \left( \frac{\Delta
  m_{ij}^{2}L}{2 E} \right)
\label{eq:Pab}
\end{equation}
$\mathrm{Re}$ stands for real part and $\mathrm{Im}$ for imaginary part.
The terms in eq.(\ref{eq:Pab}) clearly show
that the squared mass differences 
$\Delta m_{ij}^{2} \equiv m_{i}^{2} - m_{j}^{2}$ are
measurable from oscillation (although the overall mass scale isn't). For a more careful derivation or further details, see for example the original treatment in \cite{Kayser:1981ye}, or the neutrino mixing review in \cite{PDG} which includes extensive references.

A convenient summary of the neutrino oscillation data is given in
\cite{Maltoni:2004ei}. For ease of reference, we reproduce here the relevant table with the values (updated in June 2006 \cite{Maltoni:2004ei}).

\begin{table}[htb] \centering
      \begin{tabular}{|c||c|c|c|c|}
        \hline
        parameter & best fit & 2$\sigma$ & 3$\sigma$ & 4$\sigma$
        \\
        \hline
        $\Delta m^2_{21} [10^{-5} \mathrm{eV}]$
        & 7.9 & 7.3--8.5 & 7.1--8.9 & 6.8--9.3\\
        $\Delta m^2_{31} [10^{-3} \mathrm{eV}]$
        & 2.6 & 2.2--3.0 & 2.0--3.2 & 1.8--3.5\\
        $\sin^2\theta_{12}$
        & 0.30 & 0.26--0.36 & 0.24--0.40 & 0.22--0.44\\
        $\sin^2\theta_{23}$
        & 0.50 & 0.38--0.63 & 0.34--0.68 & 0.31--0.71 \\
        $\sin^2\theta_{13}$
        & 0.000 &  $\leq$ 0.025 & $\leq$ 0.040  & $\leq$ 0.058 \\
        \hline
\end{tabular}
\caption[Neutrino data: best-fit values, 2$\sigma$, 3$\sigma$, and 4$\sigma$ intervals.]{Neutrino data: best-fit values, 2$\sigma$, 3$\sigma$, and 4$\sigma$ intervals (from \protect\cite{Maltoni:2004ei}).}
\label{ta:data}
\end{table}
From \cite{Maltoni:2004ei}, the angles of table \ref{ta:data} refer to
the standard Particle Data Group \cite{PDG} parametrisation of
the unitary mixing matrix:

\begin{equation} \label{eq:mix}
    U=\left(
    \begin{array}{ccc}
        c_{12} c_{13}
        & s_{12} c_{13}
        & s_{13} e^{-i \delta_{13}} \\
        -s_{12} c_{23} -  c_{12} s_{23} s_{13} e^{i \delta_{13}}
        & c_{12} c_{23} - s_{12} s_{23} s_{13} e^{i \delta_{13}}
        & s_{23} c_{13} \\
        s_{12} s_{23} - c_{12} c_{23} s_{13} e^{i \delta_{13}}
        & -c_{12} s_{23} - s_{12} c_{23} s_{13} e^{i \delta_{13}}
        & c_{23} c_{13}
    \end{array} \right)
\end{equation}
$c_{ij} \equiv \cos\theta_{ij}$ and $s_{ij} \equiv \sin\theta_{ij}$
 (furthermore, $\theta_{12}$ is the solar angle
$\theta_{\odot}$, $\theta_{23}$ is the atmospheric angle $\theta_{@}$ and $\theta_{13}$ is the reactor angle). $\delta_{13}$ is a CP-violating phase that hasn't been measured yet, and we didn't include here the Majorana phases (usually denoted as $\alpha_{1}$ and $\alpha_{2}$), which only have physical consequences if the neutrinos are Majorana.

It is important to note that the large angles of table \ref{ta:data} contrast with the small angles of the CKM matrix (the largest of which, the Cabibbo angle, has $\sin (\theta_{C}) < 0.23 $). The angles are very close to (and consistent with) the special \TBM \ values \cite{Wolfenstein:1978uw,Harrison:1999cf,Harrison:2002er,Harrison:2002kp,Harrison:2003aw,Low:2003dz}:
\begin{equation}
s_{12_{TBM}}^{2}=\frac{1}{3}
\end{equation}
\begin{equation}
s_{23_{TBM}}^{2}=\frac{1}{2}
\end{equation}
\begin{equation}
s_{13_{TBM}}^{2}=0
\end{equation}

\begin{figure}[htb]
\centering
      \resizebox{12 cm}{!}{
        \includegraphics{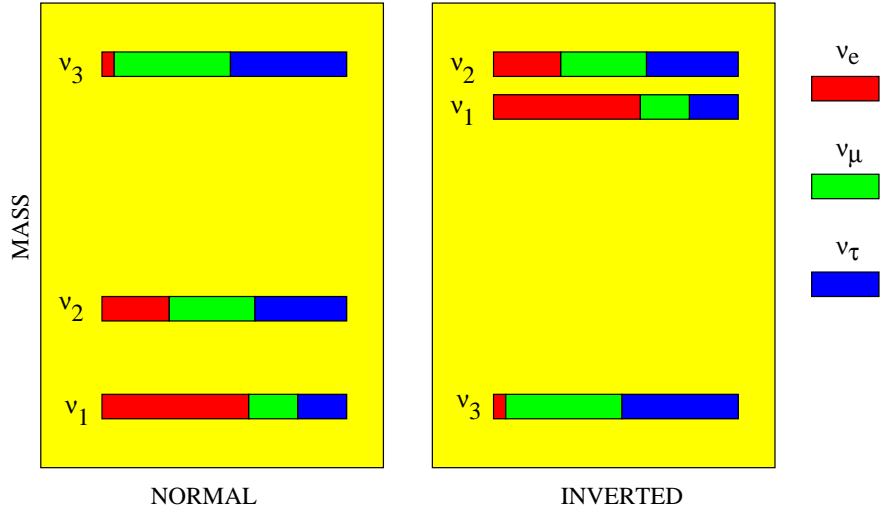}}
\caption[Neutrino mixing summary.]{Neutrino mixing summary (from \protect\cite{Smirnov:2007pw}).}
\label{fig:TBM}
\end{figure}

The experimental data is conveniently displayed in a graphical manner by use of coloured or shaded bars, as in \cite{Smirnov:1996jt}, \cite{Mena:2003ug}.
Figure \ref{fig:TBM} features the two possible mass hierarchies (due to the ambiguity in the sign of the atmospheric squared mass difference), and shows the peculiar situation described by \TBM \ quite clearly: one neutrino mass eigenstate ($\nu_{3}$) is approximately comprised of equal parts $\nu_{\mu}$ and $\nu_{\tau}$, and another ($\nu_{2}$) is approximately equal parts of all three flavour eigenstates.

\section{Family symmetry review \label{sec:fsreview}}

Including $\nu^{c}$, $U(3)^{6}$ is the largest \FS \ that commutes with the \SM \ gauge group in the absence of the mass terms, the maximum possible symmetry preserved by the kinetic terms. The $U(3)^{6}$ corresponds to one independent $U(3)_{f}$ \FS \ for each of the listed families: the \LH \ quark doublet $Q$, the quark singlets $u^{c}$, $d^{c}$; the \LH \ lepton doublet $L$, and the lepton singlets $\nu^{c}$, $e^{c}$. If the \FS \ is to commute also with an underlying \GUT , then the maximum possible \FS \ is reduced. With an $SO(10)$ \GUT , all the families of the \SM \ belong to the same fundamental representation, thus reducing the \FS \ that commutes with the gauge group to a single $U(3)_{f}$.

Any \FS \ that is introduced has to be broken in order to be consistent with the observed \FMM \ - the breaking is thus required by the Yukawa Lagrangian (eq.(\ref{eq:L_Y})). We designate the fields that break the \FS \ as ``flavons'', due to their connection to flavour \footnote{We use
  ``flavon'' consistently in all chapters, noting that these fields are often referred to instead as familons.}.
Because the \FS \ is not broken by the gauge interactions, which treat each member of the same family equally, the gauge interactions may be said to be ``family blind''.

In subsection \ref{sub:revs} we present an incomplete list of recent
family symmetry papers, and to conclude the \FS \ review, we provide a very simple \FS \ example in subsection \ref{sub:U1}. We use the example to
illustrate the general framework, in particular showing that the breaking of the \FS \ leads directly to the \FMM . The example is also used to motivate  the \FN \ mechanism \cite{fnielsen}.

\subsection{Recent models \label{sub:revs}}

With the intention of giving a flavour of the topics related to \FSs , we present a short (and incomplete) list of thirteen recent papers containing ``\FS '' in their title.

We start the review of recent models with the discussion of the paper about models based in Abelian groups. The remaining are all about models based on non-Abelian groups, the proportion perhaps hinting that non-Abelian groups are currently more in favour than Abelian ones. \cite{Law:2007jk} studies properties of the models of \cite{Low:2005yc}, based on the Abelian groups $Z_{2}$ and $Z_{4}$.

We now turn to the papers based on non-Abelian \FSs , starting with the continuous models. \cite{Appelquist:2006qq, Appelquist:2006xd} both use $SU(3)_1 \otimes SU(3)_2$: \cite{Appelquist:2006xd} is a small extension of \cite{Appelquist:2006qq} (which in turn extended \cite{Appelquist:2006ag} in order to include leptons).
We have then the $SO(3)$ model \cite{King:2006me}, a model with emphasis put
in having simple Yukawa operators (each of the leading operators have just a
single flavon insertion) - the simplicity then allows the study of details of the messenger sector.

\cite{Ma:2006sk, Ma:2006wm, King:2006np} rely on an $A_4$ \FS . $A_{4}$ has been widely used as a \FS : it is a very small subgroup of $SU(3)$ that has a ``triplet'' representation, very convenient to explain three families ($A_4$ is featured in section \ref{sec:D12} under the guise of $\Delta(12)$, so refer to that section for more details).

The remaining papers are based on discrete non-Abelian \FSs \ that are not so commonly used and may be less familiar. We won't go into details of the groups used in each paper, although we note that they are all (directly or indirectly) subgroups of $SU(3)$. Two of the papers, \cite{Ma:2006ht,  Ma:2007ku}, use $\Sigma(81)$, a subgroup of $SU(3)$: \cite{Ma:2007ku} extends \cite{Ma:2006ht} in order to obtain \TBM . We have also \cite{Kajiyama:2006ww} based on $D(6)$, a subgroup of $SO(3)$, and \cite{Kajiyama:2007pr} based on $Q_6$, a subgroup of $SU(2)$ (and as both $SO(3)$ and $SU(2)$ are continuous subgroups of $SU(3)$, it then follows that $D(6)$ and $Q_6$ are discrete subgroups of $SU(3)$).

For the discussion of the last two non-Abelian \FS \ models \cite{Ivo3, Luhn:2007sy}, we add again the $A_4$ model of \cite{King:2006np}, as it shares a common feature with them.
\cite{Ivo3} is the original paper presenting the $\Delta(27)$ \FS \ model discussed in full detail in section \ref{sec:D27}. \cite{King:2006np} uses its $A_4$ \FS \ as a subgroup of $SO(3)$ similarly to how $\Delta(27)$ is implemented as a subgroup of $SU(3)_{f}$ in \cite{Ivo3} (see chapter \ref{ch:fs} and chapter \ref{ch:fsd}). In turn, \cite{Luhn:2007sy} uses as \FS \ $Z_7 \rtimes Z_3$ - a discrete subgroup of $SU(3)$ only slightly smaller than $\Delta(27)$ (in terms of number of elements - 21 as opposed to 27 - and consequently slightly larger than $\Delta(27)$ in number of allowed invariants).
The link connecting the papers is their \VEV \ alignment mechanism - both \cite{King:2006np} and \cite{Luhn:2007sy} rely on a \VEV \ alignment mechanism similar to the one introduced in \cite{Ivo3} - using quartic terms arising from $D$-terms (see subsection \ref{sec:VEVs and masses} for details). Further, as the group $Z_7 \rtimes Z_3$ shares the same relevant allowed invariants  with $\Delta(27)$, the \VEV \ alignment mechanism of the two models can actually be identical.

\subsection{A simple $U(1)_{f}$ toy model \label{sub:U1}}

In this supersymmetric toy model we introduce a $U(1)_{f}$ \FS \ commuting with the \SM \ gauge group. We introduce only one flavon field $\phi$, charged under $U(1)_{f}$. $\phi$ will acquire a non-vanishing \VEV \ $\langle \phi \rangle$ (through an unspecified mechanism), thus breaking the \FS .

We will concentrate solely on the three generations of down-type quarks (for simplicity). The three generations of down quarks are represented as $d_{i}$ for the \LH \ fields and as $d^{c}_{i}$ for the \RH \ fields ($i$ is the family index, so for example $d_{2}$ is the \LH \ strange quark).

The charges under the \FS \ ($U(1)_{f}$) of the Higgs giving mass to the down quarks ($H_{d}$), of the down quarks ($d_{i}$, $d^{c}_{i}$) and of the flavon ($\phi$) are shown in table \ref{Ta:U1}.

\begin{table}[htb] \centering
\begin{tabular}{|c||c|}
\hline
Field & $U(1)_{f}$ \\ \hline \hline
$H_{d}$ & $\mathbf{0}$ \\
$\phi$ & $-\mathbf{1}$ \\ \hline
$d_{1}$ & $\mathbf{2}$ \\
$d_{2}$ & $\mathbf{1}$ \\
$d_{3}$ & $\mathbf{0}$ \\ \hline
$d^{c}_{1}$ & $\mathbf{2}$ \\
$d^{c}_{2}$ & $\mathbf{1}$ \\
$d^{c}_{3}$ & $\mathbf{0}$ \\ \hline
\end{tabular}
\caption{$U(1)_{f}$ charge assignments.} \label{Ta:U1}
\end{table}

Due to the \FS , nearly all the gauge invariant mass
terms require inclusion of some power of the flavon field $\phi$ (this is comparable to how all Dirac mass terms must include $H$ to be invariant under $SU(2)_{L}$). It is straightforward to obtain the effective Yukawa superpotential for the down type quarks:

\begin{equation}
P_{d} = d_{3} d^{c}_{3} H
+ \frac{\phi}{M} d_{3} d^{c}_{2} H
+ \frac{\phi}{M} d_{2} d^{c}_{3} H
\label{eq:u1_23}
\end{equation}
\begin{equation}
+ \left( \frac{\phi}{M} \right)^{2} d_{2} d^{c}_{2} H
+ \left( \frac{\phi}{M} \right)^{3} d_{2} d^{c}_{1} H
+ \left( \frac{\phi}{M} \right)^{3} d_{1} d^{c}_{2} H
\end{equation}
\begin{equation}
+ \left( \frac{\phi}{M} \right)^{4} d_{1} d^{c}_{1} H
+ \left( \frac{\phi}{M} \right)^{2} d_{1} d^{c}_{3} H
+ \left( \frac{\phi}{M} \right)^{2} d_{3} d^{c}_{1} H
\end{equation}
The first term of eq.(\ref{eq:u1_23}) generates $(M^{d})_{33}$ -
effectively the bottom mass (there are sub-leading corrections from the
other terms). When $\phi$ acquires its \VEV \ $\langle \phi \rangle$, the
remaining entries of $M^{d}$ are filled out by increasing powers
of the ratio $\frac{\langle \phi \rangle}{M} = \epsilon$. Note the presence of the yet unspecified mass scale $M$. For now we simply take $M$ to be some large mass scale such that $\epsilon$ is a small parameter. The justification lies in considering $M$ to be the mass of a \FN \ messenger as described in more detail in subsection \ref{sub:fnreview}. With small $\epsilon$, we can generate a
strong hierarchy in $M^{d}$:

\begin{equation}
M^{d} \propto \left( 
\begin{array}{ccc}
\epsilon^{4} & \epsilon^{3} & \epsilon^{2} \\ 
\epsilon^{3} & \epsilon^{2} & \epsilon \\ 
\epsilon^{2} & \epsilon & 1%
\end{array}%
\right)  \label{eq:u1 d_matrix}
\end{equation}%

Although very simple (and not phenomenologically viable), this model
clearly illustrates the philosophy of using \FSs \ together
with the \FN \ mechanism (see subsection \ref{sub:fnreview}) to control the \FMM \ (and obtain hierarchies) through expansions of small parameters: the ratio of the flavon \VEV s with superheavy messenger masses (of what will be identified as \FN \
messenger fields in subsection \ref{sub:fnreview}).

\section{Mass generation mechanisms \label{sec:mass mechs}}

\subsection{\FN \ mechanism \label{sub:fnreview}}

The \FN \ mechanism allows the generation of masses through higher
order tree-level diagrams involving superheavy fields: the \FN \ messenger fields \cite{fnielsen}.

A simple example of the \FN \ mechanism is the diagram in figure
\ref{fig:fnsimple}, where the arrows denote the chirality of the fields (like in other diagrams).

\begin{figure}[htb]
\begin{center}
\fcolorbox{white}{white}{
  \begin{picture}(195,121) (135,-179)
    \SetWidth{0.5}
    \SetColor{Black}
    \Text(150,-170)[lb]{\Large{\Black{$\psi$}}}
    \Text(180,-74)[lb]{\Large{\Black{$\langle H \rangle$}}}
    \SetWidth{0.5}
    \COval(195,-89)(2.83,2.83)(45.0){Black}{White}\Line(193.59,-90.41)(196.41,-87.59)\Line(193.59,-87.59)(196.41,-90.41)
    \Vertex(195,-134){2.83}
    \ArrowLine(150,-134)(195,-134)
    \DashArrowLine(195,-89)(195,-134){10}
    \ArrowLine(330,-134)(285,-134)
    \Vertex(285,-134){2.83}
    \COval(285,-89)(2.83,2.83)(45.0){Black}{White}\Line(283.59,-90.41)(286.41,-87.59)\Line(283.59,-87.59)(286.41,-90.41)
    \Line(238,-136)(242,-132)\Line(238,-132)(242,-136)
    \Text(300,-170)[lb]{\Large{\Black{$\psi^{c}$}}}
    \DashArrowLine(285,-89)(285,-134){10}
    \Text(275,-74)[lb]{\Large{\Black{$\langle \phi \rangle$}}}
    \ArrowLine(240,-134)(195,-134)
    \ArrowLine(240,-134)(285,-134)
    \Text(255,-170)[lb]{\Large{\Black{$A$}}}
    \Text(210,-170)[lb]{\Large{\Black{$\bar{A}$}}}
  \end{picture}
}
\caption{Simple \FN \ diagram.}
\label{fig:fnsimple}
\end{center}
\end{figure}

The superheavy fields $\bar{A}$, $A$ are the \FN \ messenger fields, and have a mass term $M_{A} \bar{A} A$ (corresponding to the mass insertion represented by ``$\times$'' in figure \ref{fig:fnsimple}).
Note that the messengers must have appropriate \SM \ and \FS \ charge assignments - namely, it is relevant to consider the placement of the $H$ insertion (as it carries $SU(2)_{L}$ charge) and likewise $\phi$ will carry family charge. Consider specifically the generation of $M^{d}_{23}$ in eq.(\ref{eq:u1 d_matrix}): it can proceed precisely through a simple \FN \ diagram with just one flavon insertion, with $d_{2}$ and $d_{3}^{c}$ as the external fields. If the ordering of $H$ and $\phi$ are as displayed in figure \ref{fig:fnsimple}, then $A$ must have $U(1)_{f}$ charge $+1$ (and respectively, $\bar{A}$ has $-1$).

When the messengers are integrated out, the superpotential term respective to figure \ref{fig:fnsimple} becomes:
\begin{equation}
P = \frac{\langle \phi \rangle}{M_{A}} \psi \psi^{c} \langle H \rangle
= m_{\psi} \psi \psi^{c}
\end{equation}
The effective mass is $m_{\psi} \equiv \frac{\langle \phi \rangle}{M_{A}}
\langle H \rangle$.

A more general diagram is displayed in figure \ref{fig:fngen},
featuring more than one superheavy mass insertion ($\bar{A}$ and $A$, $\bar{B}$ and $B$, $\bar{C}$ and $C$ with
mass terms $M_{A} \bar{A} A$, $M_{B} \bar{B} B$, $M_{C} \bar{C} C$ respectively).

\begin{figure}[htb]
\begin{center}
\fcolorbox{white}{white}{
 \begin{picture}(390,106) (150,-179)
    \SetWidth{0.5}
    \SetColor{Black}
    \COval(210,-104)(2.83,2.83)(45.0){Black}{White}\Line(208.59,-105.41)(211.41,-102.59)\Line(208.59,-102.59)(211.41,-105.41)
    \Text(195,-89)[lb]{\Large{\Black{$\langle H \rangle$}}}
    \Vertex(300,-149){2.83}
    \Vertex(210,-149){2.83}
    \DashArrowLine(210,-104)(210,-149){10}
    \Line(253,-151)(257,-147)\Line(253,-147)(257,-151)
    \Line(343,-151)(347,-147)\Line(343,-147)(347,-151)
    \Vertex(390,-149){2.83}
    \Vertex(480,-149){2.83}
    \Line(433,-151)(437,-147)\Line(433,-147)(437,-151)
    \ArrowLine(255,-149)(300,-149)
    \ArrowLine(255,-149)(210,-149)
    \ArrowLine(345,-149)(390,-149)
    \ArrowLine(345,-149)(300,-149)
    \ArrowLine(435,-149)(390,-149)
    \ArrowLine(435,-149)(480,-149)
    \DashArrowLine(300,-104)(300,-149){10}
    \COval(300,-104)(2.83,2.83)(45.0){Black}{White}\Line(298.59,-105.41)(301.41,-102.59)\Line(298.59,-102.59)(301.41,-105.41)
    \COval(390,-104)(2.83,2.83)(45.0){Black}{White}\Line(388.59,-105.41)(391.41,-102.59)\Line(388.59,-102.59)(391.41,-105.41)
    \COval(480,-104)(2.83,2.83)(45.0){Black}{White}\Line(478.59,-105.41)(481.41,-102.59)\Line(478.59,-102.59)(481.41,-105.41)
    \DashArrowLine(390,-104)(390,-149){10}
    \DashArrowLine(480,-104)(480,-149){10}
    \Text(465,-89)[lb]{\Large{\Black{$\langle \phi_{3} \rangle$}}}
    \Text(375,-89)[lb]{\Large{\Black{$\langle \phi_{2} \rangle$}}}
    \Text(285,-89)[lb]{\Large{\Black{$\langle \phi_{1} \rangle$}}}
    \ArrowLine(150,-149)(210,-149)
    \ArrowLine(540,-149)(480,-149)
    \Text(150,-179)[lb]{\Large{\Black{$\psi$}}}
    \Text(510,-179)[lb]{\Large{\Black{$\psi^{c}$}}}
    \Text(270,-179)[lb]{\Large{\Black{$A$}}}
    \Text(360,-179)[lb]{\Large{\Black{$B$}}}
    \Text(450,-179)[lb]{\Large{\Black{$C$}}}
    \Text(405,-179)[lb]{\Large{\Black{$\bar{C}$}}}
    \Text(225,-179)[lb]{\Large{\Black{$\bar{A}$}}}
    \Text(315,-179)[lb]{\Large{\Black{$\bar{B}$}}}
  \end{picture}
}
\caption{Generic \FN \ diagram.}
\label{fig:fngen}
\end{center}
\end{figure}

The generalisation is simple, but one should note again that the charges of the messengers must be such that the diagram is allowed.
In order to consider another specific case, consider for simplicity the following $U(1)_{f}$ charge assignments: $\phi_{1}$ has family charge $-1$, $\phi_{2}$ has $-2$ and $\phi_{3}$ has $+3$, with all other non-messenger fields neutral (note this is not the toy model discussed in subsection \ref{sub:U1}). With the ordering of figure \ref{fig:fngen}, the charges of the messengers would be $0$, $0$, $1$, $-1$, $+3$ and $-3$ respectively for $\bar{A}$, $A$, $\bar{B}$, $B$, $\bar{C}$ and $C$. The effective superpotential term is invariant as required:
\begin{equation}
P = \frac{
\langle \phi_{1} \rangle
\langle \phi_{2} \rangle
\langle \phi_{3} \rangle}
{M_{A} M_{B} M_{C}} \psi \psi^{c} \langle H \rangle
\end{equation} 
Corresponding to an effective mass $m_{\psi} \equiv \frac{\langle \phi_{1} \rangle
  \langle \phi_{2} \rangle \langle \phi_{3} \rangle}{M_{A} M_{B} M_{C}}
\langle H \rangle$.

\subsection{Seesaw mechanism \label{sub:ssm}}

\begin{figure}[htb]
\begin{center}
\fcolorbox{white}{white}{
\begin{picture}(225,121) (240,-179)
    \SetWidth{0.5}
    \SetColor{Black}
    \Text(285,-74)[lb]{\Large{\Black{$\langle H \rangle$}}}
    \Text(375,-74)[lb]{\Large{\Black{$\langle H \rangle$}}}
    \Text(315,-179)[lb]{\Large{\Black{$\nu^c$}}}
    \Text(360,-179)[lb]{\Large{\Black{$\nu^c$}}}
    \SetWidth{0.5}
    \Vertex(300,-149){2.83}
    \Line(343,-151)(347,-147)\Line(343,-147)(347,-151)
    \Vertex(390,-149){2.83}
    \DashArrowLine(300,-89)(300,-149){10}
    \DashArrowLine(390,-89)(390,-149){10}
    \ArrowLine(345,-149)(300,-149)
    \ArrowLine(345,-149)(390,-149)
    \COval(300,-89)(2.83,2.83)(45.0){Black}{White}\Line(298.59,-90.41)(301.41,-87.59)\Line(298.59,-87.59)(301.41,-90.41)
    \COval(390,-89)(2.83,2.83)(45.0){Black}{White}\Line(388.59,-90.41)(391.41,-87.59)\Line(388.59,-87.59)(391.41,-90.41)
    \ArrowLine(240,-149)(300,-149)
    \ArrowLine(450,-149)(390,-149)
    \Text(240,-179)[lb]{\Large{\Black{$\nu$}}}
    \Text(435,-179)[lb]{\Large{\Black{$\nu$}}}
  \end{picture}
}
\caption{Type I seesaw diagram.}
\label{fig:seesaw}
\end{center}
\end{figure}
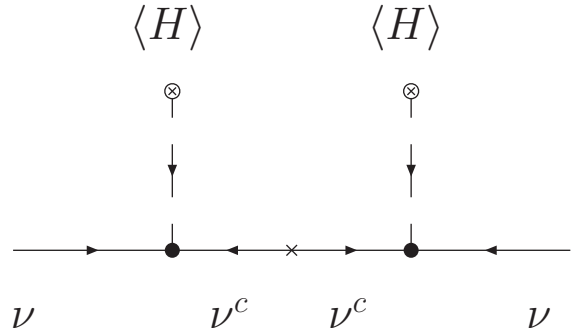

The seesaw mechanism \cite{Minkowski:1977sc,Yanagida:1979as,GellMann:1980vs,Glashow:1979nm,Mohapatra:1979ia} is similar to the \FN \ mechanism. It generates
effective masses for the light neutrinos by integrating out heavy
neutrino states. Figure \ref{fig:seesaw} is a typical type I seesaw diagram (with the ``$\times$'' in the $\nu^{c}$ propagator denoting the Majorana mass insertion).

The \RH \ neutrino masses (in general, eigenvalues
of a mass matrix $M_{M}$) are naturally expected to be much larger than the neutrino Dirac masses (eigenvalues of
$M_{D}^{\nu} \propto Y^{\nu}$ of eq.(\ref{eq:L_Y})). The reason for this is that $\nu^{c}$ are singlets of the \SM , and the mass term $M_{M} \nu^{c} \nu^{c}$ is automatically invariant. In other words, $M_{M}$ is not protected by the \SM \ gauge group, unlike $M_{D}^{\nu}$ which is only non-vanishing after $SU(2)_{L}$ is spontaneously broken by $\langle H \rangle$ (see the Yukawa Lagrangian in eq.(\ref{eq:L_Y})). If indeed the Majorana masses have large values, that enables us to integrate out the $\nu^{c}$ fields and obtain the formula for the effective light neutrino masses approximately given by \cite{Mohapatra:2005wg, Mohapatra:2006gs}:

\begin{equation}
M_{\nu }= -(M_{D}^{\nu}) (M_{M}^{-1}) (M_{D}^{\nu})^{T}
\label{eq:-seesaw}
\end{equation}

The minus sign can be absorbed by redefinition of the fields (it is however relevant if type II seesaw \cite{Lazarides:1980nt, Mohapatra:1980yp} is present).

The general structure of eq.(\ref{eq:-seesaw}) can be obtained by considering the
simpler one family case. Doing so we clearly obtain the minus sign (although we won't get the transpose, for obvious reasons). We introduce only one \LH \ neutrino, $\nu$ and one \RH \
neutrino, $\nu^{c}$. Due to the symmetry of the \SM, the $M_{L} \nu \nu$
mass term is not invariant. However, we can have in general a Dirac
mass term $M_{D}^{\nu}\nu \nu^{c}$ as well as a Majorana mass term
$M_{R} \nu^{c} \nu^{c}$. We can express this as a $2 \times 2$
neutrino mass matrix $M^{\nu}$:

\begin{equation}
L = \left( \begin{array}{cc} \nu^{\dagger} & \nu^{c \dagger}
  \end{array} \right)
M^{\nu}
\left( \begin{array}{c} \nu \\ \nu^{c} \end{array} \right)
\end{equation}

\begin{equation}
M^{\nu} = \left( \begin{array}{cc}
0 & M_{D}^{\nu} \\
M_{D}^{\nu} & M_{R}
\end{array} \right)
\end{equation} 
If $M_{D}^{\nu} \ll M_{R}$ (a natural condition, as $M_{R}$ isn't protected by $SU(2)_{L}$), one can readily identify the approximate
eigenvalues of $M^{\nu}$ by using its matrix invariants, namely the
determinant $Det \left( M^{\nu} \right) = - \left( M_{D}^{\nu} \right) ^{2}$ and the trace
$Tr \left( M^{\nu} \right) = M_{R}$. The largest eigenvalue must then be
approximately $M_{R}$ (keeping the 
trace invariant), and the smallest eigenvalue must then be $-\left( M_{D}^{\nu} \right) ^{2}/M_{R}$ to
preserve the determinant: this is precisely the result one gets from
eq.(\ref{eq:-seesaw}) applied to the particular case of one generation
of each type of neutrino. The exact eigenvalues in this one generation
case are:
\begin{equation}
M^{\nu}_{+} = \frac{1}{2} \left( M_{R} + \sqrt{M_{R}^{2} + 4 \left( M_{D}^{\nu} \right) ^{2}} \right)
\end{equation}
\begin{equation}
M^{\nu}_{-} = \frac{1}{2} \left( M_{R} - \sqrt{M_{R}^{2} + 4 \left( M_{D}^{\nu} \right) ^{2}} \right)
\end{equation}
By expanding the square root we verify that the approximate values
obtained based on the matrix invariants are indeed correct to leading order.
Generalising to three generations of each type of neutrino, we obtain eq.(\ref{eq:-seesaw}), where $M_{D}^{\nu}$ and $M_{R}$ are now $3 \times 3$ matrices.

We can also intuitively understand eq.(\ref{eq:-seesaw}) by inspection of figure \ref{fig:seesaw}, now with the three generations and using the appropriate mass matrices. When $H$ acquires the \VEV \ $\langle H \rangle = h$, the vertex on the left becomes a neutrino Dirac mass matrix proportional to $h$, corresponding to $M_{D}^{\nu}$: the mass term is $M_{D}^{\nu} \nu \nu^{c}$, suppressing the family indices. The mass insertion is the Majorana mass matrix $M_{M}$ of the $\nu^{c}$ fields: the mass term is $M_{M} \nu^{c} \nu^{c}$, suppressing the family indices. The \RH \ neutrino propagates in the internal line and by integrating it out we get the inverse matrix, $M_{M}^{-1}$. Finally the vertex on the right likewise corresponds to $\left( M_{D}^{\nu} \right) ^{T}$ (the transpose due to the inverted ordering), and the resulting $M_{\nu}$ mass matrix for the effective neutrinos is proportional to $h^{2}$.

\begin{figure}[htb]
\begin{center}
\fcolorbox{white}{white}{
  \begin{picture}(165,136) (225,-149)
    \SetWidth{0.5}
    \SetColor{Black}
    \ArrowLine(225,-119)(300,-119)
    \Vertex(300,-119){2.83}
    \ArrowLine(375,-119)(300,-119)
    \ArrowLine(300,-74)(300,-119)
    \DashArrowLine(300,-74)(270,-44){10}
    \DashArrowLine(300,-74)(330,-44){10}
    \COval(270,-44)(2.83,2.83)(45.0){Black}{White}\Line(268.59,-45.41)(271.41,-42.59)\Line(268.59,-42.59)(271.41,-45.41)
    \COval(330,-44)(2.83,2.83)(45.0){Black}{White}\Line(328.59,-45.41)(331.41,-42.59)\Line(328.59,-42.59)(331.41,-45.41)
    \Text(250,-29)[lb]{\Large{\Black{$\langle H \rangle$}}}
    \Text(320,-29)[lb]{\Large{\Black{$\langle H \rangle$}}}
    \Text(225,-149)[lb]{\Large{\Black{$\nu$}}}
    \Text(360,-149)[lb]{\Large{\Black{$\nu$}}}
    \Text(315,-104)[lb]{\Large{\Black{$\Delta$}}}
  \end{picture}
}
\caption{Type II seesaw diagram.}
\label{fig:seesaw2}
\end{center}
\end{figure}
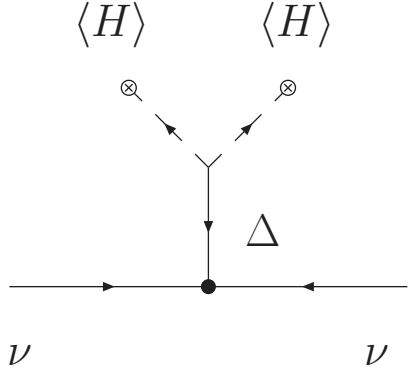

Although we make no use of it in the following chapters, we note for completeness the existence of the type II seesaw mechanism \cite{Lazarides:1980nt, Mohapatra:1980yp}, requiring an $SU(2)_{L}$ triplet Higgs $\Delta$, of which a typical diagram is shown in figure \ref{fig:seesaw2}.

If the type II seesaw is present in a model and the contribution can't be neglected, an extra term $M_L$ must be added to the seesaw formula in eq.(\ref{eq:-seesaw}):

\begin{equation}
M_{\nu }= M_{L} - (M_{D}^{\nu}) (M_{M}^{-1}) (M_{D}^{\nu})^{T}
\label{eq:-seesaw2}
\end{equation}
Note that now the minus sign is relevant, and indeed due to the presence of $M_{L}$ one can't freely redefine the fields any longer (the relative sign between type I and type II terms is important).
$M_{L}$ occupies the formerly vanishing top left quadrant: the left-left quadrant of the neutrino matrix $M^{\nu}$ (hence the subscript $L$ in $M_{L}$). Again taking the simpler example of one family of each neutrino type (\LH \ and \RH ), $M_{L}$ occupies the $11$ entry of the $2 \times 2$ matrix (where there used to be a zero): 

\begin{equation}
M^{\nu} = \left( \begin{array}{cc}
M_{L} & M_{D}^{\nu} \\
M_{D}^{\nu} & M_{R}
\end{array} \right)
\end{equation}

Refer for example to \cite{Mohapatra:2005wg, Mohapatra:2006gs} for detailed reviews of neutrino physics.

\chapter{Continuous \FSs \label{ch:fs}}

\section{Framework and outline \label{sec:fsintro}}

As discussed in section \ref{sec:mot},
explaining the observed pattern of quark and lepton masses and
mixing angles remains a central issue in our attempt to construct a theory
beyond the \SM. Perhaps the most conservative possible
explanation is that the symmetry of the \SM \ is extended to
include a \FS \ which orders the Yukawa
couplings responsible for the mass matrix structure. In this chapter,
we consider this possibility, presenting as example a specific
model with a continuous $SU(3)_{f}$ \FS \ \cite{Ivo1}. In this section we begin to establish the framework introduced in \cite{Ivo1}, used not only in the continuous \FS \ model of this chapter, but also used for models based in
discrete \FSs \ (presented in chapter \ref{ch:fsd}).

If one restricts the discussion to the quark sector it is possible to build
quite elegant examples involving a spontaneously broken \FS \
which generates the observed hierarchical structure of quark masses and mixing
angles. However, attempts to extend this to the leptons has proved very
difficult, mainly because the large mixing angles needed to explain neutrino
oscillation contrast with the small mixing angles observed in the
quark sector. As established in subsection \ref{sub:summary}, the present experimental values for lepton mixing are consistent and actually well described by
the Harrison, Perkins and Scott ``\TBM''
scheme \cite{Wolfenstein:1978uw,Harrison:1999cf,Harrison:2002er,Harrison:2002kp,Harrison:2003aw,Low:2003dz}
in which the atmospheric neutrino mixing angle is maximal
($\sin ^{2}(\theta_{@}) = 1/2)$ and the solar neutrino mixing is
``tri-maximal'' ($\sin ^{2}(\theta_{\odot}) = 1/3$), which corresponds
to the PMNS leptonic mixing matrix taking
the following special form:

\begin{equation}
U_{PMNS}\propto \left[ 
\begin{array}{ccc}
-\sqrt{\frac{2}{6}} & \sqrt{\frac{1}{3}} & 0 \\ 
\sqrt{\frac{1}{6}} & \sqrt{\frac{1}{3}} & \sqrt{\frac{1}{2}} \\ 
\sqrt{\frac{1}{6}} & \sqrt{\frac{1}{3}} & -\sqrt{\frac{1}{2}}%
\end{array}%
\right]  \label{eq:HPS}
\end{equation}

If the mixing is indeed very
close to \TBM, it represents a strong indication that an
existing \FS \ should be non-Abelian to be able to
relate the magnitude of the Yukawa couplings of different families
(something an Abelian symmetry cannot do). Several models based on 
non-Abelian symmetries have been constructed that account for
this structure of leptonic mixing (examples include
\cite{Grimus:2005rf,Ma:2005qf,Altarelli:2005yx,Haba:2006dz,Mohapatra:2006pu}).
It is possible to construct models in a different class, where the
underlying family symmetry provides a full description of 
the complete fermionic structure (e.g. \cite{Babu:2004tn,King:2005bj}) -  see
\cite{Mohapatra:2005wg,Mohapatra:2006gs} or \cite{King:2003jb,Zee:2005ut, Ma:2007ia, Altarelli:2007cd} for
review papers with extensive references of both types of models. In models describing not just the leptons, the \FS \ explains also why the
quarks have a strongly 
hierarchical structure with small mixing (in contrast to the large leptonic mixing),
and the Yukawa coupling
matrices at the \GUT \ scale take the form given by fits to the data \cite{Roberts:2001zy, Mario}:

\begin{equation}
   Y^u \propto \left( \begin{array}{ccc} 0 & i\,\epsilon_u^3 & \epsilon_u^3 \cr
                 \cdot & \epsilon_u^2 & \epsilon_u^2 \cr
                 \cdot & \cdot &  1   \end{array}
                 \right)
\label{eq:Yu2}
   \end{equation}

     \begin{equation}
   Y^d \propto \left( \begin{array}{ccc} 0 & 1.7\epsilon_d^3 & (0.8)e^{-i\,(45)^o}
\epsilon_d^3 \cr
                 \cdot & \epsilon_d^2 & {(2.1)} \epsilon_d^2 \cr
                 \cdot & \cdot &  1   \end{array}
                 \right)
\label{eq:Yd2}
  \end{equation}
The expansion parameters are:

\begin{equation}
  \epsilon_d \simeq {0.13}, \epsilon_u
  \simeq 0.048
 \label{eq:u,d eps2}
   \end{equation}
Following \cite{Mario}, we represent the Yukawa matrices in the $\overline{MS}$ scheme below the $Z^{0}$ mass $M_{Z}$, and in the $\overline{DR}$ scheme above $M_{Z}$.

In this chapter, we consider in detail 
how the hierarchical quark structures together with lepton
\TBM \ can emerge in theories with an $SU(3)_{f}$ \FS \ (and in chapter \ref{ch:fsd}, how those can emerge in theories with non-Abelian discrete subgroups of $SU(3)_{f}$).

The use of $SU(3)_{f}$ is of particular
interest: $U(3)$ is the largest \FS \ that commutes
with $SO(10)$ and thus one can fit the \FS \ group together
with promising \GUT \ 
extensions of the \SM .
We consider this to be a desirable feature of a complete model of
quark and lepton masses and 
mixing angles, and choose to include an $SO(10)$ symmetry, aiming to preserve the respective phenomenologically
successful \GUT \ relations between quark and 
lepton masses. To do so, we require that the models be consistent with
the underlying
unified structure, either at the field theory level or at the level of the
superstring. 
This requirement is very constraining because all the \LH \
\footnote{As in chapter \ref{ch:intro}, we limit ourselves to
  referring to \LH \ states (here $\psi$) with their charge conjugates ($\psi^{c}$) transforming the same way as \RH \ states.}
members of a single family must have the same family charge (or
multiplet assignments). 
Despite these strong constraints, it is possible to build
models capable of describing all quark and lepton masses and mixing
angles, featuring \TBM \ in the lepton sector. Due to the \GUT, there is a
close relation between quark and lepton masses, and the \GJ \
relations between charged lepton and down-type quark masses
\cite{Georgi:1979df} are 
obtained. Likewise, the symmetric structure of the mass matrices is
motivated by the $SO(10)$
\GUT, reproducing the phenomenologically successful
\GST \ relation \cite{Gatto:1968ss} for the $(1,2)$ sector
mixing:
\begin{equation}
\left| V_{us} \right| = \left| \sqrt{\frac{m_{d}}{m_{s}}} - e^{i \phi_{1}}\sqrt{\frac{m_{u}}{m_{c}}} \right|
\label{eq:GST}
\end{equation}
The phase $\phi_{1}$ can be in some cases a good approximation to the CKM complex phase $\delta$, as discussed in \cite{Mario} (see also \cite{Roberts:2001zy}).We note for completeness that the corrections induced by the two schemes used ($\overline{MS}$ scheme below $M_{Z}$ and $\overline{DR}$ scheme above) is smaller than the accuracy of the \GST \ relation.

Finally, because the charged lepton structure is related to
down quark structure, it is possible to relate the quark mixing
angles with the predicted
deviations of leptonic mixing from the tri-bi-maximal (neutrino
mixing) values \cite{Plentinger:2005kx, Antusch:2005kw}.

By itself, such a unified implementation of $SU(3)_{f}$ does not
explain why the mixing angles are small in the quark sector while they
are large in the lepton sector.
If these contrasting observations are to be consistent with a
spontaneously broken \FS \ there must be a
mismatch between the symmetry breaking pattern in the quark and
charged lepton sectors and the symmetry breaking pattern in the
neutrino sector. In the quark sector and charged lepton sectors the
first stage of \FS \ breaking, $SU(3)_{f} \rightarrow
SU(2)_{f}$, generates the third generation masses while the remaining
masses are only generated by the second stage of breaking of the
residual $SU(2)_{f}$. However, in the neutrino mass sector the
dominating breaking must be rotated by $\pi /4$ relative to this, so
that an equal combination of $\nu_{\tau}$ and $\nu_{\mu}$ receives
mass at the first stage of mass generation. The subsequent breaking
generating the light masses must be misaligned by approximately
the tri-maximal angle in order to describe solar neutrino oscillation ($\sin ^{2}(\theta_{\odot}) = 1/3$).

One can obtain \TBM \ through the effective Lagrangian:
\begin{equation}
L_{\nu } = \lambda_{@} (\nu_{\mu}-\nu_{\tau})^{2} + \lambda_{\odot} (\nu_{e}+
\nu_{\mu}+\nu_{\tau})^{2}
\label{eq:effL}
\end{equation}
$\lambda_{@}$ and $\lambda_{\odot}$ need to have the appropriate
values (to account for the observed mass squared differences). This Lagrangian 
represents a normal hierarchy scheme of $m_1 < m_2 < m_3$ (see figure
\ref{fig:TBM}) where the lightest neutrino mass
eigenstate is approximately massless ($m_1 \simeq 0$, and thus its term is not shown
in eq.(\ref{eq:effL})). In this case, the masses are given to good approximation by taking the square root of the squared mass differences: $\lambda_{@,\odot} = \sqrt{\Delta m^{2}_{@,\odot}}$. The effective Lagrangian in eq.(\ref{eq:effL}) clearly shows the solar and atmospheric eigenstates feature \TBM.

The main difficulty in realising \TBM \ in this class of models with underlying unification is the need to explain why the
dominant breaking leading to the generation of third generation masses in
the quark sector is not also the dominant effect in the neutrino sector. At
first sight it appears quite unnatural. However, if neutrino masses are
generated by the seesaw mechanism (see subsection \ref{sub:ssm})
this difficulty can be 
overcome, and one can obtain neutrino \TBM \ as shown in
eq.(\ref{eq:effL}) even if
all quark and lepton Dirac mass matrices, including those of 
the neutrinos, have similar forms up to \GJ \ type factors. To see
this consider the form of the seesaw mechanism (neglecting here the
minus sign of eq.(\ref{eq:-seesaw})):
\begin{equation}
M_{\nu }= (M_{D}^{\nu}) (M_{M}^{-1}) (M_{D}^{\nu})^{T}
\label{eq:seesaw}
\end{equation}
As in subsection \ref{sub:ssm}, $M_{\nu }$ is the effective mass matrix for the light neutrino
states coupling $\nu$ to $\nu$,
$M_{D}^{\nu }$ is the Dirac mass matrix coupling $\nu $ to $\nu ^{c}$
and $M_{M}$ is the Majorana mass matrix coupling $\nu ^{c}$ to $\nu ^{c}$. We
consider the case where the Majorana mass matrix has an hierarchical
structure of the form:
\begin{equation}
M_{M}\simeq \left( {{\begin{array}{*{20}c} {M_1 } \hfill & \hfill & \hfill
\\ \hfill & {M_2 } \hfill & \hfill \\ \hfill & \hfill & {M_3 } \hfill \\
\end{array}}}\right) \quad M_{1}<<M_{2}<<M_{3}.
\end{equation}%
For a sufficiently strong hierarchy this gives rise to \SD \
\cite{King:1998jw,King:1999cm,King:1999mb,King:2002nf,Altarelli:1999dg,Smirnov:1993af}: the heaviest of the three light eigenstates gets its
mass from the exchange of the lightest (\RH) singlet neutrino
with Majorana mass $M_{1}$. In this case the contribution to the light
neutrino mass matrix of the field responsible for the dominant terms
in the Dirac mass matrix, $(M^{\nu}_{D})_{33}$, is suppressed by the
relative factor $M_{2}/M_{3}$ (or $M_{1}/M_{3}$) and can readily be
sub-dominant in the neutrino sector. The key point is that any
underlying quark-lepton symmetry is necessarily broken in the neutrino
sector due to the Majorana masses of the \RH \ neutrino states and,
through the seesaw mechanism, this feeds into the neutrino masses and the
lepton mixing angles. The simpler case where we take the Majorana
masses in the diagonal basis clearly illustrates how this effect can hide an
existing quark-lepton symmetry in the Dirac mass sector.

In the $SU(3)_{f}$ model detailed in this chapter (and in the models of chapter \ref{ch:fsd}) we implement this structure to achieve near
\TBM \ for the leptons. We consider only the case of supersymmetric
extensions of the \SM, because we rely on SUSY to keep the
hierarchy problem associated with a high-scale \GUT \ under control. Rather
than work with a complete $SO(10) \otimes G_{f}$ theory \footnote{Here $G_{f} = SU(3)_{f}$, but this equally applies for chapter \ref{ch:fsd}, where $G_{f}$ is instead a discrete subgroup of $SU(3)_{f}$.} (which, in a
string theory, may only be relevant above the string scale) we consider here
the case where the gauge symmetry is $G_{PS}\otimes G_{f}$ where $%
G_{PS}$ is the Pati-Salam group $G_{PS}\equiv SU(4)_{PS}\otimes
SU(2)_{L}\otimes SU(2)_{R}$ described in subsection \ref{sub:GUTs}. The $G_{f}$ representation assignments are
chosen in a way consistent with this being a subgroup of $SO(10)\otimes
G_{f}$. The construction of the models closely follows that of
\cite{King:2003rf}
and \cite{Ross:2004qn}, and we proceed by identifying simple auxiliary
symmetries capable of restricting the allowed Yukawa couplings to give
viable mass matrices for the quarks and leptons. For a particular model,
we need to pay particular attention to an analysis of the scalar
potential which is ultimately responsible for the vacuum alignment generating \TBM.

The Majorana mass matrices are generated by the lepton number violating
sector. To fulfil the suppression of the otherwise
dominant contribution arising from the Dirac masses,
$(M^{\nu}_{D})_{33}$, the dominant contribution to the 
Majorana mass matrix for the neutrinos is also going to be aligned
along the $3$rd direction (as is also the case for all the fermion Dirac
matrices).
We will show that by doing so, it is possible to achieve \TBM \ very closely,
with the small (but significant) deviations coming from the charged
lepton sector. This type of situation is described in some detail in
\cite{Plentinger:2005kx, Antusch:2005kw}.

In section \ref{sec:Symmetries}, we present the specific charge
assignments of the continuous model, continuing to outline the general strategy that is implemented not only in the continuous model featured in this chapter but also in the discrete models of chapter \ref{ch:fsd}.
The respective spontaneous symmetry breaking discussion is presented in subsection \ref{sec:SSB}. Subsection \ref{sec:Masses} features the leading superpotential terms generating
the fermions masses as well as the discussion of the messenger sector - features which are going to be used (with small changes) also in chapter \ref{ch:fsd}.
The phenomenology of the continuous model is presented in subsection
\ref{sec:Phenomenology}.
The summary of the continuous model results, in
subsection \ref{sec:Confmm1}, serves also as motivation for the
discrete models presented in chapter \ref{ch:fsd} (namely the $\Delta(27)$ model of section \ref{sec:D27}).

\section{$SU(3)_{f}$ \FS \ model \label{sec:Symmetries}}

As discussed in section \ref{sec:fsintro}, we keep the assignment of $SU(3)_{f}$ representations consistent with an underlying $SO(10) \otimes G_{f}$ symmetry, even though
we will effectively use only the Pati-Salam subgroup of $SO(10)$ in
constructing models ($G_{PS} \equiv
SU(4)_{PS}\otimes SU(2)_{L}\otimes SU(2)_{R}$). We denote the
\SM \ fermions as $\psi _{i}$ and $\psi _{j}^{c}$ (where $i,j=1,2,3$ are family indices). $\psi_{i}$ and $\psi_{i}^{c}$ are assigned to a $(\mathbf{16},\mathbf{3})$ representation of $SO(10)\otimes
SU(3)_{f}$. The Higgs doublets of the \SM \ (of which we require two due to SUSY) are part of 
$(\mathbf{10},\mathbf{1})$ representations, labelled jointly as $H$ for simplicity. In addition we
introduce the Higgs $H_{45}$ in the adjoint representation of $SO(10)$, as  $(\mathbf{45},\mathbf{1})$. In our effective theory
$H_{45}$ has a \VEV \ consistent with the residual
$G_{PS}\otimes G_{f}$ symmetry which leaves the hypercharge $Y_{k}= k T_{3R}+(B-L)/2$ unbroken, as discussed in \cite{Ross:2002fb} (see also \cite{King:2003rf}; note that the expression we use for $Y_k$ differs by an overall multiplicative factor of $2$ from the hypercharge used in those references). $T_{3R}$ is the right isospin associated with $SU(2)_{R}$ (so for example, $T_{3R} (\nu^{c})=+1/2$). Although $Y_{k}$ is the most general expression, the phenomenology of the models requires a factor of magnitude $3$ ($-3$ or $+3$) between the $Y_{k}$ of charged leptons and of down quarks and as such, the possible choices for $k$ are $0$ or $1$ as we will see. We choose $k=1$ so that $Y(\nu^{c})=0$, which will be helpful in separating the neutrino terms from the charged leptons:

\begin{equation}
Y_{k} = k T_{3R}+(B-L)/2
\label{eq:Yk45}
\end{equation}

\begin{equation}
Y_{k=1} \equiv Y = T_{3R}+(B-L)/2
\label{eq:Y45}
\end{equation}

To successfully recover the \SM \ the \FS \ must be completely
broken. We will do
so in steps, first with a dominant breaking $SU(3)_{f}\rightarrow
SU(2)_{f}$, followed by the breaking of the remaining $SU(2)_{f}$. This
spontaneous symmetry breaking will be achieved by additional \SM \
singlet scalar fields, which in the models discussed here are typically (but not always) either triplets ($\mathbf{3}_{i}$) or anti-triplets
($\bar{\mathbf{3}}^{i}$) of the \FS \ $SU(3)_{f}$. The alignment
of their \VEV s is extremely relevant to the results and
is discussed in subsection \ref{sec:SSB} (as well as in more detail in appendix \ref{app_A}). To construct a
realistic model it is necessary to further extend the symmetry in
order to eliminate 
unwanted terms that would otherwise show up in the effective Lagrangian. The
construction of a specific model 
requires the specification of the full multiplet content together with its
transformation properties under $G_{PS}\otimes SU(3)_{f}$ and under the
additional symmetry needed to limit the allowed couplings. In the $SU(3)_{f}$ model we consider in this chapter \cite{Ivo1}, the additional symmetry \footnote{$R$ is an $R-$ symmetry.} is
$G = R \otimes U(1)\otimes U^{\prime }(1)$. The multiplet content and transformation
properties for the model are 
given in table \ref{Ta:Table 1}. In addition to the fields already discussed, table \ref{Ta:Table 1}
includes the fields $\theta $ and $\bar{\theta }$ whose \VEV s break
$SU(4)_{PS}$, breaking also lepton number and thus generating the Majorana
masses (as described in subsection \ref{sec:Masses}). Table \ref{Ta:Table 1} also features  additional $G_{PS}$ singlet fields required for \VEV \ alignment, as discussed in appendix \ref{app_A}.


\begin{table}[htbp] \centering%

\begin{tabular}{|c||c||c|c|c||c|c|c|}
\hline
Field & $SU(3)_{f}$ & $SU(4)_{PS}$ & $SU(2)_{L}$ & $SU(2)_{R}$ & $R$ & $U(1)$
& $U(1)^{ \prime }$ \\ \hline\hline
$\psi $ & $\mathbf{3}$ & $\mathbf{4}$ & $\mathbf{2}$ & $\mathbf{1}$ & $%
\mathbf{1}$ & $\mathbf{0}$ & $\mathbf{0}$ \\ 
$\psi ^{c}$ & $\mathbf{3}$ & $\bar{\mathbf{4}}$ & $\mathbf{1}$ & $\bar{\mathbf{2}}$
& $\mathbf{1}$ & $\mathbf{0}$ & $\mathbf{0}$ \\ \hline\hline
$\theta$ & $\mathbf{3}$ & $\bar{\mathbf{4}}$ & $\mathbf{1}$ & $\bar{\mathbf{2%
}}$ & $\mathbf{0}$ & $\mathbf{0}$ & $\mathbf{0}$ \\ 
$\bar{\theta}$ & $\bar{\mathbf{3}}$ & $\mathbf{4}$ & $\mathbf{1}$ & $%
\mathbf{2}$ & $\mathbf{0}$ & $\mathbf{0}$ & $\mathbf{0}$ \\ \hline\hline
$H$ & $\mathbf{1}$ & $\mathbf{1}$ & $\mathbf{2}$ & $\mathbf{2}$ & $\mathbf{0}
$ & $-\mathbf{4}$ & $-\mathbf{4}$ \\ 
$H_{45}$ & $\mathbf{1}$ & $\mathbf{15}$ & $\mathbf{1}$ & $\mathbf{3}$ & $%
\mathbf{0}$ & $\mathbf{2}$ & $\mathbf{2}$ \\ \hline\hline
$\phi_{3}$ & $\mathbf{3}$ & $\mathbf{1}$ & $\mathbf{1}$ & $\mathbf{1}$ & $%
\mathbf{0}$ & $-\mathbf{2}$ & $-\mathbf{3}$ \\ 
$\bar{\phi}_{3}$ & $\mathbf{\bar{3}}$ & $\mathbf{1}$ & $\mathbf{1}$ & $%
\mathbf{3}\oplus \mathbf{1}$ & $\mathbf{0}$ & $\mathbf{2}$ & $\mathbf{2}$ \\ 
$\phi_{2}$ & $\mathbf{3}$ & $\mathbf{1}$ & $\mathbf{1}$ & $\mathbf{1}$ & $%
\mathbf{0}$ & $-\mathbf{1}$ & $\mathbf{1}$ \\ 
$\bar{\phi}_{2}$ & $\bar{\mathbf{3}}$ & $\mathbf{1}$ & $\mathbf{1}$ & $\mathbf{1}$
& $\mathbf{0}$ & $-\mathbf{1}$ & $\mathbf{1}$ \\ 
$\phi_{23}$ & $\mathbf{3}$ & $\mathbf{1}$ & $\mathbf{1}$ & $\mathbf{1}$
& $\mathbf{0}$ & $-\mathbf{4}$ & $-\mathbf{3}$ \\ 
$\bar{\phi}_{23}$ & $\bar{\mathbf{3}}$ & $\mathbf{1}$ & $\mathbf{1}$ & $%
\mathbf{1}$ & $\mathbf{0}$ & $\mathbf{1}$ & $\mathbf{1}$ \\ 
$\phi_{123}$ & $\mathbf{3}$ & $\mathbf{1}$ & $\mathbf{1}$ & $\mathbf{1}$ & $%
\mathbf{0}$ & $\mathbf{0}$ & $\mathbf{1}$ \\ 
$\bar{\phi}_{123}$ & $\mathbf{\bar{3}}$ & $\mathbf{1}$ & $\mathbf{1}$ & $%
\mathbf{1}$ & $\mathbf{0}$ & $\mathbf{3}$ & $\mathbf{3}$ \\ \hline\hline
$X_{3}$ & $\mathbf{1}$ & $\mathbf{1}$ & $\mathbf{1}$ & $\mathbf{1}$ & $%
\mathbf{2}$ & $\mathbf{0}$ & $\mathbf{1}$ \\ 
$Y_{2}$ & $\mathbf{1}$ & $\mathbf{1}$ & $\mathbf{1}$ & $\mathbf{1}$ & $%
\mathbf{2}$ & $-\mathbf{1}$ & $-\mathbf{3}$ \\ 
$X_{23}$ & $\mathbf{1}$ & $\mathbf{1}$ & $\mathbf{1}$ & $\mathbf{1}$ & $%
\mathbf{2}$ & $\mathbf{1}$ & $\mathbf{0}$ \\ 
$Y_{23}$ & $\mathbf{1}$ & $\mathbf{1}$ & $\mathbf{1}$ & $\mathbf{1}$ & $%
\mathbf{2}$ & $\mathbf{3}$ & $\mathbf{2}$ \\ 
$X_{123}$ & $\mathbf{1}$ & $\mathbf{1}$ & $\mathbf{1}$ & $\mathbf{1}$ & $%
\mathbf{2}$ & $-\mathbf{3}$ & $-\mathbf{4}$ \\ 
$Y_{123}$ & $\mathbf{1}$ & $\mathbf{1}$ & $\mathbf{1}$ & $\mathbf{1}$ & $%
\mathbf{2}$ & $-\mathbf{1}$ & $-\mathbf{2}$ \\ 
$Z_{123}$ & $\mathbf{1}$ & $\mathbf{1}$ & $\mathbf{1}$ & $\mathbf{1}$ & $%
\mathbf{4/3}$ & $-\mathbf{3}$ & $-\mathbf{4}$ \\ \hline\hline
$S_{3}$ & $\mathbf{1} $ & $\mathbf{1}$ & $\mathbf{1}
$ & $\mathbf{1}$ & $ \mathbf{0} $ & $\mathbf{0}$ & $ -\mathbf{1}$ \\
$\Sigma$ & $\mathbf{3} \otimes \bar{\mathbf{3}}$ & $\mathbf{1}$ & $\mathbf{1}
$ & $\mathbf{1}$ & $\mathbf{2/3}$ & $\mathbf{0}$ & $\mathbf{0}$ \\ \hline
\end{tabular}
\caption{Field and representation content of continuous model.} \label{Ta:Table 1}%
\end{table}%

\section{Spontaneous symmetry breaking \label{sec:SSB}}

We now summarise the pattern of \FS \ breaking in the
continuous model. The 
detailed minimisation of the effective potential which gives this structure
is addressed in appendix \ref{app_A}.

The dominant breaking of $SU(3)_{f}$ responsible for the third generation
quark and charged lepton masses is provided by the $\bar{\phi}_{3}$ \VEV :
\begin{equation}
\langle \bar{\phi}_{3}\rangle =\left( 
\begin{array}{ccc}
0 & 0 & 1%
\end{array}%
\right) \otimes \left( 
\begin{array}{cc}
a_{u} & 0 \\ 
0 & a_{d}%
\end{array}%
\right)  \label{eq:P3B vev}
\end{equation}%
The $SU(3)\times SU(2)_{R}$ structure is explicitly exhibited in eq.(\ref{eq:P3B vev}). Note that $\bar{\phi}_{3}$ also
breaks $SU(2)_{R}$; at this stage, the residual symmetry is $(SU(4)_{PS}\otimes
SU(2)_{L}\otimes U(1)_{R}) \otimes SU(2)_{f}$. To preserve $D$-flatness, another field, $\phi _{3}$, also acquires a large \VEV :
\begin{equation}
\langle \phi _{3}\rangle = \left( 
\begin{array}{c}
0 \\ 
0 \\ 
\sqrt{a_{u}^{2}+a_{d}^{2}}%
\end{array}%
\right)  \label{eq:P3 vev}
\end{equation}%

The fields $\theta$ and $\bar{\theta}$,
responsible for breaking $SU(4)_{PS}$ also acquire \VEV s \ along the same
direction (see appendix \ref{app_A}):
\begin{equation}
\langle \bar{\theta}\rangle \propto \left( 
\begin{array}{ccc}
0 & 0 & 1%
\end{array}%
\right)  \label{eq:T3B vev}
\end{equation}

\begin{equation}
\langle \theta \rangle \propto \left( 
\begin{array}{c}
0 \\ 
0 \\ 
1%
\end{array}%
\right)  \label{eq:T3 vev}
\end{equation}%

The breaking of the remaining $SU(2)_{f}$ \FS \ is achieved when a
triplet $\phi _{2}$ acquires the \VEV :
\begin{equation}
\langle \phi _{2}\rangle = \left( 
\begin{array}{c}
0 \\ 
y \\ 
0%
\end{array}%
\right)  \label{eq:P2 vev}
\end{equation}
Due to the allowed couplings in the superpotential (see appendix \ref{app_A}) this \VEV \ is orthogonal to $\langle \bar{\phi%
}_{3}\rangle$.

Further fields acquire \VEV s \ constrained by the allowed couplings in the
theory. As detailed in appendix \ref{app_A} the field
$\bar{\phi}_{23}$ acquires a \VEV :
\begin{equation}
\langle \bar{\phi}_{23}\rangle = \left( 
\begin{array}{ccc}
0 & b & -b%
\end{array}%
\right)  \label{eq:P23B vev}
\end{equation}
It is the underlying $SU(3)_{f}$ that forces the \VEV s \ in the $2$nd
and the $3$rd directions to be equal in magnitude, so that the $\bar{\phi}_{23}$ is
rotated by $\pi /4$ relative to the $\bar{\phi}_{3}$ \VEV . This is
important in generating an acceptable pattern for quark masses and in
generating bi-maximal mixing in the lepton sector. Finally the fields
$\bar{\phi}_{123}$ and $\phi_{123}$ acquire the \VEV s: 
\begin{equation}
\langle \bar{\phi}_{123}\rangle =\left( 
\begin{array}{ccc}
\bar{c} & \bar{c} & \bar{c}%
\end{array}%
\right)  \label{eq:P123B vev}
\end{equation}%

\begin{equation}
\langle \phi_{123}\rangle =\left(%
\begin{array}{c}
c \\ 
c \\ 
c%
\end{array}%
\right)  \label{eq:P123 vev}
\end{equation}
The magnitudes are related by $c=\bar{c}e^{i\gamma }$. We note again that even
though $SU(3)_{f}$ is spontaneously broken by the \VEV s, it is the
\FS \ that is responsible for aligning  $\bar{\phi}_{123}$, $\phi_{123}$  in these particular
directions (namely, all the components have equal magnitude). This
structure will prove to be essential in 
obtaining tri-maximal mixing for the solar neutrino.

\section{Mass terms and messengers \label{sec:Masses}}

\subsection{Effective superpotential}

Having specified the multiplet content and the symmetry properties it is now
straightforward to write down all terms in the superpotential allowed by the
symmetries of the theory. 
In all the superpotential terms we omit the overall coupling
associated with each term. These are not determined by the symmetries
alone and are expected to be of $O(1)$. In this work we don't consider explicitly the K\"ahler potential corrections to the structures arising from the superpotential. These corrections have been shown to be sub-leading for hierarchical structures \cite{King:2004tx} (which is the case with the models being considered). The corrections depend on powers of the small expansion parameters $\langle \phi \rangle / M$ and typically leave the structure given by the superpotential terms essentially unchanged (the corrections can be absorbed into the $O(1)$ coefficients).

\begin{figure}[htb]
\begin{center}
\fcolorbox{white}{white}{
  \begin{picture}(300,106) (150,-179)
    \SetWidth{0.5}
    \SetColor{Black}
    \COval(210,-104)(2.83,2.83)(45.0){Black}{White}\Line(208.59,-105.41)(211.41,-102.59)\Line(208.59,-102.59)(211.41,-105.41)
    \Text(195,-89)[lb]{\Large{\Black{$\langle H \rangle$}}}
    \Vertex(300,-149){2.83}
    \Vertex(210,-149){2.83}
    \DashArrowLine(210,-104)(210,-149){10}
    \Line(253,-151)(257,-147)\Line(253,-147)(257,-151)
    \Line(343,-151)(347,-147)\Line(343,-147)(347,-151)
    \Vertex(390,-149){2.83}
    \ArrowLine(255,-149)(300,-149)
    \ArrowLine(255,-149)(210,-149)
    \ArrowLine(345,-149)(390,-149)
    \ArrowLine(345,-149)(300,-149)
    \DashArrowLine(300,-104)(300,-149){10}
    \COval(300,-104)(2.83,2.83)(45.0){Black}{White}\Line(298.59,-105.41)(301.41,-102.59)\Line(298.59,-102.59)(301.41,-105.41)
    \COval(390,-104)(2.83,2.83)(45.0){Black}{White}\Line(388.59,-105.41)(391.41,-102.59)\Line(388.59,-102.59)(391.41,-105.41)
    \DashArrowLine(390,-104)(390,-149){10}
    \Text(285,-89)[lb]{\Large{\Black{$\langle \bar{\phi}_{3} \rangle$}}}
    \ArrowLine(150,-149)(210,-149)
    \Text(150,-179)[lb]{\Large{\Black{$\psi$}}}
    \Text(270,-179)[lb]{\Large{\Black{$X^{1}$}}}
    \Text(225,-179)[lb]{\Large{\Black{$\bar{X}^{1}$}}}
    \ArrowLine(450,-149)(390,-149)
    \Text(420,-179)[lb]{\Large{\Black{$\psi^{c}$}}}
    \Text(375,-89)[lb]{\Large{\Black{$\langle \bar{\phi}_{3} \rangle$}}}
    \Text(315,-179)[lb]{\Large{\Black{$\bar{X}^{2}$}}}
    \Text(360,-179)[lb]{\Large{\Black{$X^{2}$}}}
  \end{picture}
}
\caption{Leading contribution to third generation Dirac mass.}
\label{fig:P3BP3B}
\end{center}
\end{figure}
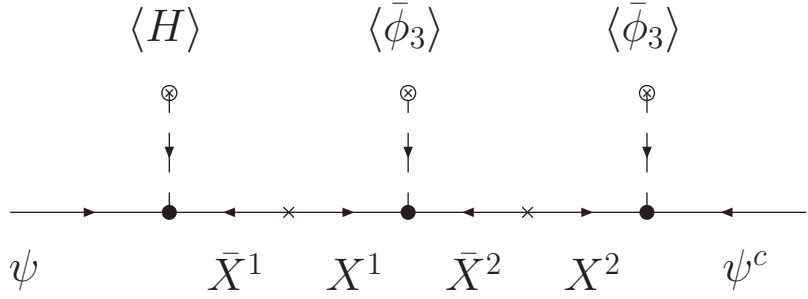

We focus on terms responsible for generating the fermion mass
matrices. Since we are working 
with an effective field theory in which the heavy modes associated with the
various stages of symmetry breaking have been integrated out we must include
terms of arbitrary dimension. In practise, to evaluate the form of the
mass matrices, it is only necessary to keep the
leading terms that give the \FMM . The leading order terms generating the quark,
charged lepton and neutrino Dirac masses are:
\begin{equation}
P_{Y} = \frac{1}{M^{2}}\bar{\phi}_{3}^{i}\psi _{i}\bar{\phi}_{3}^{j}\psi
_{j}^{c}H  \label{eq:Y33}
\end{equation}%
\begin{equation}
+\frac{1}{M^{3}}\bar{\phi}_{23}^{i}\psi _{i}\bar{\phi}_{23}^{j}\psi
_{j}^{c}HH_{45}  \label{eq:Y_P23_P23}
\end{equation}

\begin{equation}
+\frac{1}{M^{2}}\bar{\phi}_{23}^{i}\psi _{i}\bar{\phi}_{123}^{j}\psi
_{j}^{c}H  \label{eq:Y_P23_P123}
\end{equation}%
\begin{equation}
+\frac{1}{M^{2}}\bar{\phi}_{123}^{i}\psi _{i}\bar{\phi}_{23}^{j}\psi
_{j}^{c}H  \label{eq:Y_P123_P23}
\end{equation}%

\begin{equation}
+\frac{1}{M^{7}}\bar{\phi}_{2}^{i}\psi _{i}\bar{\phi}_{123}^{j}\psi
_{j}^{c}HH_{45} (\bar{\phi}_{3}^{k} \phi_{3_{k}})^2 \label{eq:Y_P2_P123}
\end{equation}
These terms arise through the \FN \ mechanism \cite{fnielsen} (see
subsection \ref{sub:fnreview}); for example, the diagram in figure
\ref{fig:P3BP3B} corresponds to the superpotential term in eq.(\ref{eq:Y33}).

For simplicity we display the superpotential terms as suppressed by
inverse powers of a mass scale which we have generically denoted by
$M$. In figure eq.(\ref{fig:P3BP3B}), $M$ actually corresponds to the mass of the generic messengers $\bar{X}^{a}$ and $X^{a}$ ($a$ is a label, not an index - note that in general each messenger pair has to be different and carry appropriate charge assignments, as the flavons themselves carry charge, as seen in subsection \ref{sub:fnreview}). The identification of the specific messengers and the associated mass 
scale for each fermion sector is important in studying the
phenomenology. To do so one has to discuss how these
non-renormalisable terms arise: it occurs at the stage where
the superheavy messenger field are integrated out. In
subsection \ref{sec:Messengers} we consider this in more detail.

The symmetry allowed terms responsible for the Majorana mass matrix
involve the $\bar{\theta}^{i}$ anti-triplet, whose \VEV \ breaks lepton
number and $SU(4)_{PS}$. The \VEV \ is aligned along the third direction,
similarly to $\bar{\phi}_{3}$ (see
appendix \ref{app_A}). The leading terms are:

\begin{equation}
P_{M} = \frac{1}{M}\bar{\theta}^{i}\psi _{i}^{c}\bar{\theta}^{j}\psi _{j}^{c}
\label{eq:M_th_th}
\end{equation}%
\begin{equation}
+\frac{1}{M^{5}}\bar{\phi}_{23}^{i}\psi _{i}^{c}\bar{\phi}_{23}^{j}\psi
_{j}^{c}\bar{\theta}^{k}\phi _{123_{k}}\bar{\theta}^{l}\phi _{3_{l}}+\frac{1%
}{M^{5}}\bar{\theta}^{i}\psi _{i}^{c}\bar{\phi}_{23}^{j}\psi _{j}^{c}\bar{%
\theta}^{k}\phi _{123_{k}}\bar{\phi}_{23}^{l}\phi _{3_{l}}
\label{eq:M_P23_P23}
\end{equation}%
\begin{equation}
+\frac{1}{M^{5}}\bar{\phi}_{123}^{i}\psi _{i}^{c}\bar{\phi}_{123}^{j}\psi
_{j}^{c}\bar{\theta}^{k}\phi _{23_{k}}\bar{\theta}^{l}\phi _{3_{l}}
\end{equation}
\begin{equation}
+\frac{1}{M^{5}}\bar{\theta}^{i}\psi _{i}^{c}\bar{\phi}_{123}^{j}\psi
_{j}^{c}\bar{\phi}_{123}^{k}\phi _{23_{k}}\bar{\theta}^{l}\phi
_{3_{l}}
\end{equation}
\begin{equation}
+\frac{1}{M^{5}}\bar{\theta}^{i}\psi _{i}^{c}\bar{\phi}_{123}^{j}\psi _{j}^{c}\bar{\theta}^{k}\phi _{23_{k}}\bar{\phi}_{123}^{l}\phi _{3_{l}}  \label{eq:M_P123_P123}
\end{equation}

\subsection{Messenger sector\label{sec:Messengers}}

The scale generically denoted as $M$ seen in the effective
superpotential terms of subsection \ref{sec:Masses} is set by the heavy
messenger states in the tree level \FN \ diagrams giving rise to the
higher dimension terms. There are two classes of diagrams,
corresponding either to heavy messenger
states that transform as $\mathbf{4}$s under $SU(4)_{PS}$ (vector-like
families) and those that transform otherwise (heavy Higgs).
Which class of diagram dominates depends on the massive 
multiplet (messenger) spectrum, which in turn is specified by the
details of the theory at the high scale. For our purposes, we assume
that the heavy vector-like families are the lightest states and thus
their contributions to the \FN \ diagrams dominate.
In the \FN \ diagrams generating the masses, the vector-like states
carry the same quantum numbers as the external states - quark
or lepton fields. As the \FN \ messengers carry quark or lepton quantum numbers, we will label the messengers (and their mass scales) according to the specific \SM \ fermions they are associated with.

Due to $SU(2)_{L}$, $M_{Q_{L}}$ (the \LH \
quark messenger mass scale) will be the same for the \LH \ up and down
quarks. With $SU(2)_{R}$ being broken (possibly by $\langle \bar{\phi}_{3} \rangle$, although we won't specify details), the messenger
mass $M_{u_{R}}$ (the \RH \ up quark messenger mass scale)
need not be the same as $M_{d_{R}}$ (the \RH \ down quark
messenger mass scale) - in fact if the $SU(2)_{R}$ breaking contribution to the \RH \ quark messengers depends linearly on $\langle \bar{\phi}_{3} \rangle$ and dominates over the $SU(2)_{R}$ preserving contributions, we obtain phenomenologically viable masses for these messengers.
The lepton messenger mass scales have a similar
structure, with $M_{L_{L}}$ (the 
\LH \ lepton messenger mass scale) contributing equally to the charged
lepton and neutrino Dirac couplings, but with $M_{e_{R},\nu _{R}}$ (the
\RH \ charged lepton and \RH \ neutrino messenger mass scale respectively) having different scales due to $SU(2)_{R}$ breaking effects.
This splitting of the messenger masses is important because it is
responsible for the different hierarchies of the different fermion
sectors. As we noted before, the underlying $SO(10)$ structure forces all
matter states to have the same family charges and so the leading terms
in the superpotential contribute equally to all sectors. However the
soft messenger masses which enter the effective Lagrangian are
sensitive in leading order to $SO(10)$ breaking effects and thus can
differentiate between these sectors by fixing different expansion
parameters in the different sectors.

To see what choice for the messenger masses is necessary phenomenologically
we refer again to the \GUT \ scale fits of up and down quark mass
matrices, of the form displayed in eq.(\ref{eq:Yu2}) and
eq.(\ref{eq:Yd2}) \cite{Roberts:2001zy, Mario}, having the expansion
parameters of eq.(\ref{eq:u,d eps2}).

From eq.(\ref{eq:Y_P23_P23}) it may be seen that in the quark sector the
expansion parameters in the $(2,3)$ block are essentially determined
by the  $\bar{\phi}_{23}$ \VEV \ divided by the relevant messenger mass
scale. If the
expansion parameters are to differ, we require $M_{Q_{L}}$ to be
larger than the other relevant messenger masses, in which case:
\begin{equation}
\epsilon _{u,d}\simeq \frac{b}{M_{u_{R},d_{R}}}
\end{equation}%
To generate the form of eq.(\ref{eq:u,d eps2}) we require:
\begin{equation}
M_{d_{R}}\simeq 0.37 \ M_{u_{R}} < M_{Q_{L}}
\label{eq:eps3}
\end{equation}

In the lepton sector we know that the \GJ \ relation
$m_{b}\simeq m_{\tau} $ at the unification scale
(including radiative corrections) is in
good agreement with the measured masses. For this to be the case in our model, we
require the $SU(4)_{PS}$ breaking contribution to the down sector messenger masses to not be dominant. The required condition is to be expected
in our model, as $SU(4)_{PS}$ is broken in the lepton
number breaking sector, which does not couple in leading order to the the
\RH \ charged lepton messenger states. The dominant
messenger mass scales associated with (charged) leptons and (down-type) quarks are related by $SU(4)_{PS}$:
\begin{eqnarray}
M_{e_{R}} &\simeq &M_{d_{R}} \\
M_{Q_{L}} &\simeq &M_{L_{L}}
\end{eqnarray}%
The lighter \RH \ messengers dominate over the \LH , and the relation $m_{b}\simeq m_{\tau }$
follows from $M_{e_{R}} \simeq M_{d_{R}}$. However, the \RH \ neutrino messengers do couple in leading order to the $SU(4)_{PS}$ breaking fields (like $\bar{\theta}$) and so may be
anomalously heavy. This is helpful, because a small \RH \ neutrino
expansion parameter $\epsilon _{\nu _{R}}$ naturally explains the required large
hierarchical structure of Majorana masses that leads to a
\SD \ scenario - allowing the model to overcome the
large Dirac neutrino mass in the $(3,3)$ direction.
To summarise, the
different expansion parameters in the lepton sector are given by:
\begin{equation}
\epsilon _{\nu _{L},\nu _{R},e_{R}}\simeq \frac{b}{M_{L_{L},\nu _{R},e_{R}}}
\label{eq:L eps}
\end{equation}
We note that it is possible that $M_{\nu_{R}} \ll M_{L_{L}}$, in which case it is $\epsilon_{\nu_{R}}$ (and not $\epsilon_{\nu_{L}}$) that governs the hierarchy of the neutrino Dirac masses. Bounds on the messenger masses of the neutrinos will be presented in subsection \ref{sec:Phenomenology}. The other expansion parameters are
chosen to fit the masses, as described by eq.(\ref{eq:u,d eps2}), eq.(\ref{eq:eps3}).

Note that the contribution to the $(3,3)$ entries of the quark and charged
lepton mass matrices involves the combination $a_{u}/M_{u_{R}}$ and
$a_{d}/M_{d_{R}} \simeq a_{d}/M_{e_{R}}$ for the up and down sectors
respectively.
In general, the \RH \ messengers (generically denoted $\bar{X}_{R}$ and $X_{R}$) have masses $M_{u_{R}}$ and $M_{d_{R}}$ with a contribution that preserves $SU(2)_{R}$ and a contribution that doesn't.
If the dominant $SU(2)_{R}$ breaking component arises through some superpotential term like $\bar{X}_{R} \bar{\phi}_{3} X_{R}$, we may have $M_{u,d_{R}}=\alpha \langle \bar{\phi}_{3} \rangle + \beta$ (where $\alpha \gtrsim 1$).
As long as the $SU(2)_{R}$ preserving contribution $\beta$ is negligibly small (or entirely absent), we have  $M_{u,d_{R}} \propto \langle \bar{\phi}_{3} \rangle$ and obtain $a_{u}/M_{u_{R}}\simeq a_{d}/M_{d_{R}}\lesssim 1$, which is indeed the phenomenologically desirable choice \cite{King:2003rf}.

\subsection{Dirac mass matrix structure \label{sec:Dirac}}

Using the expansion parameters introduced above we can now write the
approximate quark mass matrices for the second and third generations
(following from eq.(\ref{eq:Y33}) and eq.(\ref{eq:Y_P23_P23})):
\begin{equation}
Y^{u}\propto \left( 
\begin{array}{cc}
-2\epsilon _{u}^{2} \frac{\epsilon_{u}}{\epsilon_{d}} & 2\epsilon _{u}^{2} 
\frac{\epsilon_{u}}{\epsilon_{d}} \\ 
2\epsilon _{u}^{2} \frac{\epsilon_{u}}{\epsilon_{d}} & 1%
\end{array}%
\right) ,Y^{d}\propto \left( 
\begin{array}{cc}
\epsilon _{d}^{2} & -\epsilon _{d}^{2} \\ 
-\epsilon _{d}^{2} & 1%
\end{array}%
\right)  \label{eq:Y_Q_23 matrices}
\end{equation}
The $-2 \epsilon_{u}/\epsilon_{d}$ factors in $Y^{u}$ come about due to $\langle H_{45} \rangle / M$: in writing eq.(\ref{eq:Y_Q_23 matrices}) we have made an implicit choice for the $H_{45}$
\VEV, as it appears
in terms contributing to these elements. With the choice $k=1$ leading to eq.(\ref{eq:Y45}) (from eq.(\ref{eq:Yk45})), $\langle H_{45}
\rangle$ preserves $G_{PS}$ and is proportional to the
hypercharge $Y=T_{3R}+(B-L)/2$. To fit the strange quark mass
\cite{Roberts:2001zy, Mario} we take its magnitude to be such that :
\begin{equation}
\frac{ \langle H_{45} \rangle}{M}|_{d}\equiv
\frac{Y\left(d^{c}\right) h_{45}}{M_{d_{R}}} \simeq O(1)
\end{equation}
With $Y$ taken as in eq.(\ref{eq:Y45}) the factor $Y\left( u^{c}\right)
/Y\left(d^{c}\right) =-2$ appears in $Y^{u}$, and the extra $M_{d_{R}} / M_{u_{R}}$ produces the ratio of expansion parameters.

It is important to note that the leading order terms present in the model naturally lead to $Y^{d}_{22}=-Y^{d}_{23}$ as shown in eq.(\ref{eq:Y_Q_23 matrices}). This was favoured by earlier fits (see \cite{Roberts:2001zy}), but is now disfavoured by the data, as seen in the more recent fits presented in eq.(\ref{eq:Yd2}) \cite{Mario} that prefer a relative factor $\left| Y^{d}_{23}/Y^{d}_{22} \right|$ of about $2$. While this doesn't rule out the model, it is something that has to be obtained from sub-leading operators and thus makes the model less appealing. One way this can occur is through a specific term with coefficient that is relatively large (greater than $O(1)$); otherwise, one can have a suitable combination of terms that lower the magnitude of $Y^{d}_{22}$ and terms that raise the magnitude of $Y^{d}_{23}$, in such a way that the magnitude between the entries is close to $2$.

Because the charged lepton messengers have the same messenger mass scale as
the down quarks, the charged lepton mass matrix is similar to $Y^{d}$, taking
the form:
\begin{equation}
Y^{l}\propto \left( 
\begin{array}{cc}
3\epsilon _{d}^{2} & -3\epsilon _{d}^{2} \\ 
-3\epsilon _{d}^{2} & 1
\end{array}
\right)  \label{eq:Y_c_23 matrix}
\end{equation}
With the choice $k=1$ of eq.(\ref{eq:Y45}), the hypercharge $Y=T_{3R}+(B-L)/2$ leads to the correct \GJ \  \cite{Georgi:1979df} factor $Y\left( e^{c}\right) /Y\left( d^{c}\right)=+3$ arising through $\langle H_{45} \rangle$. This factor gives $m_{\mu
}\simeq 3m_{s}$ required from the \GUT \ scale fits
\cite{Roberts:2001zy, Mario}, which are in good agreement with the
measured (low scale) masses after 
including radiative corrections - an obvious advantage to having an underlying \GUT.
At this stage it is relevant to add that one would obtain instead the equally viable ratio $Y\left( e^{c}\right) /Y\left( d^{c}\right)=-3$ if we had chosen $k=0$ instead of $k=1$ in eq.(\ref{eq:Yk45}), leading to $Y_{k=0}=(B-L)/2$. The ratio of $-3$ would also be obtained (regardless of $k$) if the dominant messenger states were \LH , due to the vanishing $T_{3R}$ (this option is not phenomenologically viable for quarks, as it would lead to the same hierarchy for the up and down quarks - we will however considered the possibility of dominating \LH \ messengers for neutrinos).

Having explained the origin of the structure in the $(2,3)$ block it is
straightforward to follow the origin of the full three generation Yukawa
matrices for the quarks and leptons. Including the effect of the terms in
eq.(\ref{eq:Y_P23_P123}) and eq.(\ref{eq:Y_P123_P23}), we have
\begin{equation}
Y^{u}\propto \left( 
\begin{array}{ccc}
0 & g_{\odot }\epsilon _{u}^{2}\epsilon _{d} & -g_{\odot }\epsilon
_{u}^{2}\epsilon _{d} \\ 
g_{@}\epsilon _{u}^{2}\epsilon _{d} & -2\epsilon _{u}^{2}\frac{\epsilon _{u}%
}{\epsilon _{d}} & 2\epsilon _{u}^{2}\frac{\epsilon _{u}}{\epsilon _{d}} \\ 
-g_{@}\epsilon _{u}^{2}\epsilon _{d} & 2\epsilon _{u}^{2}\frac{\epsilon _{u}%
}{\epsilon _{d}} & 1%
\end{array}%
\right)  \label{eq:Y_u matrix}
\end{equation}%
\begin{equation}
Y^{d}\propto \left( 
\begin{array}{ccc}
0 & g_{\odot }\epsilon _{d}^{3} & -g_{\odot }\epsilon _{d}^{3} \\ 
g_{@}\epsilon _{d}^{3} & \epsilon _{d}^{2} & -\epsilon _{d}^{2} \\ 
-g_{@}\epsilon _{d}^{3} & -\epsilon _{d}^{2} & 1%
\end{array}%
\right)  \label{eq:Y_d matrix}
\end{equation}%
\begin{equation}
Y^{l}\propto \left( 
\begin{array}{ccc}
0 & g_{\odot }\epsilon _{d}^{3} & -g_{\odot }\epsilon _{d}^{3} \\ 
g_{@}\epsilon _{d}^{3} & 3\epsilon _{d}^{2} & -3\epsilon _{d}^{2} \\ 
-g_{@}\epsilon _{d}^{3} & -3\epsilon _{d}^{2} & 1%
\end{array}%
\right)  \label{eq:Y_c matrix}
\end{equation}

\begin{equation}
Y^{\nu }\propto \left( 
\begin{array}{ccc}
0 & g_{\odot }\epsilon _{\nu}^{2}\epsilon _{d} & -g_{\odot }\epsilon
_{\nu}^{2}\epsilon _{d} \\ 
g_{@}\epsilon _{\nu}^{2}\epsilon _{d} & (g_{@}+g_{\odot })\epsilon
_{\nu}^{2}\epsilon _{d} & (g_{@}-g_{\odot })\epsilon _{\nu}^{2}\epsilon _{d} \\ 
-g_{@}\epsilon _{\nu}^{2}\epsilon _{d} & (-g_{@}+g_{\odot })\epsilon
_{\nu}^{2}\epsilon _{d} & \frac{\epsilon _{\nu}^{2}}{\epsilon
_{d}^{2}}%
\end{array}%
\right)  \label{eq:Y_nu matrix}
\end{equation}%
In this form we have restored the dependence on some of the unknown Yukawa
couplings of $O(1)$, $g_{\odot }$ and $g_{@}$. As we assume a
symmetric form (consistent with an underlying $SO(10)$ commuting with the \FS ), we consider the case $g_{\odot}=g_{@}$, noting also that this equality is required in order to obtain the \GST \ relation \cite{Gatto:1968ss} displayed in eq.(\ref{eq:GST}).

The structure of the Dirac neutrino mass matrix $Y^{\nu}$ follows from the
terms in eq.(\ref{eq:Y_P23_P123}) and eq.(\ref{eq:Y_P123_P23}). The form shown
in eq.(\ref{eq:Y_nu matrix}) displays as expansion parameter the unspecified
$\epsilon_{\nu}$. As such it applies in either case: if the limit where the
dominant carriers are \LH \ (in which case $\epsilon_{\nu} = \epsilon_{\nu_{L}}$), or if instead the dominant carriers are \RH \ (in which case $\epsilon_{\nu} = \epsilon_{\nu_{R}}$). In the latter case the Dirac and Majorana neutrino
mass matrices share the same expansion parameter. 
It is phenomenologically possible to have either situation, as long as an
important requirement is verified: the term involving 
$H_{45}$ in eq.(\ref{eq:Y_P23_P23}) must not contribute
significantly to the neutrino Dirac mass matrix (or it spoils \TBM ).

If $\epsilon_{\nu_{R}}$ is the relevant expansion parameter, the respective
\FN \ diagram proceeds through heavy messenger
states $\bar{X}^{a}_{R}$ and $X^{a}_{R}$, sharing the quantum numbers of \RH \ neutrinos (note the position of $\langle H \rangle$ in figure \ref{fig:H45R}), and
the $H_{45}$ contribution exactly decouples due to its \VEV : $Y(\nu^{c})=0$ due to choice $k=1$ made for eq.(\ref{eq:Y45}).

\begin{figure}[htb]
\begin{center}
\fcolorbox{white}{white}{
\begin{picture}(390,106) (150,-179)
    \SetWidth{0.5}
    \SetColor{Black}
    \COval(210,-104)(2.83,2.83)(45.0){Black}{White}\Line(208.59,-105.41)(211.41,-102.59)\Line(208.59,-102.59)(211.41,-105.41)
    \Text(195,-89)[lb]{\Large{\Black{$\langle H \rangle$}}}
    \Vertex(300,-149){2.83}
    \Vertex(210,-149){2.83}
    \DashArrowLine(210,-104)(210,-149){10}
    \Line(253,-151)(257,-147)\Line(253,-147)(257,-151)
    \Line(343,-151)(347,-147)\Line(343,-147)(347,-151)
    \Vertex(390,-149){2.83}
    \ArrowLine(255,-149)(300,-149)
    \ArrowLine(255,-149)(210,-149)
    \ArrowLine(345,-149)(390,-149)
    \ArrowLine(345,-149)(300,-149)
    \DashArrowLine(300,-104)(300,-149){10}
    \COval(300,-104)(2.83,2.83)(45.0){Black}{White}\Line(298.59,-105.41)(301.41,-102.59)\Line(298.59,-102.59)(301.41,-105.41)
    \COval(390,-104)(2.83,2.83)(45.0){Black}{White}\Line(388.59,-105.41)(391.41,-102.59)\Line(388.59,-102.59)(391.41,-105.41)
    \DashArrowLine(390,-104)(390,-149){10}
    \Text(285,-89)[lb]{\Large{\Black{$\langle \bar{\phi}_{23} \rangle$}}}
    \ArrowLine(150,-149)(210,-149)
    \Text(150,-179)[lb]{\Large{\Black{$\nu$}}}
    \Text(270,-179)[lb]{\Large{\Black{$X^{1}_{R}$}}}
    \Text(225,-179)[lb]{\Large{\Black{$\bar{X}^{1}_{R}$}}}
    \Text(375,-89)[lb]{\Large{\Black{$\langle \bar{\phi}_{23} \rangle$}}}
    \Text(315,-179)[lb]{\Large{\Black{$\bar{X}^{2}_{R}$}}}
    \Text(360,-179)[lb]{\Large{\Black{$X^{2}_{R}$}}}
    \ArrowLine(540,-149)(480,-149)
    \Vertex(480,-149){2.83}
    \DashArrowLine(480,-104)(480,-149){10}
    \COval(480,-104)(2.83,2.83)(45.0){Black}{White}\Line(478.59,-105.41)(481.41,-102.59)\Line(478.59,-102.59)(481.41,-105.41)
    \COval(480,-104)(2.83,2.83)(45.0){Black}{White}\Line(478.59,-105.41)(481.41,-102.59)\Line(478.59,-102.59)(481.41,-105.41)
    \COval(480,-104)(2.83,2.83)(45.0){Black}{White}\Line(478.59,-105.41)(481.41,-102.59)\Line(478.59,-102.59)(481.41,-105.41)
    \ArrowLine(435,-149)(390,-149)
    \ArrowLine(435,-149)(480,-149)
    \Line(433,-151)(437,-147)\Line(433,-147)(437,-151)
    \Text(510,-179)[lb]{\Large{\Black{$\nu^{c}$}}}
    \Text(445,-89)[lb]{\Large{\Black{$\langle H_{45} \rangle = 0$}}}
    \Text(450,-179)[lb]{\Large{\Black{$X^{3}_{R}$}}}
    \Text(405,-179)[lb]{\Large{\Black{$\bar{X}^{3}_{R}$}}}
  \end{picture}
}
\caption{\FN \ diagram involving $H_{45}$ and \RH \ messengers.}
\label{fig:H45R}
\end{center}
\end{figure}

If $\epsilon_{\nu_{L}}$ is the relevant expansion parameter
\footnote{The situation $\epsilon_{\nu_{R}} \ll \epsilon_{\nu_{L}}$
  can be natural as long as the $SU(2)_{R}$ breaking gives rise to a
  very heavy \RH \ neutrino messenger mass $M_{\nu_{R}} \gg M_{L_{L}}$.}, the respective \FN \ diagram proceeds
through heavy messenger states $\bar{X}^{a}_{L}$ and $X^{a}_{L}$, sharing the quantum numbers of \LH \ neutrinos (note the position of $\langle H \rangle$ in figure \ref{fig:H45L}). Regardless of the choice of $k$ made, the contribution no longer can be made to vanish, so it must be made negligible. This can be achieved through an extra suppression due to the additional messenger mass (the term involving $H_{45}$ has one extra \FN \ mass insertion - compare figure \ref{fig:H45L} with figure \ref{fig:P3BP3B}, for example).
The requirement then translates into a 
constraint on the magnitude of $\epsilon_{\nu_{L}}$: we must have 
$\frac{3\epsilon _{\nu _{L}}^{3}}{2\epsilon _{d}}\ll
\epsilon_{\nu_{L}}^{2} \epsilon_{d}$ to ensure the term involving
$H_{45}$ remains sub-dominant in the neutrino sector (in order to keep
the the leading terms to be just those shown in eq.(\ref{eq:Y_nu matrix})).
This corresponds to the upper bound
$\epsilon _{\nu _{L}}\ll \frac{2}{3}\epsilon _{d}^{2}$ (a similar suppression would be required and an associated upper bound for $\epsilon_{\nu_{R}}$ would be obtained in the dominant \RH \ messenger case, if we didn't have $Y(\nu^{c})=0$).

\begin{figure}[htb]
\begin{center}
\fcolorbox{white}{white}{
\begin{picture}(390,106) (150,-179)
    \SetWidth{0.5}
    \SetColor{Black}
    \COval(210,-104)(2.83,2.83)(45.0){Black}{White}\Line(208.59,-105.41)(211.41,-102.59)\Line(208.59,-102.59)(211.41,-105.41)
    \Vertex(300,-149){2.83}
    \Vertex(210,-149){2.83}
    \DashArrowLine(210,-104)(210,-149){10}
    \Line(253,-151)(257,-147)\Line(253,-147)(257,-151)
    \Line(343,-151)(347,-147)\Line(343,-147)(347,-151)
    \Vertex(390,-149){2.83}
    \ArrowLine(255,-149)(300,-149)
    \ArrowLine(255,-149)(210,-149)
    \ArrowLine(345,-149)(390,-149)
    \ArrowLine(345,-149)(300,-149)
    \DashArrowLine(300,-104)(300,-149){10}
    \COval(300,-104)(2.83,2.83)(45.0){Black}{White}\Line(298.59,-105.41)(301.41,-102.59)\Line(298.59,-102.59)(301.41,-105.41)
    \COval(390,-104)(2.83,2.83)(45.0){Black}{White}\Line(388.59,-105.41)(391.41,-102.59)\Line(388.59,-102.59)(391.41,-105.41)
    \DashArrowLine(390,-104)(390,-149){10}
    \ArrowLine(150,-149)(210,-149)
    \Text(150,-179)[lb]{\Large{\Black{$\nu$}}}
    \Text(270,-179)[lb]{\Large{\Black{$X^{1}_{L}$}}}
    \Text(225,-179)[lb]{\Large{\Black{$\bar{X}^{1}_{L}$}}}
    \Text(315,-179)[lb]{\Large{\Black{$\bar{X}^{2}_{L}$}}}
    \Text(360,-179)[lb]{\Large{\Black{$X^{2}_{L}$}}}
    \ArrowLine(540,-149)(480,-149)
    \Vertex(480,-149){2.83}
    \DashArrowLine(480,-104)(480,-149){10}
    \COval(480,-104)(2.83,2.83)(45.0){Black}{White}\Line(478.59,-105.41)(481.41,-102.59)\Line(478.59,-102.59)(481.41,-105.41)
    \COval(480,-104)(2.83,2.83)(45.0){Black}{White}\Line(478.59,-105.41)(481.41,-102.59)\Line(478.59,-102.59)(481.41,-105.41)
    \COval(480,-104)(2.83,2.83)(45.0){Black}{White}\Line(478.59,-105.41)(481.41,-102.59)\Line(478.59,-102.59)(481.41,-105.41)
    \ArrowLine(435,-149)(390,-149)
    \ArrowLine(435,-149)(480,-149)
    \Line(433,-151)(437,-147)\Line(433,-147)(437,-151)
    \Text(510,-179)[lb]{\Large{\Black{$\nu^{c}$}}}
    \Text(450,-179)[lb]{\Large{\Black{$X^{3}_{L}$}}}
    \Text(405,-179)[lb]{\Large{\Black{$\bar{X}^{3}_{L}$}}}
    \Text(465,-89)[lb]{\Large{\Black{$\langle H \rangle$}}}
    \Text(195,-89)[lb]{\Large{\Black{$\langle \bar{\phi}_{23} \rangle$}}}
    \Text(285,-89)[lb]{\Large{\Black{$\langle \bar{\phi}_{23} \rangle$}}}
    \Text(375,-89)[lb]{\Large{\Black{$\langle H_{45} \rangle$}}}
  \end{picture}
}
\caption{\FN \ diagram involving $H_{45}$ and \LH \ messengers.}
\label{fig:H45L}
\end{center}
\end{figure}

The differences between the $(1,2)$ and $(1,3)$ elements of $Y^{d}$,
needed to fit the data (eq.(\ref{eq:Yd2})), arise from the term in
eq.(\ref{eq:Y_P2_P123}).
Thus, due to $H_{45}$ this contribution is either decoupled or
sub-dominant in the neutrino sector for the reasons given
above. To summarise, eq.(\ref{eq:Y_nu matrix}) is essentially unchanged by the
contribution from eq.(\ref{eq:Y_P23_P23}) and eq.(\ref{eq:Y_P2_P123}) (although if the dominant messengers are \LH , we must impose a bound on $\epsilon_{\nu_{L}}$).

\subsection{Majorana masses\label{sub:Majorana}}

The heavy \RH \ neutrino Majorana mass matrix has its largest
contribution coming from the operator in eq.(\ref{eq:M_th_th}), which gives rise to the
dominant $(M_{N_{R}})_{33}$ component: 
\begin{equation}
\left( M_{N_{R}}\right) _{33}\simeq M_{3}\simeq \frac{\langle \bar{%
\theta}\rangle ^{2}}{M_{\nu _{R}}}  \label{eq:M3}
\end{equation}
The terms of eq.(\ref{eq:M_P23_P23}) and eq.(\ref{eq:M_P123_P123}) give the
Majorana mass matrix of the form:
\begin{equation}
M_{N_{R}}\simeq M_{3}\left( 
\begin{array}{ccc}
\lambda _{1}\left( \frac{\epsilon _{\nu _{R}}}{\epsilon _{d}}\right)
^{4}\epsilon _{d}^{5} & \lambda _{1}\left( \frac{\epsilon _{\nu _{R}}}{%
\epsilon _{d}}\right) ^{4}\epsilon _{d}^{5} & \lambda _{3}\left( \frac{%
\epsilon _{\nu _{R}}}{\epsilon _{d}}\right) ^{4}\epsilon _{d}^{5} \\ 
\lambda _{1}\left( \frac{\epsilon _{\nu _{R}}}{\epsilon _{d}}\right)
^{4}\epsilon _{d}^{5} & \lambda _{2}\left( \frac{\epsilon _{\nu _{R}}}{%
\epsilon _{d}}\right) ^{4}\epsilon _{d}^{4} & \lambda _{4}\left( \frac{%
\epsilon _{\nu _{R}}}{\epsilon _{d}}\right) ^{4}\epsilon _{d}^{4} \\ 
\lambda _{3}\left( \frac{\epsilon _{\nu _{R}}}{\epsilon _{d}}\right)
^{4}\epsilon _{d}^{5} & \lambda _{4}\left( \frac{\epsilon _{\nu _{R}}}{%
\epsilon _{d}}\right) ^{4}\epsilon _{d}^{4} & 1%
\end{array}%
\right)  \label{eq:M matrix}
\end{equation}%
In eq.(\ref{eq:M matrix}) we explicitly show the $O(1)$ factors coming from
the couplings associated with the contributions of different operators such as those of eq.
(\ref{eq:M_P23_P23}) and eq.(\ref{eq:M_P123_P123}): the $\lambda _{i}$. This makes it easy to see
the equality of entries, particularly relevant in the (1,2) quadrant. This quadrant has a rather specific structure that comes about due to eq.(\ref{eq:M_P123_P123}) being the only contribution to the three entries proportional to $\lambda_{1}$: when combining the Dirac matrix of eq.(\ref{eq:Y_nu matrix}) with the Majorana matrix of eq.(\ref{eq:M matrix}) through the seesaw mechanism (eq.(\ref{eq:seesaw})), one obtains precisely the effective
neutrino Lagrangian shown in eq.(\ref{eq:effL}) that leads to neutrino \TBM .

\section{Phenomenological implications \label{sec:Phenomenology}}

By construction, the forms of the up quark masses in eq.(\ref{eq:Y_u matrix}) and of the down quark masses in eq.(\ref{eq:Y_d matrix}) are in
agreement with the phenomenological fits of eq.(\ref{eq:Yu2}) and
eq.(\ref{eq:Yd2}). If we further have $g_{\odot }=g_{@}$, giving a
symmetric mass structure
\footnote{As stated in section \ref{sec:Masses}, symmetric mass matrices are expected from $SO(10)$ so we assume $g_{\odot }=g_{@}$.},
then the $(1,1)$ texture zero enables the
successful \GST \ relation \cite{Gatto:1968ss} relating the light quark
masses to the mixing angle in the $(1,2)$
sector (eq.(\ref{eq:GST})).

In subsection \ref{sec:Dirac} we established that the charged lepton
mass matrix in eq.(\ref{eq:Y_c matrix}) gives the phenomenologically successful relations
$m_{b}\simeq m_{\tau}$ and $m_{\mu}\simeq 3m_{s}$ at the unification
scale. Moreover, the $(1,1)$ texture zero 
also implies that $Det[Y^{e}]\simeq Det[Y^{d}]$ so that $m_{e}\simeq
m_{d}/3$ at the unification scale, 
again in excellent agreement with experiment once one includes the radiative
corrections to the masses \cite{Roberts:2001zy, Mario}. The
contribution to the mixing angles in the lepton sector is given by:
\begin{eqnarray}
\theta _{12}^{l} &\simeq &\sqrt{\frac{m_{e}}{m_{\mu }}} \\
\theta _{23}^{l} &\simeq &\frac{m_{\mu }}{m_{\tau }}   \\
\theta _{13}^{l} &\simeq &\frac{\sqrt{m_{e}m_{\mu }}}{m_{\tau }}  
\end{eqnarray}
The charged lepton mixing is given to good approximation by ratios of charged lepton masses. In turn, we know that the lepton masses are related by the model to the down quark masses (due to the
 \GJ \  \GUT \ scale relations). Finally, as the up quarks have an even stronger hierarchy than the down quarks, the down quark mixing contributes dominantly to the CKM angles, and the ratios of down quark masses give to good approximation the CKM angles. This shows that in this model, the charged lepton angles are connected to the CKM angles. 

The neutrino masses and mixing angles can also be determined. The Majorana
mass matrix has mass ratios given by: 
\begin{eqnarray}
\frac{M_{1}}{M_{3}} &\simeq&\left( \frac{\epsilon _{\nu _{R}}}{\epsilon _{d}}%
\right) ^{4}\epsilon _{d}^{5} \\
\frac{M_{2}}{M_{3}} &\simeq&\left( \frac{\epsilon _{\nu _{R}}}{\epsilon _{d}}%
\right) ^{4}\epsilon _{d}^{4}  
\end{eqnarray}%
Due to the large hierarchy in the Majorana mass matrix between $M_{1}$, $M_{2}$ and $M_{3}$, the contribution to the light neutrino masses from the
exchange of the heaviest \RH \ neutrino is
negligible. This is despite the fact that the dominant Yukawa
couplings in the Dirac mass matrix are to that \RH \
neutrino: this is the realisation of the \SD \ strategy discussed in section \ref{sec:fsintro}, and explains the
mismatch in the \FS \ breaking patterns in the charged
fermions and neutrino sector.

The light neutrino masses are given by:
\begin{equation}
m_{@} \simeq \frac{\epsilon _{\nu}^{4}\epsilon _{d}^{2}h^{2}}{M_{1}}
\end{equation}
\begin{equation}
m_{\odot } \simeq \frac{\epsilon _{\nu}^{4}\epsilon _{d}^{2}h^{2}}{M_{2}}
\label{eq:mn2}
\end{equation}%
\begin{equation}
m_{1} \simeq \frac{\left( \frac{\epsilon _{\nu}}{\epsilon _{d}}\right)
^{4}h^{2}}{M_{3}} \label{eq:mn3}
\end{equation}%
$h$ is the \VEV \ of the doublet $H$ Higgs that generates the Dirac neutrino masses (and thus also the up quark masses). $\epsilon_{\nu}$ can be either $\epsilon_{\nu_{L}}$ or $\epsilon_{\nu_{R}}$ as discussed in subsection \ref{sec:Dirac}, although note that the ratio between the light neutrino
masses does not depend on $\epsilon_{\nu}$. We have absorbed the $O(1)$ couplings, and up to these $O(1)$ factors the light mass ratios are given by:
\begin{eqnarray}
\frac{m_{\odot }}{m_{@}} &\simeq&\epsilon _{d} \\
\frac{m_{1}}{m_{\odot }} &\simeq&\left( \frac{\epsilon _{\nu _{R}}}{
\epsilon _{d}}\right) ^{4}\epsilon _{d}^{-2} \ll 1  
\end{eqnarray}

In such a hierarchical mass structure (with $m_1 \simeq 0$), the observed squared mass differences
relevant for atmospheric and solar oscillations are approximately
given by $m_{@}^{2}$ and $m_{\odot }^{2}$ respectively. Up to the
$O(1)$ coefficients, $m_{@}=\epsilon _{d}\left( \frac{\epsilon
_{\nu}}{\epsilon _{\nu _{R}}}\right) ^{4}\frac{1}{M_{3}}$,
and a fit to atmospheric oscillation is
readily obtained by a suitable choice of
$\left( \frac{\epsilon _{\nu}}{\epsilon_{\nu _{R}}}\right)
^{4}\frac{1}{M_{3}}$ (if $\epsilon_{\nu}=\epsilon_{\nu_{R}}$, we have directly constrained $M_3$). Having fitted these
parameters, the solar oscillation mass squared difference is predicted
by this model to be $m_{\odot }^{2} \simeq \epsilon
_{d}^{2}m_{@}^{2}$. With the $\epsilon_{d}$ expansion parameter given in
eq.(\ref{eq:u,d eps2}), fixed by fitting the down type quark and
charged lepton mass hierarchy, we obtain excellent agreement with the
magnitude of the mass difference found in solar neutrino oscillation.

The neutrino mixing angles are readily obtained. To understand the results it is
instructive first to neglect the off-diagonal terms in the Majorana mass
matrix. The dominant exchange term in the seesaw mechanism is $\nu _{1}^{c}$. From eq.(\ref{eq:Y33}) to eq.(\ref{eq:Y_P123_P23}) we see that $\nu _{1}^{c}$
only couples via eq.(\ref{eq:Y_P23_P123}) to the combination
$\bar{\phi}_{23}^{i}\nu _{i} \propto (\nu _{\mu }-\nu _{\tau }) \equiv \nu_{@}$ (defining $\nu_{@}$). As a result the
most massive neutrino is close to bi-maximally mixed. The exchange of $\nu
_{2}^{c}$ is responsible for generating the next most massive neutrino. From
eq.(\ref{eq:Y33}) to eq.(\ref{eq:Y_P123_P23}) we see that it couples by both eq.(\ref{eq:Y_P23_P123}) and eq.(\ref{eq:Y_P123_P23}) to the combination $\nu
_{@}+\nu_{\odot}$, with $\bar{\phi}_{123}^{i}\nu _{i} \propto (\nu _{e}+\nu _{\mu }+\nu _{\tau }) \equiv \nu_{\odot}$ (defining $\nu_{\odot}$). Diagonalising
the masses the effect of this term is to introduce mixing at
$O(\frac{m_{\odot }}{m_{@}})$ in the most massive state between the
combinations $\nu_{@}$
and $\nu _{\odot}$.
However we have not yet introduced the effect of the off-diagonal terms in
the Majorana mass matrix, notably the entries $(M_{N_{R}})_{12}$ and  $(M_{N_{R}})_{21}$, which also
introduce such mixing. Taking the off-diagonal terms into account we find that, due to the underlying symmetry
of the theory, these mixing terms cancel between the two
contributions.

It is perhaps easier to understand the exact cancellation between the two contributions by using the following effective symmetry reasoning: the effective Lagrangian of eq.(\ref{eq:effL}) doesn't have any terms mixing $\nu_{@}$ and $\nu_{\odot}$. This follows from an effective $Z_{2}$ symmetry, that the neutrino Dirac and Majorana terms possess, under which only one of the flavons transforms non-trivially, say $\bar{\phi}_{23} \rightarrow - \bar{\phi}_{23}$. Under this effective symmetry the Yukawa terms in eq.(\ref{eq:Y_P23_P123}) and eq.(\ref{eq:Y_P123_P23}) are no longer invariant, unless we have also $\psi^{c} \rightarrow - \psi^{c}$. The term in eq.(\ref{eq:Y_P23_P23}) would violate the effective symmetry, but it is decoupled from the neutrino sector due to $H_{45}$. With these $Z_{2}$ assignments, the cross terms $\bar{\phi}_{23} \psi \bar{\phi}_{123} \psi$ and $\bar{\phi}_{23} \psi \bar{\phi}_{123} \psi$ are not allowed in the effective Lagrangian: $L_{\nu}=\lambda_{23} \left( \bar{\phi}_{23} \psi \right)^{2}+\lambda_{123} \left( \bar{\phi}_{123} \psi \right)^{2}$ (precisely the effective Lagrangian of eq.(\ref{eq:effL})). 
Notice that the allowed Majorana terms (eq.(\ref{eq:M_P23_P23}) and eq.(\ref{eq:M_P123_P123})) are automatically invariant as they only include pairs of the fields charged under the effective $Z_{2}$. Due to the symmetry, when the heavy neutrinos $\nu^{c}$ are integrated out, the effective neutrino states $\nu_{@}$ and $\nu_{\odot}$ don't mix.
In any case, the end result is that the effective Majorana neutrinos
have the effective Lagrangian of eq.(\ref{eq:effL}) that leads to
exact neutrino \TBM:
\begin{eqnarray}
\sin ^{2}\theta _{12}^{\nu } &=&\frac{1}{3} \\
\sin ^{2}\theta _{23}^{\nu } &=&\frac{1}{2} \\
\sin ^{2}\theta _{13}^{\nu } &=&0
\end{eqnarray}%
It is the underlying \FS \ that is responsible
for these predictions, predominantly by shaping the \VEV s \ in
eq.(\ref{eq:P23B vev}) and eq.(\ref{eq:P123B vev}).

Finally, to obtain the measurable PMNS angles we must take into account also the contributions from the charged lepton sector
 (the neutrino angles are only
equivalent to the PMNS angles in the basis where the charged leptons
are diagonal, which is not true in our basis as shown in eq.(\ref{eq:Y_c matrix})).
We should stress that the actual value of the corrections arising from
the charged leptons depends on the value of the CP violating phase of
the lepton sector, as shown explicitly in 
\cite{Plentinger:2005kx, Antusch:2005kw}. The model doesn't allow us
to predict this phase independently (it originates from unknown phases
of the fields involved in the vacuum 
alignment).
Considering the values of the CP violating phase that predict the
largest deviations from tri-bi-maximal values, we obtain a range of
possible values for the angles given by:

\begin{eqnarray}
\sin \theta _{12} &\simeq &\frac{1}{\sqrt{3}}\left( 1\pm \theta
_{12}^{l}\right) \\
\sin \theta _{23} &\simeq &\frac{1}{\sqrt{2}}\left( 1\pm \theta
_{12}^{l}\right) \\
\sin \theta _{13} &\simeq &\frac{1}{\sqrt{2}}\theta _{12}^{l}
\end{eqnarray}%
thus we have:

\begin{equation}
\sin^{2}\theta_{12} = \frac{1}{3}\pm_{0.048}^{0.052}
\label{eq:st12}
\end{equation}

\begin{equation}
\sin^{2}\theta_{23} = \frac{1}{2}\pm_{0.058}^{0.061}
\label{eq:st23}
\end{equation}

\begin{equation}
\sin^{2}\theta_{13} = 0.0028
\label{eq:st13}
\end{equation}
Values that are in good agreement with the experimentally measured
ones. Eq.(\ref{eq:st13}) can be written as
$\theta_{13} \simeq \theta_C /(3\sqrt{2}) \simeq 3^o$, where
$\theta_C$ is the Cabibbo angle. This prediction demonstrates a
relation with the quark Cabibbo angle, with the leptonic angle getting a relative factor of $1/3$  
due to the respective \GJ \ relation and a relative factor of $1 / \sqrt{2}$ due to
commutation through the maximal atmospheric angle. $\theta_{12}$ can
also be related to $\theta_{13}$ and the CP violating phase $\delta$ of the CKM matrix,
via the so called neutrino sum rule \cite{King:2005bj}: 
\begin{equation}
\theta_{12}+\theta_{13}\cos(\delta -\pi)\simeq 35.26^o
\label{eq:Sum_rule}
\end{equation}
The near \TBM \ values of eq.(\ref{eq:st12}), eq.(\ref{eq:st23}) and eq.(\ref{eq:st13}), as well as the relation in eq.(\ref{eq:Sum_rule}) are general predictions that apply to a whole class of models. They are valid provided that the model features both exact neutrino \TBM \ and quark-lepton unification. If these two features are verified in a model, then it will predict that the PMNS parameters deviate from the \TBM \ values by small corrections that are related to the CKM parameters, and relations like eq.(\ref{eq:st12}), eq.(\ref{eq:st23}), eq.(\ref{eq:st13}), and eq.(\ref{eq:Sum_rule}) can be obtained (as seen in the original models belonging to this class, \cite{King:2005bj} and \cite{Ivo1}).

It is relevant to consider other phenomenological implications of the
theory that go beyond the mixing angles: in particular, there is a longstanding problem associated with having a gauged \FS \ in a supersymmetric theory. The problem is due
to the fact that the $D$-terms are typically non-vanishing and contribute to
the soft masses of the sfermions in a family dependent way (different for each generation). This is
potentially disastrous as non-degenerate sfermion masses can lead to unacceptably large
\FCNC s. From
\cite{Kawamura:1994ys} one can see that the effect is proportional to the
difference in the mass squared of the two fields developing large \VEV s 
along the $D$-flat direction. As a result the effect can be suppressed
if these masses are closely degenerate. In \cite{King:2003rf} it is
shown how this condition could naturally arise in an $SU(3)_{f}$ model and the
same structure can be used here. This is not the only way the
$D$-term contribution may be negligibly small. A specific example
follows when the SUSY breaking mediator mass is less 
than the family breaking scale because the radiative graphs generating
the dangerous soft masses are suppressed by the ratio of the two
scales. This will be the case in this model for gauge mediated
supersymmetry breaking. A more extensive discussion of the suppression
of $D$-term soft mass contributions is presented in Chapter \ref{ch:fcnc}.

\section{Anomalies \label{sec:Anom}}

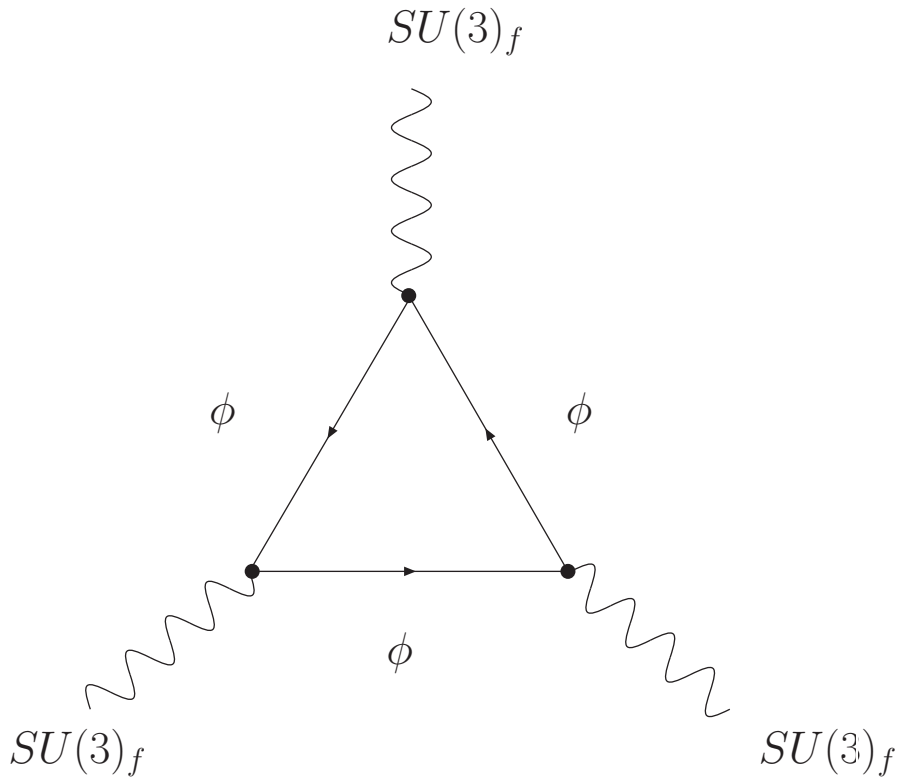
\begin{figure}[htb]
\begin{center}
\fcolorbox{white}{white}{
  \begin{picture}(314,289) (262,-244)
    \SetWidth{0.5}
    \SetColor{Black}
    \Vertex(472,-166){2.83}
    \ArrowLine(472,-166)(412,-62)
    \ArrowLine(413,-62)(352,-166)
    \ArrowLine(352,-166)(473,-166)
    \Vertex(412,-62){2.83}
    \Vertex(353,-166){2.83}
    \Photon(413,-62)(413,16){7.5}{4}
    \Photon(352,-166)(292,-218){7.5}{4}
    \Photon(473,-166)(533,-218){7.5}{4}
    \Text(338,-114)[lb]{\Large{\Black{$\phi$}}}
    \Text(405,-205)[lb]{\Large{\Black{$\phi$}}}
    \Text(546,-244)[lb]{\Large{\Black{$SU(3)_{f}$}}}
    \Text(262,-244)[lb]{\Large{\Black{$SU(3)_{f}$}}}
    \Text(405,29)[lb]{\Large{\Black{$SU(3)_{f}$}}}
    \Text(473,-114)[lb]{\Large{\Black{$\phi$}}}
  \end{picture}
}
\end{center}
\caption{$SU(3)_{f}^{3}$ triangle graph.}
\label{fig:Anom}
\end{figure}

The fermions belong to complex representations of the $SU(3)_{f}$ gauged \FS , and as such contribute to triangle graph anomalies exemplified by figure \ref{fig:Anom}. 

In order to have a ``safe'' model, it is necessary to arrange a cancellation of this type of anomalies by arranging the field content. Although we haven't considered this in detail, it is always possible to cancel outstanding anomalies associated with the $SU(3)_{f}$ by adding suitable \SM \ singlets (like the flavons) as necessary. For example, regarding the $SU(3)_{f}^{3}$ anomaly displayed in figure \ref{fig:Anom} we note that the field content in table \ref{Ta:Table 1} does not lead to such a cancellation: we have the contribution coming from the \SM \ fields: $16$  fermions $\psi$ and $16$ fermions $\psi^{c}$ - all are triplets under $SU(3)_{f}$. Then there are the $\theta$ and $\bar{\theta}$ fields, the same number of triplet and anti-triplet respectively, which cancel between themselves. The flavons are arranged as triplet and anti-triplet pairs as well (with the exception of $\bar{\phi}_{3}$, transforming non-trivially under $SU(2)_{R}$). In order to cancel the outstanding contribution of the \SM \ fermions, as triplets, to the $SU(3)_{f}$ anomaly, we require an appropriate number of \SM \ singlet, $SU(3)_{f}$ anti-triplet fields.

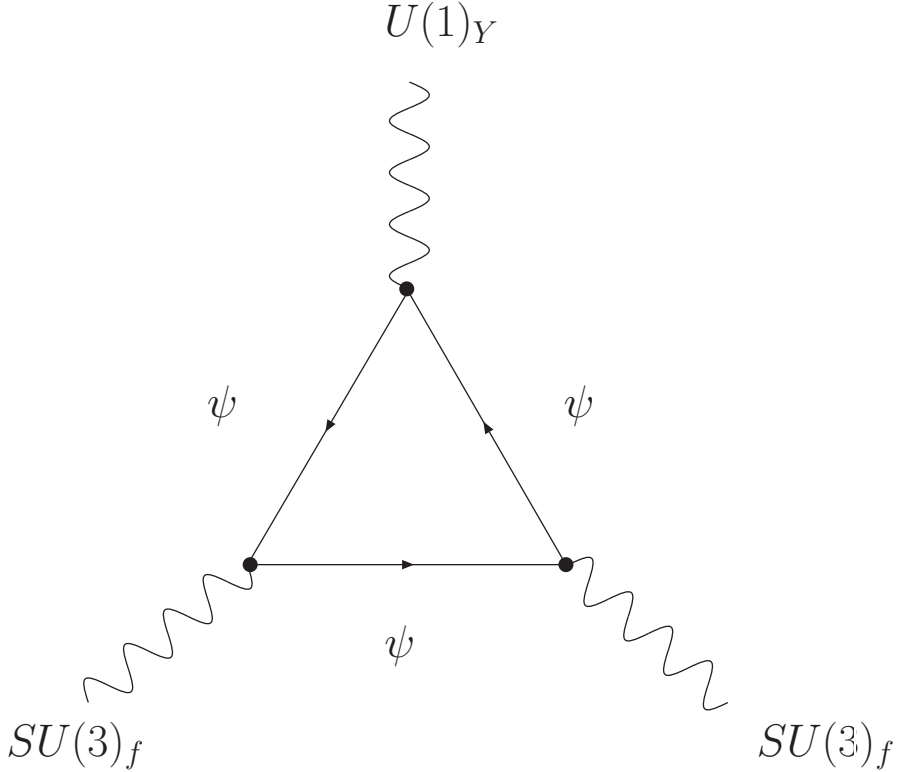
\begin{figure}[htb]
\begin{center}
\fcolorbox{white}{white}{
  \begin{picture}(314,289) (262,-244)
    \SetWidth{0.5}
    \SetColor{Black}
    \Vertex(472,-166){2.83}
    \ArrowLine(472,-166)(412,-62)
    \ArrowLine(413,-62)(352,-166)
    \ArrowLine(352,-166)(473,-166)
    \Vertex(412,-62){2.83}
    \Vertex(353,-166){2.83}
    \Photon(413,-62)(413,16){7.5}{4}
    \Photon(352,-166)(292,-218){7.5}{4}
    \Photon(473,-166)(533,-218){7.5}{4}
    \Text(338,-114)[lb]{\Large{\Black{$\psi$}}}
    \Text(405,-205)[lb]{\Large{\Black{$\psi$}}}
    \Text(546,-244)[lb]{\Large{\Black{$SU(3)_{f}$}}}
    \Text(262,-244)[lb]{\Large{\Black{$SU(3)_{f}$}}}
    \Text(473,-114)[lb]{\Large{\Black{$\psi$}}}
    \Text(405,29)[lb]{\Large{\Black{$U(1)_{Y}$}}}
  \end{picture}
}
\end{center}
\caption{$U(1)_{Y}$ triangle graph.}
\label{fig:Anom2}
\end{figure}

Other triangle graphs involve the \SM \ gauge bosons and are safe. The one in figure \ref{fig:Anom2} has an $U(1)_Y$ vertex.
Naturally, only the \SM \ fermions contribute to these diagrams. Factoring out the $SU(3)_{f}$ part of the diagram common to all \SM \ fermions (all triplets), the charges are known to precisely cancel the anomalies involving the \SM \ gauge group between the \LH \ and the \RH \ \SM \ fermions (particularly in figure \ref{fig:Anom2}, the hypercharge assignments arrange the cancellation).

\section{Summary and conclusions \label{sec:Confmm1}}

In this continuous model we have shown how near \TBM \
in the lepton sector arises from a spontaneously broken $SU(3)_{f}$ \FS . This comes about through a very specific \VEV \
alignment. The model has a phenomenologically 
acceptable pattern of \FMM . It generates the successful
\GST \ relation between the mixing angles and masses of the first two
generations and, primarily due to the underlying \GUT, it generates
the \GJ \ relations between down-type quarks and
charged leptons. The neutrino sector features precise \TBM \
and $\theta^{\nu}_{13}=0$. The charged lepton sector generates
small corrections to \TBM \ and a non-zero value for
$\theta_{13}$ - as the charged lepton sector is connected to the
down-type quark sector, these corrections are related also to the CKM
angles (namely $\theta_{C}$). The seesaw mechanism plays a very important role in explaining
the striking difference between quark and lepton mixing angles:
because the dominant \FS \ breaking contribution is along
the $3$ direction, it dominates the quark and 
charged lepton masses; however, in the effective light neutrino sector
it is suppressed by the dominant heaviest \RH \ neutrino mass
(also in the $3$ direction) through the seesaw and \SD . The model has a
non-trivial multiplet content, 
particularly due to the symmetry breaking sector. However it
represents only one of a large class of such models and more elegant
versions should exist. In any case our example demonstrates the existence of a
phenomenologically satisfactory model in this general class, and also
establishes the general strategy and the framework that is going to be followed in the
model based on $\Delta(27)$, discussed in section \ref{sec:D27}.

\chapter{Discrete \FSs \label{ch:fsd}}

\section{Outline \label{sec:fsdintro}}
In chapter \ref{ch:fs} we presented a model based on a continuous \FS . We now turn to the use of discrete \FSs .

We start by studying a simple (but incomplete) model based on the non-Abelian discrete group $A_4$ in section \ref{sec:D12}. We denote the group $A_4$ as $\Delta(12)$ (of the ``Dihedral-like'' groups $\Delta(3 n^{2})$, with $n=2$ \cite{FFK, Muto:1998na, Luhn:2007uq}) and approach it as the semi-direct product group $Z_2 \rtimes Z_3^{\prime }$, as in \cite{Ivo2} (it is a semi-direct product due to $Z_{3}^{\prime}$ not being simultaneously diagonal with $Z_{2}$, see table \ref{ta:D12_trans}). This approach more closely generalises to the approach used in section \ref{sec:D27} (and in \cite{Ivo3}) for $\Delta(27)$ as a semi-direct product group (see table \ref{transformation}). 
Our aim is to show merely how to obtain the required \VEV s similar to eq.(\ref{eq:P3B vev}), eq.(\ref{eq:P23B vev}) and eq.(\ref{eq:P123B vev}); we won't dwell on the Yukawa superpotential and thus won't derive \FMM \ from these special \VEV s (in \cite{King:2006np}, an $A_4$ model with underlying $SO(3)_{f}$ is used to obtain \TBM \ with analogous \VEV s, which are however obtained by using an alignment mechanism similar to the one of \cite{Ivo3}).

We proceed by presenting a model based on $\Delta(27)$ ($\Delta(3 n^{2})$, with $n=3$ \cite{FFK, Muto:1998na, Luhn:2007uq}) in section \ref{sec:D27}. This model follows the framework described in section \ref{sec:fsintro}. The main
distinguishing feature that the discrete implementation (based on $\Delta(27)$) has over the continuous implementation (based on $SU(3)_{f}$)
lies in the symmetry breaking discussion, 
which is described in subsection \ref{sec:VEVs and masses}. The
superpotential terms leading to the \FMM \ are similar to the
ones in the continuous model, and are briefly
discussed in subsection \ref{sub:Yukawa}.

Section \ref{sec:Confmm2} concludes with a summary 
of the results and features of the discrete models.

\section{$\Delta(12)$ \FS \ model \label{sec:D12}}

$\Delta(12)$ has only one representation with dimension greater than one, and it is a representation with dimension $3$: a ``triplet''. The transformation of a triplet $\phi_{i}$ under the group (expressed through its $Z_{2}$ and $Z_{3}^{\prime}$ factors) is shown in table \ref{ta:D12_trans}.

\begin{table}[htb] \centering%
\begin{tabular}{|c||c|c|}
\hline
Field & $Z_{2}$ & $Z_{3}^{\prime}$ \\ \hline \hline
$\phi _{1}$ & $\phi _{1}$ & $\phi _{2}$ \\
$\phi _{2}$ & $- \phi _{2}$ & $\phi _{3}$ \\
$\phi _{3}$ & $- \phi _{3}$ & $\phi _{1}$ \\ \hline
\end{tabular}
\caption{Triplet field $\phi$ transforms under $\Delta(12)$.}
\label{ta:D12_trans}
\end{table}

The invariants that will be relevant for the following discussion are the two triplet invariant $\phi_{i} \phi_{i} \equiv \phi_{1} \phi_{1}+\phi_{2} \phi_{2}+\phi_{3} \phi_{3} $ and the cyclic three triplet invariant $\phi_{i} \phi_{i} \phi_{i} \equiv  \phi_{1} \phi_{2} \phi_{3} +  \phi_{3} \phi_{1} \phi_{2} + \phi_{2} \phi_{3} \phi_{1}$.

In order to illustrate how the \VEV \ alignment can proceed, we introduce the triplet flavons $\phi_{3_{i}}$, $\phi_{23_{i}}$ and $\phi_{123_{i}}$,  use three $\Delta(12)$ triplet alignment fields $X_{3_{i}}$, $X_{23_{i}}$ and $X_{123_{i}}$, and use also two $\Delta(12)$ (trivial representation) singlet alignment fields, $Y_{3}$ and $Y_{23}$. The fields transform under $\Delta(12)$ as shown in table \ref{ta:D12_fields}. Table \ref{ta:D12_fields} also shows how the fields transform under the additional auxiliary symmetry $Z_{3} \otimes U(1)_{R}$.

\begin{table}[htb] \centering
\begin{tabular}{|c||c|c|c|}
\hline
Field & $\Delta(12)$ & $Z_{3}$ & $U(1)_{R}$ \\ \hline \hline
$\phi_{3}$ & $\mathbf{3}$ & $\mathbf{2}$ & $\mathbf{0}$ \\
$\phi_{23}$ & $\mathbf{3}$ & $\mathbf{1}$ & $\mathbf{1}$ \\
$\phi_{123}$ & $\mathbf{3}$ & $\mathbf{0}$ & $\mathbf{0}$ \\
\hline
$X_{3}$ & $\mathbf{3}$ & $\mathbf{2}$ & $\mathbf{2}$ \\
$Y_{3}$ & $\mathbf{1}$ & $\mathbf{2}$ & $\mathbf{2}$ \\
\hline
$X_{23}$ & $\mathbf{3}$ & $\mathbf{0}$ & $\mathbf{1}$ \\
$Y_{23}$ & $\mathbf{1}$ & $\mathbf{2}$ & $\mathbf{1}$ \\
\hline
$X_{123}$ & $\mathbf{3}$ & $\mathbf{0}$ & $\mathbf{2}$ \\
\hline
\end{tabular}
\caption{$\Delta(12)$ model: symmetries and charges.} \label{ta:D12_fields}
\end{table}
With these fields and symmetry assignments, it is convenient to divide the allowed terms in the alignment superpotential $P_{A}$ among $P_{3}$, $P_{23}$ and $P_{123}$ depending on which field they are relevant to:

\begin{equation}
P_{A} = P_{3} + P_{23} + P_{123}
\label{eq:P_A}
\end{equation}

Focusing on $\phi_{3}$ first, the respective superpotential includes two terms:

\begin{equation}
P_{3} =  X_{3_{i}} \phi_{3_{i}} \phi_{3_{i}} + Y_{3} (\phi_{3_{i}} \phi_{3_{i}} + \mu_{3})
\label{eq:P_A3}
\end{equation}
$\mu_{3}$ is a mass scale that can arise through radiative breaking of another field with appropriate symmetry assignments (see the discussion in appendix \ref{app_A}, where $S_3$ plays a similar role in the $SU(3)_{f}$ model). The $F$-term associated with $Y_{3}$ forces a non-vanishing \VEV \ with fixed magnitude for $\phi_{3}$. The direction of the \VEV \ is then decided through its coupling with the triplet alignment field $X_{3_{i}}$, with the $F$-terms associated with its three components forcing two of the three entries of $\langle \phi_{3} \rangle$ to vanish: for example, the $F$-term associated with the first component ($X_{3_{1}}$) leads to the minimisation condition $\langle \phi_{3_{2}} \rangle \langle \phi_{3_{3}} \rangle = 0$, the other two conditions being cyclic permutation of the indices. Without loss of generality, we define the third direction to be the non-vanishing direction of $\langle \phi_{3} \rangle$.

$\phi_{123}$ is aligned by the coupling to the triplet $X_{123_{i}}$:

\begin{equation}
P_{123} = g_{123} X_{123_{i}} \phi_{123_{i}} + X_{123_{i}} \phi_{123_{i}} \phi_{123_{i}}
\label{eq:P_A123}
\end{equation}
If $\phi_{123}$ acquires a non-vanishing \VEV , $F$-terms associated with the three components of $X_{123}$ force all three entries of $\langle \phi_{3} \rangle$ to be non-vanishing. The first component of $X_{123}$ leads to the minimisation condition $\langle \phi_{123_{1}} \rangle =- \frac{ \langle \phi_{123_{2}} \rangle \langle \phi_{123_{3}} \rangle}{g_{123}}$, and the remaining conditions are again cyclic permutations. To solve all three conditions it is required that all components of $\langle \phi_{123} \rangle$ have magnitude $\left| g_{123} \right|$. Importantly, the phases of the three components must also be related.

To finalise, we want $\langle \phi_{23} \rangle$ to be aligned relative to the previous \VEV s. The terms in the superpotential unsurprisingly include terms where $\phi_{23}$ is mixed with the other flavons:

\begin{equation}
P_{23} =  X_{23_{i}} \phi_{3_{i}} \phi_{23_{i}}
+ Y_{23}  (g_{23} \phi_{23_{i}} \phi_{123_{i}} + \phi_{23_{i}} \phi_{123_{i}} \phi_{123_{i}})  \label{eq:P_A23}
\end{equation}
If $\phi_{23}$ acquires a non-vanishing \VEV , the coupling with $Y_{23}$ ensures $\langle \phi_{23} \rangle$ is orthogonal to $\langle \phi_{123} \rangle$: the minimisation condition is $g_{23} (\langle \phi_{23_{1}} \rangle \langle \phi_{123_{1}} \rangle + ...) + (\langle \phi_{23_{1}} \rangle \langle \phi_{123_{2}} \rangle \langle \phi_{123_{3}} \rangle + ...) = 0$, where the ``$...$'' stand for cyclic permutation of the indices. Although this single condition may seem complicated, when taking into consideration the special form of $\langle \phi_{123} \rangle$, the equation simplifies greatly: for example, $\langle \phi_{23_{1}} \rangle \langle \phi_{123_{2}} \rangle \langle \phi_{123_{3}} \rangle = - g_{123} \langle \phi_{23_{1}} \rangle \langle \phi_{123_{1}} \rangle$ (as we have the $\langle \phi_{123} \rangle$ conditions from $P_{123}$). We conclude that the  minimisation condition associated with $Y_{23}$ simplifies into $(1 - g_{123}) (\langle \phi_{23_{1}} \rangle \langle \phi_{123_{1}} \rangle + ...) = 0$, and the desired orthogonality between the two triplets \VEV s follows. Two components of $\langle \phi_{23} \rangle$ must have equal magnitude, with the other component vanishing (and with the phases of the non-vanishing components being related to each other and with the phases of $\langle \phi_{123} \rangle$).
To complete the \VEV \ alignment discussion we still need to ensure that the vanishing entry isn't $\langle \phi_{23_{3}} \rangle$ - precisely what the minimisation conditions coming from $X_{23}$ prevent (namely, $\langle \phi_{3_{3}} \rangle \langle \phi_{23_{1}} \rangle = 0$ associated with $X_{23_{1}}$).

\section{$\Delta(27)$ \FS \ model \label{sec:D27}}

We consider now a complete model based on a discrete non-Abelian symmetry
\footnote{Such non-Abelian discrete symmetries often occur in
compactified string models.}.
The group we use is $\Delta (27)$, the semi-direct product group 
$(Z_{3} \times Z_{3}) \rtimes Z_{3}^{\prime }$, which is a subgroup of
$SU(3)_{f}$ \cite{FFK, Muto:1998na, Ma:2006ip, Luhn:2007uq, Ma:2007ia}.

The dominant terms of the Lagrangian leading to the Yukawa
coupling matrices of the form of eq.(\ref{eq:Yu2}) and
eq.(\ref{eq:Yd2}) are symmetric under $SU(3)_{f}$, so much of the
structure of the continuous model discussed in section \ref{sec:Symmetries} is
maintained. However the appearance of additional terms allowed by
$\Delta(27)$ (but not by $SU(3)_{f}$) determines the \VEV \ 
structure and generates \TBM . The choice of
the multiplet structure ensures that the model is consistent with a stage of
unification (\GUT \ or superstring) and the resulting model is much simpler
than the model based on the continuous $SU(3)_{f}$ symmetry.

\subsection{Field content and symmetries\label{sec:Fields and symmetries}}

The symmetry of the model is $SU(3)_{C} \otimes SU(2)_{L} \otimes
U(1)_{Y} \otimes G_{f} \otimes G$. The additional symmetry group $G$
is needed to restrict the form of the allowed couplings of the theory
and is chosen to be as simple as possible. The
family group $G_{f}$ is chosen as a non-Abelian discrete group of
$SU(3)_{f}$ in a manner that preserves the structure of the fermion
Yukawa couplings of the associated $SU(3)_{f}$ model discussed in detail in chapter \ref{ch:fs}. This means that $G_{f}$ should be a
non-Abelian subgroup of $SU(3)_{f}$ of sufficient size that it
approximates $SU(3)_{f}$ in the 
sense that most of the leading terms responsible for the fermion mass
structure in the $SU(3)_{f}$ model are still the leading terms 
allowed by $G_{f}$ (which being a subgroup, allows further terms which we
require to be sub-leading). The smallest group we have found that achieves
our goals is $\Delta (27)$, generated by $Z_{3}$ factors. We denote
one of the generators as $Z_{3}^{\prime}$ for
convenience: $(Z_{3} \times Z_{3}) \rtimes Z_{3}^{\prime} = \Delta(27)$.

The main change that results from using this smaller, discrete symmetry group is
in the alignment of \VEV s. One reason for the difference is the appearance of additional
invariants, which drive the desired \VEV \ 
structure. Another important reason for this difference is the absence of $D$-terms (we are no longer dealing with a continuous symmetry). The $D$-terms played a very important role in
determining the \VEV s in the $SU(3)_{f}$ model presented in chapter \ref{ch:fs} (see the detailed discussion in appendix \ref{app_A}).
The absence of $D$-terms leads to further differences: we are able to reduce the total field
content of the discrete model. In turn, the reduced field content only requires the introduction of an additional
$G= R \otimes U(1)\otimes Z_{2}$ in order to control the allowed terms in the
superpotential (we required an additional $R \otimes U(1)\otimes U(1)^{\prime }$ for the same
purpose in the continuous model - see table \ref{Ta:Table 1}).
Another benefit of the absent $D$-terms is that the model automatically avoids the associated \FCNC \ problem discussed in chapter \ref{ch:fcnc}.

In choosing the representation content of the theory we are guided by the
structure of the $SU(3)_{f}$ model: the framework described in chapter \ref{ch:fs} generates a 
viable form of \FMM . Since
$\Delta(27)$ is a discrete subgroup of $SU(3)_{f}$, all invariants of
$SU(3)_{f}$ are also invariant under $\Delta(27)$, and we can readily arrange suitable \FMM \ in the $\Delta(27)$ model by using superpotential terms similar to the respective Yukawa superpotential terms of the $SU(3)_{f}$ model. To implement the framework used in the continuous model, we find it convenient to label
the representation of the fields of our model by their transformation
properties under the approximate $SU(3)_{f}$ \FS . For
example, the \SM \ fermions are again denoted as $\psi _{i},\psi _{j}^{c}$ and assigned to
transform as triplets under $\Delta(27)$ (essentially, we keep most of
the notation and assignments used in chapter \ref{ch:fs}, if applicable).

The transformation properties of $\Delta(27)$ anti-triplet $\bar{\phi}^{i}$ and triplet $\phi_{i}$ under the $Z_{3}$ and $Z_{3}'$ factors that generate $\Delta(27)$ are shown in table
\ref{transformation} ($\alpha = e^{i \frac{2 \pi}{3}}$ is the complex cube root of unity, $\alpha^{3}=1$).

\begin{table}[htb] \centering%
\begin{tabular}{|c||c|c|}
\hline
Field & $Z_{3}$ & $Z_{3}^{\prime }$ \\ \hline \hline
$\phi_{1}$ & $\phi_{1}$ & $\phi_{2}$ \\ 
$\phi_{2}$ & $\alpha \phi_{2}$ & $\phi_{3}$ \\ 
$\phi_{3}$ & $\alpha ^{2}\phi_{3}$ & $\phi_{1}$ \\
& & \\
$\bar{\phi} ^{1}$ & $\bar{\phi} ^{1}$ & $\bar{\phi} ^{2}$ \\ 
$\bar{\phi} ^{2}$ & $ \alpha^{2}\bar{\phi} ^{2}$ & $\bar{\phi}%
^{3}$ \\
$\bar{\phi} ^{3}$ & $\alpha \bar{\phi} ^{3}$ & $\bar{\phi} ^{1}$ \\ \hline
\end{tabular}%
\caption{Anti-triplet $\bar{\phi}^{i}$
and triplet $\phi_i$ transform under factors of $\Delta(27)$.}
\label{transformation}
\end{table}

Although the gauge group used is that of the \SM \ it is also
instructive, in considering how the model can be embedded in a unified
structure, to display the properties of the states under the
$SU(4)_{PS}\otimes SU(2)_{L}\otimes SU(2)_{R}$ subgroup of $SO(10)$
and this is done in table \ref{fields} (again, closely following the framework
of the $SU(3)_{f}$ model). The transformation properties of the fields
under the additional symmetry group $G= R  \otimes U(1) \otimes Z_{2}$ are also shown in table \ref{fields}.

The \SM \ Higgs doublets, $H$, responsible for
electroweak breaking \footnote{Two Higgs are required due to SUSY,
both represented as $H$.} transform as singlets under the \FS . Their assignments are listed together with those of the \SM \ fermions, in table
\ref{fields}.

In a complete unified theory, quark and lepton masses will be related. A
particular question that arises in such unification is what generates the
difference between the down quark and charged lepton masses. In
the $SU(3)_{f}$ model this was done through a variant of
the \GJ \ mechanism \cite{Georgi:1979df} via the
introduction of another Higgs field $H_{45}$, which 
transforms as a $45$ of an underlying $SO(10)$ \GUT. $H_{45}$ has a \VEV \
which breaks $SO(10)$ but leaves the \SM \ gauge group 
unbroken. In this model we include $H_{45}$ to demonstrate that the model
readily unifies into a \GUT, but in practise we only use its \VEV. This does not necessarily imply that there is an underlying stage
of unification below the string scale but, if not, the
underlying theory should provide an alternative explanation for the
existence of the pattern of low energy couplings implied by terms
involving $H_{45}$.

At the stage where the \FS \ is unbroken, the \FMM \ are not
generated. To complete the model we break $\Delta(27)$
through the introduction of ``flavons'' that acquire 
\VEV s. Following the $SU(3)_{f}$
model framework, we choose a similar, but simplified flavon structure:
 $\bar{\theta}^{i}$, $\bar{\phi}_{3}^{i}$, $\bar{\phi}_{23}^{i}$
and $\bar{\phi}_{123}^{i}$ are assigned as anti-triplet fields under the
approximate $SU(3)_{f}$, and $\phi _{3_{i}}$, $\phi
_{3_{i}}^{\prime }$ and $\phi_{1_{i}}$ assigned as
triplet fields of the approximate $SU(3)_{f}$ ($\phi_{1_{i}}$ is introduced
for \VEV \ alignment purposes). These assignments are 
shown in table \ref{fields}.

\begin{table}[htbp] \centering
\begin{tabular}{|c||c||c|c|c||c||c|c|}
\hline
Field & $SU(3)_{f}$ & $SU(4)_{PS}$ & $SU(2)_{L}$ & $SU(2)_{R}$ & $R$ & $U(1)$
& $Z_{2}$ \\ \hline\hline
$\psi $ & $\mathbf{3}$ & $\mathbf{4}$ & $\mathbf{2}$ & $\mathbf{1}$ & $%
\mathbf{1}$ & $\mathbf{0}$ & $\mathbf{1}$ \\ 
$\psi ^{c}$ & $\mathbf{3}$ & $\bar{\mathbf{4}}$ & $\mathbf{1}$ & $\bar{\mathbf{2}}$
& $\mathbf{1}$ & $\mathbf{0}$ & $\mathbf{1}$ \\ 
$\bar{\theta} $ & $\bar{\mathbf{3}}$ & $\mathbf{4}$ & $\mathbf{1}$ & $\mathbf{2}$
& $\mathbf{0}$ & $\mathbf{0}$ & $-\mathbf{1}$ \\ \hline
$H$ & $\mathbf{1}$ & $\mathbf{1}$ & $\mathbf{2}$ & $\mathbf{2}$ & $\mathbf{0}
$ & $\mathbf{0}$ & $\mathbf{1}$ \\ 
$H_{45}$ & $\mathbf{1}$ & $\mathbf{15}$ & $\mathbf{1}$ & $\mathbf{3}$ & $%
\mathbf{0}$ & $\mathbf{2}$ & $\mathbf{1}$ \\ \hline\hline
$\phi _{123}$ & $\mathbf{3}$ & $\mathbf{1}$ & $\mathbf{1}$ & $\mathbf{1}$ & $%
\mathbf{0}$ & $\mathbf{-1}$ & $\mathbf{1}$ \\ 
$\phi _{3}$ & $\mathbf{3}$ & $\mathbf{1}$ & $\mathbf{1}$ & $\mathbf{1}$ & $%
\mathbf{0}$ & $\mathbf{3}$ & $\mathbf{1}$ \\ 
$\phi_{1}$ & $\mathbf{3}$ & $\mathbf{1}$ & $\mathbf{1}$ & $\mathbf{1}$ & $%
\mathbf{0}$ & $-\mathbf{4}$ & $-\mathbf{1}$ \\ 
$\bar{\phi}_{3}$ & $\mathbf{\bar{\mathbf{3}}}$ & $\mathbf{1}$ & $\mathbf{1}$
& $\mathbf{3}\oplus \mathbf{1}$ & $\mathbf{0}$ & $\mathbf{0}$ & $-\mathbf{1}$
\\ 
$\bar{\phi}_{23}$ & $\mathbf{\bar{\mathbf{3}}}$ & $\mathbf{1}$ & $\mathbf{1}$
& $\mathbf{1}$ & $\mathbf{0}$ & $-\mathbf{1}$ & $-\mathbf{1}$ \\ 
$\bar{\phi}_{123}$ & $\mathbf{\bar{\mathbf{3}}}$ & $\mathbf{1}$ & $\mathbf{1}
$ & $\mathbf{1}$ & $\mathbf{0}$ & $\mathbf{1}$ & $-\mathbf{1}$ \\ \hline
\end{tabular}%
\caption{$\Delta(27)$ model: symmetries and charges.}\label{fields}%
\end{table}

The symmetry assignments of all the fields lead to the Yukawa structure of
the $SU(3)_{f}$ model, as is discussed in subsection \ref{sub:Yukawa}.
The additional terms allowed by the
$\Delta(27)$ symmetry are sub-leading in this sector so the 
phenomenologically acceptable pattern of \FMM \
obtained in chapter \ref{ch:fs} is reproduced if
the flavon \VEV s are analogous to those 
given in chapter \ref{ch:fs}.
The desired \VEV \ structure is indeed obtained, as presented in subsection \ref{sec:VEVs and masses}. However, the discussion
leading to is completely changed relative to the continuous model. This is not surprising, as
the \VEV \ alignment is affected by
the main differences entailed by the use of $\Delta(27)$ instead of $SU(3)_{f}$: namely, the absence of
$D$-terms (associated only with continuous gauge symmetries) and the
appearance of additional invariants in the alignment superpotential that determines the
\VEV s (due to the smaller symmetry group used).

\subsection{Symmetry breaking \label{sec:VEVs and masses}}

The desired pattern of \VEV s, similar to that of chapter \ref{ch:fs}, is now:

\begin{equation}
\langle \bar{\phi}_{3}\rangle ^{T}=\left(%
\begin{array}{c}
0 \\ 
0 \\ 
1%
\end{array}%
\right)\otimes\left(%
\begin{array}{cc}
a_{u} & 0 \\ 
0 & a_{d}%
\end{array}%
\right)  \label{eq:P3 vev2}
\end{equation}

\begin{equation}
\langle \bar{\phi}_{23}\rangle ^{T}=\left(%
\begin{array}{c}
0 \\ 
-b \\ 
b%
\end{array}%
\right)  \label{eq:P23 vev}
\end{equation}

\begin{equation}
\langle \phi_{123} \rangle \propto \langle \bar{\phi}%
_{123}\rangle ^{T}=\left( 
\begin{array}{c}
c \\ 
c \\ 
c%
\end{array}%
\right)  \label{eq:P123 vev2}
\end{equation}

\begin{equation}
\langle \phi_{1} \rangle \propto \left( 
\begin{array}{c}
1 \\ 
0 \\ 
0%
\end{array}%
\right)  \label{eq:P1 vev}
\end{equation}

\begin{equation}
\langle \bar{\theta} \rangle ^{T} \propto \langle \phi_{3}\rangle
\propto \left( 
\begin{array}{c}
0 \\ 
0 \\ 
1%
\end{array}%
\right)  \label{eq:T vev}
\end{equation}%
The alignment of these \VEV s \ can proceed in various ways. By including
additional driving fields in the manner discussed in
chapter \ref{ch:fs} and in section \ref{sec:D12}, one can arrange $F-$terms that
lead to a scalar potential whose minimum has
the desired \VEV \ alignment. Here we show that a much simpler
mechanism introduced in \cite{Ivo3} achieves the desired alignment.

To understand how this vacuum alignment works we note that unlike in the case for a continuous symmetry, it is not possible in general
to rotate the \VEV \ of a field to a chosen
direction: instead, due to the underlying discrete
symmetry the \VEV \ will be one of a finite set of possible
minima. This may only be apparent if higher order terms in the
potential are included, if the lower order terms have the enhanced
continuous symmetry.
To make this more explicit, consider a general $\Delta(27)$ triplet
field $\phi_{i}$. It will have a SUSY breaking soft mass term in the
Lagrangian of the form $m_{\phi }^{2}\phi ^{i^{\dagger }}\phi_{i}$, invariant
under the approximate $SU(3)_{f}$ symmetry. Radiative corrections involving
superpotential couplings to massive states may drive the mass squared
negative at some scale $\Lambda $ triggering a \VEV \ for the field
$\phi$: $\langle \phi^{i^{\dagger}}\phi _{i} \rangle = v^{2}$,
with $v^{2}\leq \Lambda ^{2}$ set radiatively
\footnote{The radiative corrections to the soft mass term depend on
  the details of the underlying theory at the string or unification
  scale.}.
At this stage the
\VEV \ of $\phi $ can always be rotated to the $3$rd direction using the
approximate $SU(3)_{f}$ symmetry. However this does not remain true when
higher order terms allowed by the discrete \FS \ are included. For
the $\Delta(27)$ model considered, the leading higher order term is a quartic term $V_{q}$ of
the form:
 \begin{equation}
V_{q} \simeq m_{3/2}^{2}(\phi ^{\dagger }\phi \phi ^{\dagger }\phi )
\label{eq:quartic}
\end{equation}
The term arises as a
component of the $D$-term $\left[ \chi ^{\dag }\chi (\phi ^{\dagger }\phi
\phi ^{\dagger }\phi )\right] _{D}$. In eq.(\ref{eq:quartic}) (and in following equations of similar quartic terms) we have suppressed the $O(1)$ coupling constants and the respective messenger
mass scale (or scales) associated with these higher dimension
operators (which can even be the Planck mass $M_{P}$).
The $F$ component of the field $\chi$ drives SUSY breaking
and $m_{3/2}$ is the gravitino mass
($m_{3/2}^{2}=F_{\chi }^{\dagger}F_{\chi }/M_{P}^{2}$). This term
gives rise to two independent quartic invariants
under $\Delta(27)$, $V_{ij}$ and $V_{i}$ (the indices in the subscript of $V$ are just labels used to distinguish $V_{ij}$ from $V_{i}$ - they do not take values, unlike the true family index of $\phi_{i}$):
\begin{equation}
V_{ij} \simeq m_{3/2}^{2}(\phi ^{i^{\dagger}}\phi _{i}\phi ^{j^{\dagger}}\phi_{j})
\end{equation}

\begin{equation}
V_{i} \simeq m_{3/2}^{2}(\phi^{i^{\dagger }}\phi _{i}\phi ^{i^{\dagger}}\phi_{i})
\label{eq:quartic2}
\end{equation}
$V_{ij}$ is $SU(3)_{f}$ symmetric and does not remove the \VEV \ degeneracy.
However, $V_{i}$ in eq.(\ref{eq:quartic2}) is not $SU(3)_{f}$ symmetric
and does lead to an unique \VEV .
If the coefficient of $V_{i}$ is positive, the minimum
corresponds to:
\begin{equation}
\langle \phi_{i} \rangle ^{T} = v (1,1,1)/\sqrt{3}
\label{eq:111}
\end{equation}
Unlike in previous \VEV s of this type, here the phases of each entry are in general unrelated - for
simplicity we omit the phases as they won't be particularly relevant in obtaining \TBM . Eq.(\ref{eq:111}) is comparable for example with eq.(\ref{eq:P123 vev2}).
If instead the coefficient of $V_{i}$ is negative, the \VEV \ has the
form:
\begin{equation}
\langle \phi _{i} \rangle^{T}=v(0,0,1)
\label{eq:001}
\end{equation}
In contrast, eq.(\ref{eq:001}) is comparable for example with eq.(\ref{eq:T vev}).
We conclude that, in contrast with a continuous
symmetry case, a discrete non-Abelian symmetry leads to a finite number of
candidate vacuum states. Which one is chosen depends on the coefficients of higher dimension terms which in turn depends on the details of the underlying
theory. We do not attempt to construct the full theory and so
cannot determine these coefficients. What we will demonstrate, however, is that one
of the finite number of candidate vacua does have the correct properties to
generate a viable theory of \FMM \ (including \TBM ).

We will now obtain the \VEV \ alignment needed for the $\Delta(27)$ model.
Suppose that the soft masses $m_{\phi _{123}}^{2}$, $m_{\phi _{1}}^{2}$ and
$m_{\bar{\phi}_{3}}^{2}$ are driven negative close to the messengers
scale (denoted generically as) $M$:
\begin{equation}
\Lambda_{\phi_{123},\phi_{1}},_{\bar{\phi} _{3}}\lesssim M
\end{equation}
The symmetries of the model ensure that the
leading terms fixing their vacuum structure are of the form:
\begin{equation}
V_{123} \simeq m_{3/2}^{2}(\phi _{123}^{\dagger }\phi _{123}\phi _{123}^{\dagger }\phi
_{123})
\label{eq:qpure123}
\end{equation}
\begin{equation}
V_{1} \simeq m_{3/2}^{2}(\phi _{1}^{\dagger }\phi _{1}\phi _{1}^{\dagger
}\phi_{1})
\label{eq:qpure1}
\end{equation}
\begin{equation}
V_{m} \simeq m_{3/2}^{2}(\phi _{123}^{\dagger }\phi _{123}\phi
_{1}^{\dagger}\phi _{1})
\label{eq:qmix1}
\end{equation}
Plus similar terms involving $\bar{\phi}_{3}$.
These terms naturally dominate as these
fields acquire the \VEV s with largest magnitudes - the magnitude of their
\VEV s is determined by the scale at which their soft mass squared
becomes negative (in this model, this statement applies to all flavon \VEV s, as all of them are driven radiatively). 
In order to discern the directions of the \VEV s, we need to make
further assumptions. To obtain the desired structure, we require that the 
terms of the type of $V_{123}$ of eq.(\ref{eq:qpure123}) and $V_{1}$ of eq.(\ref{eq:qpure1})
(terms involving just one of the fields) dominate over mixed terms of the type of $V_{m}$ of eq.(\ref{eq:qmix1}).
If so, the \VEV s
will be determined by the signs of these pure 
terms: if the coefficient of eq.(\ref{eq:qpure123}) is positive,
$\phi_{123}$ will acquire a \VEV \ in the $(1,1,1)$
direction as in eq.(\ref{eq:P123 vev2}), and if the coefficient of
eq.(\ref{eq:qpure1}) is negative, $\phi _{1}$ will acquire a \VEV \ in
the $(1,0,0)$ direction as in eq.(\ref{eq:P1 vev}), the non-vanishing
entry defining the $1$st direction without loss of generality.

If the analogous quartic term involving solely
$\bar{\phi}_{3}$ also has a negative coefficient, $\langle
\bar{\phi}_{3} \rangle$ has a single non-zero entry. To resolve the
ambiguity in the position of this entry (relative to $\langle \phi_{1}
\rangle$), we need to investigate the leading $D$-terms involving both
fields, such as:
\begin{equation}
V_{m 1} \simeq m_{3/2}^{2}(\bar{\phi}^{i} _{3} \phi _{1_{i}} \phi _{1}^{\dagger j
}\bar{\phi} _{3_{j}}^{\dagger })
\label{eq:qmix13}
\end{equation}
If among the remaining quartic terms affecting $\bar{\phi}_{3}$, the mixed $V_{m1}$ of eq.(\ref{eq:qmix13})
dominates and has positive coefficient, it favours $\langle
\bar{\phi}_{3} \rangle$ to be orthogonal to $\langle \phi_{1}
\rangle$. We then simply define the $3$rd direction (implicitly
defining the remaining direction as the $2$nd) without loss
of generality: $\langle \bar{\phi}_{3} \rangle \propto (0,0,1)$, as
in eq.(\ref{eq:P3 vev2}).

In a similar manner it is straightforward to
discuss the fields $\bar{\theta}$ and $\phi_{3}$:
\begin{equation}
V_{m \theta} \simeq m_{3/2}^{2}(\bar{\phi}^{i}_{3} \bar{\theta}_{i} \bar{\theta} ^{\dagger j}\bar{\phi
}^{\dagger }_{3_{j}})
\label{eq:qmix3t}
\end{equation}

\begin{equation}
V_{m 3} \simeq m_{3/2}^{2}(\bar{\phi}^{i}_{3} \phi_{3_{i}} \phi _{3}^{\dagger
  j}\bar{\phi} _{3_{j}}^{\dagger })
\label{eq:qmix33}
\end{equation}

By having the mixed $V_{m \theta}$ of eq.(\ref{eq:qmix3t}) and the mixed $V_{m3}$ of eq.(\ref{eq:qmix33}) dominate the respective alignment and have negative coefficients, we
can respectively align $\phi_{3}$ and $\bar{\theta}$ along the $(0,0,1)$
direction - as in eq.(\ref{eq:T vev}). The alignment of $\langle \bar{\theta} \rangle$ is quite important, even though the direction of $\langle \phi
_{3}\rangle $ is not very relevant for the model. In any case, with the $V_{m \theta}$ and $V_{m 3}$ 
terms both can take the form in eq.(\ref{eq:T vev}). In the case of $\phi_{3}$ it is for simplicity and not by necessity that we take it to be so.

The relative alignment of the remaining fields follows in a similar manner.
Consider the field $\bar{\phi}_{23}$ with a soft mass squared becoming
negative at a scale $b<v$. We want the dominant term aligning its \VEV
\ to be:

\begin{equation}
V_{m 123} \simeq m_{3/2}^{2}(\bar{\phi}_{23}^{i}\phi _{123_{i}}
\phi _{123}^{\dagger j}\bar{\phi}_{23_{j}}^{\dagger})
\label{eq:qmix23123}
\end{equation}
With eq.(\ref{eq:qmix23123}) having a positive coefficient, $\langle
\bar{\phi}_{23} \rangle$ is favoured to be orthogonal to $\langle \phi
_{123} \rangle$. The choice of the particular orthogonal direction 
will be determined by terms such as:

\begin{equation}
V_{o3} \simeq m_{3/2}^{2}(\bar{\phi}_{3}^{i}\bar{\phi}_{23_{i}}^{\dagger
}\bar{\phi}_{23}^{j}\bar{\phi}_{3_{j}}^{\dagger })
\end{equation}

\begin{equation}
V_{o1} \simeq m_{3/2}^{2}(\bar{\phi}_{23}^{i}\phi _{1_{i}}\phi _{1}^{\dagger j}
\bar{\phi}_{23_{j}}^{\dagger })
\label{eq:qmix231}
\end{equation}
The form given in eq.(\ref{eq:P23 vev}) can be obtained if $V_{o1}$ of
eq.(\ref{eq:qmix231}) dominates and has a positive
coefficient - as that will favour $\langle \bar{\phi}_{23}\rangle$ to
be orthogonal to $\langle \phi_{1} \rangle$.

Finally, consider the field $\bar{\phi}_{123}$, with a soft mass squared
becoming negative at a scale $c\ll v$. The leading terms determining its
vacuum alignment are:
\begin{equation}
V_{a} \simeq m_{3/2}^{2}(\bar{\phi}_{3}^{i}\bar{\phi}_{23_{i}}^{\dagger }
\bar{\phi}_{3}^{j}\bar{\phi}_{123_{j}}^{\dagger })
\label{eq:qmix323}
\end{equation}

\begin{equation}
V_{c} \simeq m_{3/2}^{2}(\bar{\phi}_{123}^{i}\phi _{123_{i}}
\phi _{123}^{\dagger j}\bar{\phi}_{123_{j}}^{\dagger })
\label{eq:qmix123123}
\end{equation}
With $V_{c}$ of eq.(\ref{eq:qmix123123}) dominating with a negative
coefficient, $\langle \bar{\phi}_{123} \rangle$ will be aligned in the
same direction as $\langle \phi_{123} \rangle$, which is the form given in
eq.(\ref{eq:P123 vev2}). Note that $V_{a}$ in eq.(\ref{eq:qmix323})
is accidental in the sense that it is dependent on the additional $U(1)$
assignments of the fields.

In summary, we have shown that higher order $D$-terms constrained by the
discrete \FS \ lead to a discrete number of possible \VEV s. Which one is the true vacuum state depends fundamentally on the coefficients of these
higher order terms (in magnitude and sign). The coefficients are determined by the underlying
\GUT \ or string theory. Our analysis has shown that the \VEV s
needed for a viable theory of \FMM \ can emerge from
this discrete set of states.

\subsection{Yukawa terms \label{sub:Yukawa}}

We turn now to the structure of the quark and lepton mass matrices. The
leading Yukawa terms allowed by the symmetries are:

\begin{equation}
P_{Y} = \frac{1}{M^{2}}\bar{\phi}_{3}^{i}\psi_{i}
\bar{\phi}_{3}^{j}\psi_{j}^{c}H
\label{eq:Y_P3_P3} 
\end{equation}
\begin{equation}
+\frac{1}{M^{3}}\bar{\phi}_{23}^{i}\psi_{i}\bar{\phi}_{23}^{j}
\psi_{j}^{c}HH_{45}  \label{eq:Y_P23_P23 2}
\end{equation}

\begin{equation}
+\frac{1}{M^{2}}\bar{\phi}_{23}^{i}\psi_{i}\bar{\phi}_{123}^{j}\psi_{j}^{c}H
\label{eq:Y_P23_P123 2}
\end{equation}
\begin{equation}
+\frac{1}{M^{2}}\bar{\phi}_{123}^{i}\psi_{i}\bar{\phi}_{23}^{j}\psi_{j}^{c}H
\label{eq:Y_P123_P23 2}
\end{equation}

\begin{equation}
+\frac{1}{M^{5}}\bar{\phi}_{123}^{i}\psi_{i}^{c}\bar{\phi}%
_{3}^{j}\psi_{j}^{c}H H_{45}\bar{\phi}_{123}^{k}\phi_{1_{k}}
\label{eq:Y_P123_P3}
\end{equation}

\begin{equation}
+\frac{1}{M^{5}}\bar{\phi}_{3}^{i}\psi_{i}^{c}\bar{\phi}_{123}^{j}%
\psi_{j}^{c}H H_{45}\bar{\phi}_{123}^{k}\phi_{1_{k}}  \label{eq:Y_P3_P123}
\end{equation}

\begin{equation}
+\frac{1}{M^{6}}\bar{\phi}_{123}^{i}\psi_{i}^{c}\bar{\phi}%
_{123}^{j}\psi_{j}^{c}H\bar{\phi}_{3}^{k}\phi_{123_{k}}\bar{\phi}%
_{3}^{l}\phi_{123_{l}}  \label{eq:Y_P123_P123}
\end{equation}
Although of a slightly different form from the terms used in
chapter \ref{ch:fs}, these terms realise essentially the same mass
structure as the one presented there. As such, we won't
repeat a detailed analysis of the mass structure given by this
superpotential: it gives a
phenomenologically consistent description
of all \FMM ,
generating the hierarchy of masses through an expansion in the \FS \
breaking parameters (refer to subsection \ref{sec:Masses} for details).

The main differences of this Yukawa superpotential relative to that of
the $SU(3)_{f}$ model reside in eq.(\ref{eq:Y_P123_P3}), eq.(\ref{eq:Y_P3_P123}),
and eq.(\ref{eq:Y_P123_P123}). Eq.(\ref{eq:Y_P123_P3}) and eq.(\ref{eq:Y_P3_P123})
account for the $O\left( \epsilon _{d}^{3}\right) $ difference in
the $12,21$ and $13,31$ entries 
\footnote{Again we assume symmetric mass matrices, motivated by an underlying $SO(10)$.}
of the down-type quark mass matrix (as shown in eq.(\ref{eq:Yd2})
\cite{Roberts:2001zy, Mario}), replacing the term of
eq.(\ref{eq:Y_P2_P123}) which plays the same role in the $SU(3)_{f}$ model.
The term in eq.(\ref{eq:Y_P123_P123}) is undesirable, but allowed by the
symmetries nonetheless. Naively, one expects it would contribute to the $11$
element at $O\left(\epsilon _{d}^{4}\right)$ giving unwanted corrections
to the phenomenologically successful \GST \ relation
\cite{Gatto:1968ss} which results only if the $11$ entry is smaller.
Fortunately, the texture zero can be naturally preserved at that order
despite eq.(\ref{eq:Y_P123_P123}): the \VEV s \ of $\phi _{3}$ and
$\bar{\phi}_{3}$ are slightly smaller than the relevant messenger mass
scales, and in the eq.(\ref{eq:Y_P123_P123}) there are four
insertions of these fields, suppressing the unwanted contribution sufficiently. Thanks to the accumulated \VEV \ suppression, the
desired small magnitude of the $11$ element can be maintained while still keeping
the dimensionless coefficients in front of all the allowed Yukawa
terms as $O(1)$.
Finally, in the discrete model there many sub-leading operators not explicitly shown, so it is possible to accommodate the approximate factor of $2$ in $\left| Y^{d}_{23}/Y^{d}_{22} \right|$ required by the data \cite{Mario} without needing to rely on unappealing choices of the free parameters.

\subsection{Majorana terms\label{sub:Majorana2}}

The leading terms that contribute to the \RH \ neutrino Majorana
masses are:

\begin{equation}
P_{M} \sim \frac{1}{M}\bar{\theta} ^{i}\psi_{i}^{c}\bar{\theta} ^{j}\psi_{j}^{c}
\label{eq:M_th_th 2}
\end{equation}
\begin{equation}
+\frac{1}{M^{5}}\bar{\phi}_{23}^{i}\psi_{i}^{c}\bar{\phi}_{23}^{j}\psi
_{j}^{c}\bar{\theta} ^{k}\phi_{123_{k}}\bar{\theta} ^{l}\phi_{3_{l}}
\label{eq:M_P23_P23 2}
\end{equation}
\begin{equation}
+\frac{1}{M^{5}}\bar{\phi}_{123}^{i}\psi_{i}^{c}\bar{\phi}_{123}^{j}\psi
_{j}^{c}\bar{\theta} ^{k}\phi_{123_{k}}\bar{\theta} ^{l}\phi_{123_{l}}
\label{eq:M_P123_P123 2}
\end{equation}
Unlike what happened in the Yukawa superpotential, most of these terms are
different from those in the $SU(3)_{f}$ model, and consequently the
ratios of the Majorana masses derived in chapter \ref{ch:fs} do not apply.
In this model, the magnitude of the \VEV \ of $\phi_{3}$ controls the hierarchy between $M_{1}$ and
$M_{2}$ (which depends essentially on the ratio of magnitudes between
eq.(\ref{eq:M_P123_P123 2}) and eq.(\ref{eq:M_P23_P23 2})).
The magnitude of $\langle \phi_{3} \rangle$ is set by radiative breaking (note how the direction of this \VEV \ isn't very relevant). We
require $\langle \phi_{3} \rangle$ to lie close to the scale of
$\langle \bar{\phi}_{23} \rangle$, such that
after the seesaw we can fit the ratio of the neutrino squared mass
differences $\frac{\Delta m_{\odot }^{2}}{\Delta m_{@}^{2}}$. This is
different from what occurs in the $SU(3)_{f}$ model, where the ratio 
$\frac{M_{1}}{M_{2}}$ is predicted and associated with the expansion parameter
$\epsilon_{d}$ in a way that is simultaneously consistent with the
quark sector and with the experimentally
measured magnitude of $\frac{\Delta m_{\odot}^{2}}{\Delta m_{@}^{2}}$.

The hierarchy between the lightest Majorana
mass, $M_{1}$, and the heaviest,$M_{3}$, is given by:
\begin{equation}
\frac{M_{1}}{M_{3}}\simeq \epsilon _{d}^{4}\frac{M_{d}^{4}}{M_{\nu _{R}}^{4}}
\label{eq:M1 M3}
\end{equation}
$M_{d}$ is the mass of the messenger responsible for the down-type quark
masses, and $M_{\nu_{R}}$ is the mass of the messenger responsible for
right-handed neutrino masses (for details on the messenger sector, see subsection \ref{sec:Messengers}).
To obtain a viable pattern of neutrino mixing we need to ensure that
the hierarchy in eq.(\ref{eq:M1 M3}) is sufficiently strong to
suppress the contribution from $\nu _{3}^{c}$ exchange which would
otherwise give an unacceptably large $\nu_{\tau}$ component in the
atmospheric or solar neutrino eigenstates. This requirement on
the Majorana hierarchy places a lower bound 
on the mass of corresponding \RH \ neutrino messenger $M_{\nu _{R}}$, as is clear
from eq.(\ref{eq:M1 M3}). The resulting effective neutrino eigenstates obtained by the seesaw mechanism have a strongly
hierarchical mass structure, just like in the $SU(3)_{f}$ model: with $m_{1} \simeq 0$, the heaviest of the effective neutrinos
has a mass given approximately by $\sqrt{\Delta m_{@}^{2}}$. Using this,
together with eq.(\ref{eq:M1 M3}), we find:

\begin{equation}
M_{3} \simeq \epsilon _{d}^{2}\langle H \rangle ^{2}\frac{M_{\nu
    _{R}}^{4}}{M_{\nu }^{4}}\Delta m_{@}^{2}{}^{-\frac{1}{2}}\simeq
10^{13}\frac{M_{\nu_{R}}^{4}}{M_{\nu }^{4}} \mathrm{GeV}  \label{eq:M3 bound}
\end{equation}
$M_{\nu }$ is the mass of the messenger responsible for the Dirac
neutrino mass. As discussed in subsection \ref{sec:Messengers}, the \FN \ diagram can proceed by either the \LH \ or \RH \ neutrino messenger, depending on the messenger mass spectrum. If $M_{\nu} = M_{\nu_{R}}$, eq.(\ref{eq:M3 bound}) imposes a direct constrain on $M_{3}$ (similarly to what happened in the $SU(3)_{f}$ model, although with different details).

By construction, the final structure of neutrino mixing is directly comparable to the one in the $SU(3)_{f}$ model, and generates the same \TBM \ predictions
for the neutrino mixing angles. The PMNS leptonic mixing angles are
obtained after taking into account the (small) effect of the charged
leptons, leading to small deviations from \TBM \ \cite{Plentinger:2005kx, Antusch:2005kw}, just like in section \ref{sec:Phenomenology}:
\begin{equation}
\sin^{2}\theta_{12} = \frac{1}{3}\pm_{0.048}^{0.052}
\end{equation}

\begin{equation}
\sin^{2}\theta_{23} = \frac{1}{2}\pm_{0.058}^{0.061}
\end{equation}

\begin{equation}
\sin^{2}\theta_{13} = 0.0028
\end{equation}

\section{Summary and conclusions\label{sec:Confmm2}}

We briefly exemplified how the special flavon \VEV s that can lead to \TBM \ can be obtained in the context of a discrete subgroup of $SU(3)_{f}$, $\Delta(12)$ (or $A_4$). 

We have then constructed a complete \FS \ model of \FMM , based on
the spontaneous breaking of the discrete non-Abelian group
$\Delta(27)$. The model is constructed in a manner
consistent with an underlying \GUT , with all the members of
a family of fermions having the same symmetry properties under the \FS . Many of the properties of the model rely on the approximate $SU(3)_{f}$ symmetry that the discrete group possesses and the model is very
similar to the continuous $SU(3)_{f}$ \FS \ model of reference
\cite{Ivo1}, following the framework that was discussed throughout chapter \ref{ch:fs}.
The main difference is a
significant simplification in the \VEV \ alignment mechanism in which
the near \TBM \ of the lepton sector directly follows from the
non-Abelian discrete group. In addition to the prediction of near \TBM
\ the model preserves the \GST \ \cite{Gatto:1968ss} relation between the light
quark masses and the Cabibbo mixing angle, and can also accommodate the
\GUT \ relations between the down quark and lepton masses (\GJ \ relations). It provides an
explanation for the hierarchy of \FMM \ in
terms of expansions in powers of \FS \ breaking parameters.

The presence of additional invariants and the absence of $D$-terms (present in models featuring continuous \FSs ) were important features in achieving this simpler model. The use of
a discrete \FS \ then automatically avoids the \FP \ discussed in chapter
\ref{ch:fcnc}, and also avoids unwanted anomalies as discussed briefly in section \ref{sec:Anom}, two relevant concerns that needed to be considered with $SU(3)_{f}$. Finally, there is another important phenomenological difference
of models featuring discrete \FSs : with the existence of an
approximate continuous symmetry, there will be associated light
pseudo-Goldstone bosons. The mass of these states might be small enough
for them to be within the reach of future experiments (such as in the LHC).

\chapter{Family symmetry \FP \ \label{ch:fcnc}}

In this chapter we re-examine contributions to 
sfermion masses coming from $D$-terms associated with continuous \FSs . These generation dependent sfermion mass contributions in turn lead to constraints that arise from
experimental limits on \FCNC s. We show that,
for a restricted choice of the \FS \ breaking flavon sector, continuous \FSs \ are consistent even with the most restrictive experimental bounds, both for the case of gauge mediated SUSY
breaking and the case of gravity mediated SUSY breaking.

\section{SUSY \FP}

The SUSY \FP \ is a longstanding problem for SUSY theories. Due
to the introduction of additional flavoured particles (namely the sfermions) there
are extra contributions to 4-fermion flavour changing interactions, leading
to \FCNC s that are potentially too large.
For example, quark \FCNC \ processes can occur through box diagrams
involving gluino and squarks in the internal lines, as shown in figure \ref{fig:fcnc_gs}.

\begin{figure}[htb]
\begin{center}
\fcolorbox{white}{white}{
  \begin{picture}(195,151) (120,-179)
    \SetWidth{0.5}
    \SetColor{Black}
    \ArrowLine(120,-59)(165,-59)
    \Vertex(165,-59){2.83}
    \Vertex(255,-59){2.83}
    \ArrowLine(255,-59)(300,-59)
    \ArrowLine(255,-59)(165,-59)
    \DashArrowLine(165,-104)(165,-59){10}
    \Line(163,-106)(167,-102)\Line(163,-102)(167,-106)
    \Line(253,-106)(257,-102)\Line(253,-102)(257,-106)
    \DashArrowLine(255,-59)(255,-104){10}
    \DashArrowLine(255,-104)(255,-149){10}
    \DashArrowLine(165,-149)(165,-104){10}
    \Vertex(165,-149){2.83}
    \Vertex(255,-149){2.83}
    \ArrowLine(165,-149)(120,-149)
    \ArrowLine(165,-149)(255,-149)
    \ArrowLine(300,-149)(255,-149)
    \Text(210,-44)[lb]{\Large{\Black{$\tilde{g}$}}}
    \Text(210,-179)[lb]{\Large{\Black{$\tilde{g}$}}}
    \Text(285,-44)[lb]{\Large{\Black{$d$}}}
    \Text(135,-179)[lb]{\Large{\Black{$d$}}}
    \Text(135,-44)[lb]{\Large{\Black{$s$}}}
    \Text(285,-179)[lb]{\Large{\Black{$s$}}}
    \Text(135,-134)[lb]{\Large{\Black{$\tilde{d}$}}}
    \Text(135,-89)[lb]{\Large{\Black{$\tilde{s}$}}}
    \Text(285,-134)[lb]{\Large{\Black{$\tilde{s}$}}}
    \Text(285,-89)[lb]{\Large{\Black{$\tilde{d}$}}}
  \end{picture}
}
\caption{A \FCNC \ process involving superpartners.}
\label{fig:fcnc_gs}
\end{center}
\end{figure}

The contribution of diagrams like the one in figure \ref{fig:fcnc_gs} depends on the masses of
the squarks mediating the flavour change.
Considering such processes, experiments have placed rather stringent
constraints on the mass matrices of the sfermions
\cite{Gabbiani:1996hi,Endo:2003te,Foster:2006ze,
  Ciuchini:2007ha}.

The simplest way to
satisfy the experimental constraints is by requiring the three generations of sfermions to have
nearly degenerate masses, although there are other alternatives (for example, alignment of the sfermions with the fermions) \cite{Nir:1993mx,Cohen:1996vb}. If we do require sfermion mass degeneracy, there are three established ways of obtaining it: \SUGRA \ models, where
universal soft masses are generated by gravitational interactions; gauge mediated models, in which case the SUSY breaking mechanism
giving rise to the soft masses is generation blind; and \FSs, where the added symmetry explaining the \FMM \ may also be used to keep the required approximate degeneracy of sfermion masses.
Although one of these mechanisms is necessary to avoid large \FCNC s, they may not be
sufficient if there are further sources of family dependent masses. This is
the case if there is a continuous \FS \ because the
associated $D$-term 
\footnote{If the \FS \ is discrete, like in chapter \ref{ch:fsd}, there are no $D$-terms
  associated with it - see for example \cite{Kajiyama:2005rk}.}
spoils the required degeneracy of sfermion masses (see \cite{Ross:2004qn},
\cite{Kawamura:1994ys}, \cite{Murayama:1995fv,Ramage:2003pf, King:2004tx}). This apparently unavoidable loss of the required degeneracy is commonly thought to
rule out symmetries which differentiate between 
the first two families (for which the experimental bounds are tighter). In this chapter we show how this problem can
readily be avoided. The way this 
works depends on the origin of the soft masses and we discuss the cases of
both gravity and gauge mediation in subsection \ref{sub:SUGRA} and subsection
\ref{sub:Gauge} respectively.

The chapter is organised as follows: in subsection \ref{sec:SCKM} we
review how to express the sfermion mass 
matrices in the  ``Super-CKM'' (SCKM) basis 
\cite{Hall:1985dx} and compare with the experimental bounds presented
in \cite{Gabbiani:1996hi,Endo:2003te,Foster:2006ze,
  Ciuchini:2007ha}.
Section \ref{sec:Consider} shows how the $D$-term gives rise to 
generation dependent contributions which potentially violate the \FCNC \
bounds; we also present a method for obtaining an upper bound on
model dependent predictions for the \FCNC \ effects and discuss
the energy scales relevant to the analysis. In section \ref{sec:Confp}
we present ways of solving the SUSY \FP \ associated with
continuous \FSs . We conclude the chapter with a summary in section
\ref{sec:Sumfp}.

\subsection{Super-CKM basis and the experimental bounds \label{sec:SCKM}}

In SUSY models we specify not just the basis chosen for the fermion
states (as in the \SM ) but also the basis chosen for the
sfermion states. In the \SM \ it is often convenient to use either the
``mass'' basis, where each fermion state
has a well defined mass (the mass matrices are diagonal) or the
``flavour''/``weak'' basis, where each fermion state
has a well defined flavour (the weak interaction matrix is diagonal).
Similarly, it is often convenient to use a specific SUSY basis, and to study SUSY \FCNC s it is
particularly useful to use the SCKM basis \cite{Hall:1985dx}. We start by briefly reviewing the SCKM basis, illustrating its definition with the down-type quark
and squark sector. We generalise the mass matrices to a $6\times 6$ notation,
distinguishing \LH \ and \RH \ states. In an arbitrary basis
for the quarks, we have:

\begin{equation}
L_{Y^{d}} =
(\bar{d^{\prime }}_{L} , \bar{d^{\prime }}_{R}) \left[ 
\begin{array}{cc}
0 & M_{d}^{D} \\ 
M_{d}^{D^{\dagger}} & 0%
\end{array}%
\right] \left( 
\begin{array}{c}
d^{\prime }_{L} \\ 
d^{\prime }_{R}%
\end{array}
\right)  \label{eq:mdD}
\end{equation}
$d^{\prime }_{L}$, $d^{\prime }_{R}$ are 3 component columns
containing the 3 down-type quarks ($d$, $s$, $b$). The $6 \times 6$
matrix is represented by four $3 \times 3$ blocks. The prime denotes that
the states are taken in the arbitrary basis. $M_{d}^{D}$ is the Dirac mass
matrix for the down quarks.

The scalar partners have a squared mass matrix containing their
squared masses:

\begin{equation}
M_{\tilde{d}}^{2}=\left[ 
\begin{array}{cc}
M_{\tilde{d}}^{LL^{2}} & M_{\tilde{d}}^{LR^{2}} \\ 
M_{\tilde{d}}^{RL^{2}} & M_{\tilde{d}}^{RR^{2}}%
\end{array}%
\right]  \label{eq:smass}
\end{equation}%
Again we express the full $6\times 6$ matrix in terms of $3\times 3$ block
matrices. In the LR quadrant, $M_{\tilde{d}}^{LR^{2}}$ is equal to the down
quark Dirac mass matrix $M_{d}^{D}$ multiplied by a generation independent mass
factor, and in the RL quadrant, $M_{\tilde{d}}^{RL^{2}}$ is similarly
proportional to the hermitian conjugate $M_{d}^{D^{\dagger }}$.
In the LL and RR quadrants we have squark squared masses for the \LH \ and \RH
\ squarks respectively. 

Without loss of generality we can consider the basis of sfermion states where $M_{\tilde{f}}^{LL^{2}}$ and $M_{\tilde{f}}^{RR^{2}}$ are diagonal, and parametrise 
the diagonal matrices with an explicit universal contribution $m_{0}^{2}$ that is
generation blind, plus generation dependent deviations from the degenerate spectrum, parametrised as $\Delta m_{\tilde{f}}^{2}$. It is useful to use such a
parametrisation as we are interested in models where the common $m_{0}$
comes from a specific SUSY breaking messenger mechanism (for example,
gravity mediation), and
deviations arise from $D$-term contributions associated with a continuous
\FS . We note that when this is the case, as the $\Delta m_{\tilde{f}}^{2}$ are flavour dependent, the diagonalising basis corresponds to sfermion states with specific flavour - in other words, the sfermion ``flavour'' basis (in terms of gauge interactions) coincides in this case with the sfermion ``mass'' basis (in terms of the LL and RR masses). Taking this basis and this parametrisation for the down-type squarks we have:
\begin{equation}
M_{\tilde{d}}^{LL^{2};RR^{2}} = \left[ 
\begin{array}{ccc}
m_{0}^{2} + \Delta m_{\tilde{d}_{L;R}}^{2} & 0 & 0 \\ 
0 & m_{0}^{2} + \Delta m_{\tilde{s}_{L;R}}^{2} & 0 \\ 
0 & 0 & m_{0}^{2} + \Delta m_{\tilde{b}_{L;R}}^{2}%
\end{array}
\right]  \label{eq:squark mass}
\end{equation}
We now re-express this matrix in the SCKM basis. The SCKM basis consists of having the fermion states in the basis where their Dirac mass matrices are diagonalised, and the
sfermion states in the basis that has the neutral gauginos couplings
flavour diagonal. To take care of the latter, and starting from the squark basis where
$M_{\tilde{d}}^{LL;RR^{2}}$ is diagonal (as in eq.(\ref{eq:squark
  mass})), it is then convenient to start with the quarks in their flavour basis. To go into the SCKM, it is now sufficient to diagonalise the quarks without undiagonalising the neutral gaugino couplings of the squarks - so we must apply to the squark states the same
transformation we apply to the quarks. To do so we use the $6\times 6$
mixing matrix that diagonalises the quark Dirac masses:

\begin{equation}
\left( 
\begin{array}{c}
d_{L} \\ 
d_{R}%
\end{array}
\right) = \left[ 
\begin{array}{cc}
V_{L} & 0 \\ 
0 & V_{R}%
\end{array}%
\right] \left( 
\begin{array}{c}
d^{\prime }_{L} \\ 
d^{\prime }_{R}%
\end{array}
\right)
\end{equation}
The unprimed quark states now represent the mass
eigenstates. Applying it to eq.(\ref{eq:mdD}), we get:

\begin{equation}
L_{Y^{d}} = (\bar{d}_{L},\bar{d}_{R})\left[ 
\begin{array}{cc}
V_{L} & 0 \\ 
0 & V_{R}%
\end{array}%
\right] \left[ 
\begin{array}{cc}
0 & M_{d}^{D} \\ 
M_{d}^{D^{\dagger }} & 0%
\end{array}%
\right] \left[ 
\begin{array}{cc}
V_{L}^{\dagger } & 0 \\ 
0 & V_{R}^{\dagger }%
\end{array}%
\right] \left( 
\begin{array}{c}
d_{L} \\ 
d_{R}%
\end{array}%
\right)
\end{equation}%
The product of the three $6\times 6$ matrices will of course result in
diagonalised LR and RL quadrants (the diagonal down quark Dirac mass matrix
and its hermitian conjugate, respectively). We arrive to the SCKM basis by applying the same mixing matrices to the $M_{\tilde{d}}^{2}$ of eq.(\ref{eq:smass}), which in the SCKM then has the form:

\begin{equation}
M_{\tilde{d}}^{2} = \left[ 
\begin{array}{cc}
V_{L} M_{\tilde{d}}^{LL^{2}} V_{L}^{\dagger} & V_{L} M_{\tilde{d}}^{LR^{2}}
V_{R}^{\dagger} \\ 
V_{R} M_{\tilde{d}}^{RL^{2}} V_{L}^{\dagger} & V_{R} M_{\tilde{d}}^{RR^{2}}
V_{R}^{\dagger}%
\end{array}%
\right]  \label{eq:sdownLL}
\end{equation}
Note that the LR and RL blocks are now diagonal, but 
the LL and RR blocks need not be (due to the presence of the $\Delta m_{\tilde{f}}^{2}$).

The mass insertions $\Delta$
\cite{Gabbiani:1996hi,Endo:2003te,Foster:2006ze, Ciuchini:2007ha} are
defined as the components of the sfermion mass matrix in the SCKM
basis. For example, $\Delta^{\tilde{d}}_{12_{LL}}$ (the $\tilde{d}_{L}
\tilde{s}_{L}$ component) is given by the appropriate entry of the $M_{\tilde{d}}^{2}$ in eq.(\ref{eq:sdownLL}) :

\begin{equation}
\Delta^{\tilde{d}}_{12_{LL}} \equiv (m_{0}^2 + \Delta m_{\tilde{d}_{L}}^{2})
V_{L_{11}} V_{L_{21}}^{*} + (m_{0}^2 + \Delta m_{\tilde{s}_{L}}^{2})
V_{L_{12}} V_{L_{22}}^{*} + (m_{0}^2 + \Delta m_{\tilde{b}_{L}}^{2})
V_{L_{13}} V_{L_{23}}^{*}  \label{eq:m_12}
\end{equation}
Since $V_{L}$ is unitary, we have:

\begin{equation}
V_{L_{11}} V_{L_{21}}^{*} + V_{L_{12}} V_{L_{22}}^{*} + V_{L_{13}}
V_{L_{23}}^{*} = 0  \label{eq:unitary}
\end{equation}
Using eq.(\ref{eq:unitary}) immediately shows that the terms
proportional to $m_{0}^{2}$ in eq.(\ref{eq:m_12}) vanish as expected
(as would any generation independent contribution).

In \cite{Gabbiani:1996hi,Endo:2003te,Foster:2006ze, Ciuchini:2007ha}
the experimental constraints are presented in terms of dimensionless
quantities $\delta $, which are obtained by dividing the mass
insertions $\Delta $ by the average sfermion mass. To illustrate, with
down squarks in the LL block, we have (from eq.(\ref{eq:sdownLL}), eq.(\ref{eq:m_12}) and having used eq.(\ref{eq:unitary})):

\begin{eqnarray}
\delta _{12_{LL}}^{d} &\equiv &\frac{\left( V_{L}M_{\tilde{d}%
}^{LL^{2}}V_{L}^{\dagger }\right) _{12}}{\langle m_{\tilde{q}}^{2}\rangle } \\
\delta _{12_{LL}}^{d}&=&\frac{\Delta m_{\tilde{d}_{L}}^{2}V_{L_{11}}V_{L_{21}}^{\ast }+\Delta m_{%
\tilde{s}_{L}}^{2}V_{L_{12}}V_{L_{22}}^{\ast }+\Delta m_{\tilde{b}%
_{L}}^{2}V_{L_{13}}V_{L_{23}}^{\ast }}{\langle m_{\tilde{q}}^{2}\rangle }
\label{eq:delta_12}
\end{eqnarray}%
$\langle m_{\tilde{q}}^{2}\rangle $ is the geometrical average for the
squark mass (see \cite{Gabbiani:1996hi,Endo:2003te,Foster:2006ze, Ciuchini:2007ha}).

The $\delta$ are constrained by the non-observation of \FCNC . The most stringent model independent experimental upper bounds from
\cite{Gabbiani:1996hi,Endo:2003te,Foster:2006ze, Ciuchini:2007ha} are
shown in table \ref{ta:bounds} (for quarks) and table \ref{ta:lbounds}
(for leptons)
\footnote{Making some assumptions about the underlying physics stronger bounds are obtained in
\cite{Gabbiani:1996hi,Endo:2003te,Foster:2006ze, Ciuchini:2007ha}, but
we do not consider these here.}. ``Re'' stands for the real part and
``Im'' for the imaginary part.

\begin{table}[htbp]
\centering
\begin{tabular}{|c||c|c|}
\hline
$\frac{m_{\tilde{g}}^{2}}{m_{\tilde{q}}^{2}}$ & $\sqrt{\left| \mathrm{Re}
\left( \delta^{d}_{12_{LL}} \delta^{d}_{12_{LL}} \right) \right|}$ &
$\sqrt{\left| \mathrm{Re} \left( 
\delta^{d}_{12_{LL}} \delta^{d}_{12_{RR}} \right) \right|}$ \\ \hline
$0.3$ & $1.9 \times 10^{-2}$ & $2.5 \times 10^{-3}$ \\ 
$1.0$ & $4.0 \times 10^{-2}$ & $2.8 \times 10^{-3}$ \\ 
$4.0$ & $9.3 \times 10^{-2}$ & $4.0 \times 10^{-3}$ \\ \hline \hline
$\frac{m_{\tilde{g}}^{2}}{m_{\tilde{q}}^{2}}$ & $\sqrt{\left| \mathrm{Re}
\left( \delta^{d}_{13_{LL}}\delta^{d}_{13_{LL}} \right) \right|}$ & $\sqrt{\left| \mathrm{Re}
\left( \delta^{d}_{13_{LL}} \delta^{d}_{13_{RR}} \right) \right|}$ \\ \hline
$0.3$ & $4.6 \times 10^{-2}$ & $1.6 \times 10^{-2}$ \\ 
$1.0$ & $9.8 \times 10^{-2}$ & $1.8 \times 10^{-2}$ \\ 
$4.0$ & $2.3 \times 10^{-1}$ & $2.5 \times 10^{-2}$ \\ \hline \hline
$\frac{m_{\tilde{g}}^{2}}{m_{\tilde{q}}^{2}}$ & $\sqrt{\left| \mathrm{Re}
\left( \delta^{u}_{12_{LL}}\delta^{u}_{12_{LL}} \right) \right|}$ & $\sqrt{\left| \mathrm{Re}
\left( \delta^{u}_{12_{LL}} \delta^{u}_{12_{RR}} \right) \right|}$ \\ \hline
$0.3$ & $4.7 \times 10^{-2}$ & $1.6 \times 10^{-2}$ \\ 
$1.0$ & $1.0 \times 10^{-1}$ & $1.7 \times 10^{-2}$ \\ 
$4.0$ & $2.4 \times 10^{-1}$ & $2.5 \times 10^{-2}$ \\ \hline \hline
$\frac{m_{\tilde{g}}^{2}}{m_{\tilde{q}}^{2}}$ & $\sqrt{\left| \mathrm{Im}
\left( \delta^{d}_{12_{LL}} \delta^{d}_{12_{LL}} \right) \right|}$ &
$\sqrt{\left| \mathrm{Im}
\left( \delta^{d}_{12_{LL}} \delta^{d}_{12_{RR}} \right) \right|}$ \\ \hline
$0.3$ & $1.5 \times 10^{-3}$ & $2.0 \times 10^{-4}$ \\ 
$1.0$ & $3.2 \times 10^{-3}$ & $2.2 \times 10^{-4}$ \\ 
$4.0$ & $7.5 \times 10^{-3}$ & $3.2 \times 10^{-4}$ \\ \hline \hline
$\frac{m_{\tilde{g}}^{2}}{m_{\tilde{q}}^{2}}$ & \multicolumn{2}{|c|}{$\left|
\delta^{d}_{23_{LL}} \right|$} \\ \hline
$1.0$ & \multicolumn{2}{|c|}{$1.6 \times 10^{-1}$} \\  \hline
\end{tabular}%
\caption[Bounds for $\protect\delta$, assuming $m_{\tilde{q}} = 500$ GeV.]{Bounds for $\protect\delta$, assuming $m_{\tilde{q}} = 500$ GeV 
\protect\cite{Gabbiani:1996hi,Endo:2003te,Foster:2006ze, Ciuchini:2007ha}.}
\label{ta:bounds}
\end{table}

\begin{table}[htbp]
\centering
\begin{tabular}{|c||c|c|c|}
\hline
$\frac{m_{\tilde{\gamma}}^{2}}{m_{\tilde{l}}^{2}}$ &
$\left| \delta^{e}_{12_{LL}} \right|$ &
$\left| \delta^{e}_{13_{LL}} \right|$ &
$\left| \delta^{e}_{23_{LL}} \right|$ \\ \hline 
$1.0$ & $6.0 \times 10^{-4}$ & $1.5 \times 10^{-1}$ & $1.2 \times 10^{-1}$ \\ \hline
\end{tabular}%
\caption[Bounds for $\protect\delta$, assuming $m_{\tilde{l}} = 100$ GeV.]{Bounds for $\protect\delta$, assuming $m_{\tilde{l}} = 100$ GeV 
\protect\cite{Gabbiani:1996hi,Endo:2003te,Foster:2006ze, Ciuchini:2007ha}.}
\label{ta:lbounds}
\end{table}

\section{The \FS \ \FP \ \label{sec:Consider}}

\subsection{$D$-term contributions \label{sec:Dterms}}

We now illustrate how $D$-terms associated with continuous \FS \
groups can generate family dependent contributions to the sfermion masses.
We consider a simple example of $U(1)_{f}$ \FS, with the field
content extended to include two flavons, $\phi $ and $\bar{\phi}$. We
define the coupling constant so that the family charge of $\phi$ is
$+1$, meaning that the other family charges are defined relative to
the charge of this flavon. The contribution to the potential of the
$D$-term associated with the continuous \FS \ is then:

\begin{equation}
V_{D} = g_{f}^{2} \left(
  |\phi|^{2} + c |\bar{\phi}|^{2} + c_{\tilde{d}_{L}}
  \tilde{d}_{L}^{2} + c_{\tilde{d}_{R}} \tilde{d}_{R}^{2} + (...)
\right)^{2}
\label{eq:VDt}
\end{equation}
$g_{f}$ is the family coupling constant, $c$ is the family charge of
$\bar{\phi}$, $c_{\tilde{d}_{L;R}}$ are the family charges of the \LH \
and \RH \ down squarks  respectively, and ``$(...)$'' stands for
similar terms for all the other sfermions.

Expanding $V_{D}$ in eq.(\ref{eq:VDt}), we can identify terms quadratic in the down squarks which become contributions to their masses when the flavons acquire \VEV s. Using the notation of
eq.(\ref{eq:squark mass}):

\begin{equation}
\Delta m_{\tilde{f}_{L;R}}^{2}=2 c_{\tilde{f}_{L;R}}\left\langle
D^{2}\right\rangle  \label{eq:Delta_m}
\end{equation}%
The magnitude of the $D$-term is approximately expressed through the quantity $\langle D^{2} \rangle$, implicitly defined in eq.(\ref{eq:Delta_m}) to be:
\begin{equation}
\left\langle D^{2} \right\rangle = g_{f}^{2} \langle |\phi |^{2}+c|\bar{\phi}%
|^{2}\rangle
\end{equation}%
The contributions shown in eq.(\ref{eq:Delta_m}) are explicitly generation
dependent, and give rise to the \FS \ 
\FP . We quantify the problem by calculating the $\delta$
predicted by the model, allowing for a direct comparison with the
experimental bounds. For example, if we substitute eq.(\ref{eq:Delta_m}) into
eq.(\ref{eq:delta_12}) we obtain:

\begin{equation}
\delta _{12_{LL}}^{d}\simeq \frac{2 \left\langle D^{2}\right\rangle (c_{%
\tilde{d}_{L}}V_{L_{11}}V_{L_{21}}^{\ast }+c_{\tilde{s}%
_{L}}V_{L_{12}}V_{L_{22}}^{\ast }+c_{\tilde{b}_{L}}V_{L_{13}}V_{L_{23}}^{%
\ast })}{\langle m_{\tilde{q}}^{2}\rangle }  \label{eq:cdelta_12}
\end{equation}
The other off-diagonal $\delta_{ij}$ have similar expressions.

\subsection{Upper bounds on the theoretical predictions \label{sub:Unknown}}

Since the mixing matrices ($V_{L}$ and $V_{R}$) are in principle unknown, it is useful to derive a mixing matrix independent
upper bound for the $\delta$. Consider just the
part of eq.(\ref{eq:cdelta_12}) dependent on the charges and on the mixing
matrix entries. We denote it as $C$ as it is essentially an effective
charge that incorporates information from the mixing matrix:

\begin{equation}
C \equiv c_{\tilde{d}_{L}} V_{L_{11}} V_{L_{21}}^{*} + c_{\tilde{s}_{L}} V_{L_{12}}
V_{L_{22}}^{*} + c_{\tilde{b}_{L}} V_{L_{13}} V_{L_{23}}^{*}
\label{eq:mixing}
\end{equation}
Each of the terms in eq.(\ref{eq:mixing}) contains two elements of the
mixing matrix that share a common column (like $V_{L_{11}}V_{L_{21}}^{\ast }$%
). We designate these combinations as ``mixing pairs''.
The unitarity of the mixing matrix imposes restrictions on the mixing pairs:
for example eq.(\ref{eq:unitary}), where three such pairs add up to
zero on the complex plane (often referred to as a unitarity
triangle). Also because of unitarity properties, a mixing pair can be written in
the form $\frac{1}{2} \sin (2 \theta) \cos (\varphi)$,
where $\theta$ and $\varphi$ are mixing angles - we conclude then that the
maximum magnitude of any of these pairs is $\frac{1}{2}$.
Furthermore, eq.(\ref{eq:unitary}) shows
that if one of the pairs has the maximum magnitude of $\frac{1}{2}$, the
other two pairs have to point opposite in relation to the maximum magnitude
pair (thus closing the respective unitarity triangle). Because of this, in
order to maximise $C$ in eq.(\ref{eq:mixing}), we identify which two
of the three family charges produce $\mathrm{Max}\left\vert
  c_{i}-c_{j}\right\vert $ (``Max'' standing for maximum). The other charge is somewhere in the middle of the extremising values. To obtain the maximum value for $C$,
the mixing pair multiplying the middle charge must vanish, and the mixing pairs that multiply the other two charges must take the maximum magnitude of $\frac{1}{2}$.
Thus, for example, from eq.(\ref{eq:cdelta_12}):

\begin{equation}
\left\vert \delta _{12}^{LL} \right\vert < \left\vert \frac{\left\langle
D^{2}\right\rangle }{\langle m_{\tilde{q}}^{2}\rangle }\right\vert
\mathrm{Max}\left\vert c_{i}-c_{j}\right\vert  
\label{eq:maximal}
\end{equation}
A specific model can saturate the upper bounds given in
eq.(\ref{eq:maximal}) if two conditions are fulfilled: the first is that there is maximal mixing in two families (with
the other family not mixing); the second is that the two families that mix correspond
to those that maximise $\left\vert c_{i}-c_{j}\right\vert $.
Even with
the upper bound saturated, comparing with the experimental bounds only
yields the most stringent
constraints if two more conditions are verified: the two families that mix are the $(1,2)$ families
(corresponding to the most stringent experimental upper bounds); and
further, the overall phase of that $\delta$ is the
one that aligns it with the strictest experimental bound (in the case
the bounds are placed on real or imaginary parts).
Given all these requirements, it is quite unlikely that a
specific model will indeed saturate a particular bound - the conclusion is
that the bounds should be considered to be very conservative.

\subsection{Running effects \label{sub:Running}}

Before comparing theory to experiment it is important to discuss the energy
scales at which the comparison should be made. The soft SUSY breaking
masses are generated at a scale corresponding to the mediator scale $M_{X}$
communicating SUSY breaking from the hidden to the visible sector and
radiative corrections to the mass will be cutoff at this scale. For the case
of \SUGRA \ this is the Planck scale and there are substantial radiative
corrections in continuing to the electroweak scale - the scale where
the experimental bounds are obtained. For the case of gauge mediation
the gauge messenger scale can be much lower than the Planck scale and
so the radiative corrections may be much smaller.

The dominant radiative corrections are due to the gauge interactions
which are flavour blind. They have the effect of increasing $\langle
m_{\tilde{f}}^{2}\rangle $ while leaving $\Delta
m_{\tilde{f}_{L;R}}^{2}$ unchanged. As a result they systematically
reduce the \FCNC \ effects (see \cite{Ramage:2003pf}, \cite{Ellis:1981ts, Barbieri:1981gn,Inami:1982nu,Ellis:1981tv,Hisano:1992jj}).

It is more convenient, when comparing with the theoretical
expectation, to make the comparison at the messenger scale by continuing the
experimental bounds up in energy. Due to the radiative corrections just
discussed we consider $\delta$, which will depend on the scale $\mu$
at which the comparison is to be made, as the function
$\delta (\mu^{2})$:

\begin{equation}
\delta (M_{X}^{2}) \simeq \delta (M_{W}^{2})\frac{\langle m_{\tilde{f}} \rangle ^{2}(M_{W}^{2})}{\langle m_{\tilde{f}} \rangle ^{2}(M_{X}^{2})}  \label{eq:deltap}
\end{equation}
To evaluate the size of the effect, one can use the renormalisation group
equations \cite{Martin:1997ns,Allanach:2002nj,Porod:2003um}. In table
\ref{ta:compare} we display sample values of $\delta(M_{X}^{2})$ for gauge
and gravity mediation. For gravity mediation, we considered the low
energy squark masses of $m_{\tilde{q}}=500$ GeV and 
slepton masses of $m_{\tilde{l}}=100$ GeV (as used in the bounds of
\cite{Gabbiani:1996hi,Endo:2003te,Foster:2006ze, Ciuchini:2007ha}) and
running effects corresponding to a common unified gaugino mass
$m_{1/2} \sim 250$ GeV lead the sfermions masses to run to a unified
value $m_{0}=80$ GeV at the Planck scale (see figure \ref{fig:susy_mass} for an approximately equivalent situation, if we neglect the running from $M_{GUT}$ to $M_{X} = M_{P}$).
For gauge mediation, we considered the messenger scale to be
$M_{X}=200$ TeV (we use the SPS $8$ scenario in
\cite{Martin:1997ns,Allanach:2002nj,Porod:2003um}).
The slepton masses don't run significantly up to that energy range. However,
the low scale average squark mass is considerably higher than
in the gravity mediation scenario, 
taking the value of $1100$ GeV and running to $1000$ GeV at the
messenger scale $M_{X}=200$ TeV. $\delta(M_{W})$ needs to be scaled
with respect to the higher average squark mass (as prescribed in
\cite{Gabbiani:1996hi,Endo:2003te,Foster:2006ze, Ciuchini:2007ha})
before applying eq.(\ref{eq:deltap}).

In obtaining the values of table \ref{ta:compare}, we use as starting point the
$\delta(M_{W}^{2})$ corresponding to the mass ratios
$\frac{m_{\tilde{\gamma}}^{2}}{m_{\tilde{q}}^{2}}$ and 
$\frac{m_{\tilde{\gamma}}^{2}}{m_{\tilde{l}}^{2}}$ of $1.0$ in tables
\ref{ta:bounds} and \ref{ta:lbounds}.
The bounds for $|\delta |$ shown in table \ref{ta:compare} are obtained
under the most conservative assumption about the phases to make the bound as
strong as possible: when two different $\delta $ are present in the
original experimental bound (as in the $3$rd column of table
\ref{ta:bounds}), we took the value for $|\delta_{LL}|$ to be the same
as $|\delta _{RR}|$ (this leads to the same upper bound for $\delta
_{LL}$ and $\delta _{RR}$ in consecutive rows of table
\ref{ta:compare}).

\begin{table}[t]
\centering
\begin{tabular}{|c|c|c|}
\hline
$\delta(M_{X}^{2})$ & Gauge mediation ($M_{X}=200$ TeV) & Gravity mediation ($M_{X}=M_{P}$) \\ 
\hline
&  &  \\ 
$\left| \delta ^{\prime d}_{12_{LL}} \right|$ & $5.9 \times 10^{-4}$ & $8.6
\times 10^{-3}$ \\ 
&  &  \\ 
$\left| \delta ^{\prime d}_{12_{RR}} \right|$ & $5.9 \times 10^{-4}$ & $8.6
\times 10^{-3}$ \\ 
&  &  \\ 
$\left| \delta ^{\prime d}_{13_{LL}} \right|$ & $4.8 \times 10^{-2}$ & $7.0
\times 10^{-1}$ \\ 
&  &  \\ 
$\left| \delta ^{\prime d}_{13_{RR}} \right|$ & $4.8 \times 10^{-2}$ & $7.0
\times 10^{-1}$ \\ 
&  &  \\ 
$\left| \delta ^{\prime u}_{12_{LL}} \right|$ & $4.5 \times 10^{-2}$ & $6.6
\times 10^{-1}$ \\ 
&  &  \\ 
$\left| \delta ^{\prime u}_{12_{RR}} \right|$ & $4.5 \times 10^{-2}$ & $6.6
\times 10^{-1}$ \\ 
&  &  \\ 
$\left| \delta ^{\prime d}_{23_{LL}} \right|$ & $4.3 \times 10^{-1}$ & $6.25$
\\ 
&  &  \\ 
$\left| \delta ^{\prime e}_{12_{LL}} \right|$ & $ 6.0 \times 10^{-4}$ & $9.4
\times 10^{-4}$ \\ 
&  &  \\ 
$\left| \delta ^{\prime e}_{13_{LL}} \right|$ & $1.5 \times 10^{-1}$ & $2.3 \times 10^{-1}$ \\ 
&  &  \\ 
$\left| \delta ^{\prime e}_{23_{LL}} \right|$ & $1.2 \times 10^{-1}$ & $1.8 \times 10^{-1}$ \\ 
&  &  \\ \hline
\end{tabular}%
\caption{Upper bounds at typical gauge / gravity mediator scales.}
\label{ta:compare}
\end{table}

It may be seen from table \ref{ta:compare} that the most stringent limits
(applying to the first two generations) are rather tight. One should note however that we are
considering the most pessimistic case (as discussed in subsection
\ref{sub:Unknown}). For example, it is quite conceivable that the
mixing matrices feature small mixing, which implies the mixing pairs
in eq.(\ref{eq:cdelta_12}) (and similar equations) can readily take values $O(10^{-1})$
or even smaller, rather than $\frac{1}{2}$. How much suppression one
allows from the mixing depends on what is considered natural -
requiring them to be very small in order to completely solve the
\FP \ falls under the alignment solution \cite{Nir:1993mx}, 
which is only natural if explained by some specific mechanism. In this chapter
we eschew this explanation and look for a more general explanation for the
suppression of the \FCNC s.

\section{Solving the \FS \ \FP \ \label{sec:Confp}}

In this section we discuss the conditions for $\langle D^{2} \rangle$ in
eq.(\ref{eq:maximal}) to be anomalously small. $\langle D^{2} \rangle$ is fixed
when minimising the flavon potential and this relates it to the flavon
masses. We illustrate the general expectation for $\langle D^{2} \rangle$ in the context
of a $U(1)$ \FS \ with a simple flavon sector (see subsection \ref{sec:Dterms}). The important aspects of the form of the $D$-term are common to more complicated flavon sectors and even to non-Abelian \FSs : this may be seen from the fact that it is always possible to choose a basis in which the dominant $D$-term contribution to
the mixing between two particular generations corresponds to a diagonal
generator and thus has the same form as the Abelian case.

We consider only the case $c<0$ in order to allow a partial
cancellation of the $D$-term 
\footnote{$c<0$ excludes the one flavon case, which in some cases is equivalent to $c=0$.} and we assume also that the \FS \ breaking scale is much greater than the SUSY soft masses. 

\subsection{$F$-term breaking \label{sub:Fterm}}

We first consider a case where the flavons acquire non-vanishing \VEV s due
to an $F$-term. We take the real potential to be:

\begin{equation}
V = g_{f}^{2}\left( |\phi |^{2}+c|\bar{\phi}|^{2}\right) ^{2}+m^{2}|\phi |^{2}+
\bar{m}^{2}|\bar{\phi}|^{2}+g\left\vert \phi \bar{\phi}- \mu_{Z}^{2}\right\vert ^{2}
\end{equation}
We have included the $D$-term, flavon soft mass terms and also an $F$-term
$F_Z$ coming from a superpotential term $g Z \left( \phi \bar{\phi}- \mu_{Z}^{2}\right) $. By minimising the potential we find:

\begin{eqnarray}
\langle |\phi |^{2} \rangle &\simeq &-c \langle |\bar{\phi}|^{2} \rangle \\
\left\langle D^{2}\right\rangle &\equiv &\langle g_{f}^{2}\left( |\phi
|^{2}+c|\bar{\phi}|^{2}\right) \rangle \simeq \langle \frac{-m^{2}-\bar{m}%
^{2}/c}{4}\rangle  \label{eq:Dvev}
\end{eqnarray}

\subsection{Radiative breaking \label{sub:Radiative}}

As a second example, we consider a case where the \VEV s are driven
radiatively. We take $V$ to have the form:

\begin{equation}
V=g_{f}^{2}\left( |\phi |^{2}+c|\bar{\phi}|^{2}\right) ^{2}+\alpha _{\phi
}|\phi |^{2}m^{\prime 2}\mathrm{ln}\left( \frac{|\phi |^{2}}{|\Lambda |^{2}}%
\right) +\alpha _{\bar{\phi}}|\bar{\phi}|^{2}\bar{m}^{\prime 2}\mathrm{%
ln}\left( \frac{|\bar{\phi}|^{2}}{\bar{|\Lambda |}^{2}}\right)
\end{equation}%
The two last terms include the effects of radiative corrections, $\alpha _{\phi }$ and $\alpha _{\bar{\phi}}$ are the fine structure constants
associated with the interactions of $\phi $ and $\bar{\phi }$ and the
tree level contributions have been absorbed in $\Lambda $ and $\bar{%
\Lambda }$. Minimisation gives:

\begin{eqnarray}
\left\langle D^{2}\right\rangle &\equiv &\langle g_{f}^{2}\left( |\phi
|^{2}+c|\bar{\phi}|^{2}\right) \rangle \\
\left\langle D^{2}\right\rangle &\simeq &\langle \frac{-\alpha _{\phi }m^{\prime 2}\mathrm{ln}\left( \frac{%
|\phi |^{2}}{|\Lambda |^{2}}\right) -\alpha _{\bar{\phi}}\bar{m}%
^{\prime 2}\mathrm{ln}\left( \frac{|\bar{\phi}|^{2}}{\bar{|\Lambda |}^{2}}%
\right) /c}{4}\rangle  \label{drunning}
\end{eqnarray}%
Eq.(\ref{drunning}) has the form of eq.(\ref{eq:Dvev}), where $m$ and $\bar{m}$ are now
interpreted as running masses:

\begin{equation}
m^{2}\equiv \alpha _{\phi } m^{\prime 2}\mathrm{ln}\left( \frac{|\phi |^{2}}{|\Lambda |^{2}}%
\right)
\end{equation}

\begin{equation}
\bar{m}^{2}\equiv \alpha _{\bar{\phi}} \bar{m}^{\prime 2}\mathrm{ln}\left( \frac{|\bar{\phi}%
|^{2}}{\bar{|\Lambda |}^{2}}\right)  \label{eq:running_barm}
\end{equation}
With this form of $\langle D ^{2} \rangle$ we have from eq.(\ref{eq:Delta_m}):

\begin{equation}
\Delta m_{\tilde{f}_{L;R}}^{2}\simeq \frac{c_{\tilde{f}_{L;R}}}{2}(-m^{2}-%
\bar{m}^{2}/c)  \label{eq:D_combine}
\end{equation}
We can then determine $\delta $ from eq.(\ref{eq:maximal}). For example, the $(1,2)$ element is:

\begin{equation}
\left\vert \delta _{12}^{LL}\right\vert < \frac{\mathrm{Max}%
\left\vert c_{i}-c_{j}\right\vert }{4} \left\vert \frac{m^{2}+\bar{m}%
^{2}/c}{\langle m_{\tilde{q}}^{2}\rangle }\right\vert   \label{eq:fmax}
\end{equation}%
$\langle m_{\tilde{q}}^{2}\rangle $ must be evaluated at the
appropriate mediator scale (see subsection \ref{sub:Running}). To estimate the factor involving
the flavon masses we must consider the origin of the soft masses
$m$ and $\bar{m}$. To do this we consider separately the case of
gravity and of gauge mediation.

\subsection{Gravity mediated SUSY breaking \label{sub:SUGRA}}

We first consider \SUGRA \ as the origin of the flavon and sfermion masses.
The soft masses are generated at the Planck scale, so we use the estimated
values in the third column of table \ref{ta:compare}. It may be seen that the
bounds are only significant for the mixing between the first two
generations (the others are at worse $O(10^{-1})$). Are there ways these bounds can naturally be satisfied without appealing to small mixing angles?

As we have stressed, \SUGRA \ models solve the SUSY \FP \ by
taking all the soft masses to have a common value at the Planck scale,
arguing that gravity is family and flavour blind. This naturally
extends to the flavon sector too, so we expect $m^{2}=\bar{m}^{2}$ at
the Planck scale. As long as the radiative corrections are small, it is immediately 
obvious that the common soft masses
offer an elegant solution not just to the SUSY \FP \ but also to the \FS \ \FP \ if the
flavons have equal but opposite charges ($c=-1$). In this case,
comparing with eq.(\ref{eq:Dvev}), we see that $\langle D^{2} \rangle$
vanishes and the \FCNC \ bounds are automatically satisfied. The underlying
justification is that the flavon potential is symmetric 
under the interchange of $\phi $ and $\bar{\phi}$. Of course, radiative
corrections involving Yukawa couplings may spoil this symmetry, but these
radiative corrections are suppressed by loop factors and so
can satisfy the bounds even with the very conservative assumptions about
mixing angles and phases discussed in subsection \ref{sub:Unknown}. Although we
illustrated this $D$-term cancellation mechanism with a very simple flavon
sector, the mechanism applies also in the case where several flavons $\phi_{A}$ \footnote{Note that $A$ is a label, not an index.} of charge $c_{A}$
contribute significantly to the $D$-term. We can obtain the result of eq.(\ref{eq:Dvev}) whenever the charges of the flavons are of equal magnitude. In that special case we can factorise out the common soft mass and the common charge, redefining the multiple fields with positive charge into an effective $\phi$ and the fields with negative charge into an effective $\bar{\phi}$ regardless of the number of flavons, and recovering eq.(\ref{eq:Dvev}). Thus we see that $\langle D^{2} \rangle$ still vanishes. 
This occurs as the common charges and universal soft masses preserve the underlying interchange symmetry that existed already in the two flavon case.

The case of radiative \FS \ breaking is perhaps more interesting
as it does not require the introduction of the mass scale $\mu_Z $. For the
case $c=-1$, we have $m^{\prime 2}=\bar{m}^{\prime 2}$, the initial tree level
contributions cancel and we get: 
\begin{equation}
\left\langle D^{2}\right\rangle \simeq (\alpha _{\phi }-\alpha _{\bar{\phi}%
})m^{\prime 2}\ln (\frac{\left\langle |\phi |^{2}\right\rangle }{M_{P}^{2}})
\end{equation}%
Since the radiative breaking mechanism requires that the radiative
corrections to the soft masses are of the same order as the tree level
contributions each of the two terms is of $O(m^{\prime 2}).$ Thus, if the $%
D-$term is to vanish, it is necessary for $\alpha _{\phi }=\alpha _{\bar{\phi%
}}$ corresponding to the couplings driving the radiative breaking being
symmetric under the interchange of $\phi $ and $\bar{\phi}$ . The
possibility of such a symmetry is not unnatural for the class of \FS \
models discussed in
\cite{King:2003rf,Ivo1,Ivo3}
(see also chapter \ref{ch:fs} and chapter \ref{ch:fsd}). In these models, the \FMM \ are 
generated through the \FN \ mechanism by the coupling of flavons (such as $\phi$) to heavy supermultiplets which come in vector-like pairs such as $\bar{A}$,$A$, and $\bar{B}$, $B$. Being vector-like, if the coupling $\phi A \bar{B}$ is allowed then so too is the coupling $\bar{\phi }
\bar{A} B$ so it is easy to implement a symmetry connecting these terms.

The case of anomaly mediated SUSY breaking offers another way of
suppressing the \FCNC \ effects, because the soft masses are given in terms of
the anomalous dimensions of the fields. If $c=-1$, the gauge contributions to 
$m$ and $\bar{m}$ are equal. The generation dependent non-gauge contributions
are expected to be small, leading to a suppression of $\langle D^{2} \rangle$ as
discussed above. The important point is that ultraviolet effects decouple in anomaly
mediation, meaning that the anomalous dimension has only contributions from
fields light at the relevant scale, which here is the scale of \FS \ breaking. Provided the generation dependent couplings of the flavons
involve only states heavier than this scale they will not split the
degeneracy driven by the gauge coupling. This is again the case in the class of
family models discussed in
\cite{King:2003rf,Ivo1,Ivo3}
because there the vector-like 
supermultiplet mass (generically denoted in chapter \ref{ch:fs} and chapter \ref{ch:fsd} as
$M$) is necessarily heavier than the flavons \VEV s in order to 
generate small expansion parameters like $\langle \phi \rangle /M$,
which generate the required hierarchies of \FMM .

Yet another way of suppressing the $D$-term is provided by orbifold
compactification of string models where the soft masses depend on the the
modular weights of the superfields \cite{Casas:1999xj} and can be
anomalously small if their modular weights are $-3$. Thus if the
flavons have this modular weight and the squarks and sleptons do not,
the factor $(m^{2}+\bar{m}^{2}/c)/\langle m_{\tilde{q}}^{2}\rangle $
appearing in eq.(\ref{eq:fmax}) may be very small leading to the
required \FCNC s suppression.

\subsection{Gauge mediated SUSY breaking \label{sub:Gauge}}

We turn now to the case where the soft masses are due to gauge mediated SUSY
breaking. Since the mediator mass is low the radiative corrections discussed
in subsection \ref{sub:Running} are small. This may be seen by the values in
the second column of table \ref{ta:compare} where the bounds are
close to the experimental values obtained at the electroweak scale. To be
consistent with these bounds requires a larger suppression to come
from the $(m^{2}+\bar{m}^{2}/c)/\langle m_{\tilde{q}}^{2}\rangle $
factor than in the gravity mediated case.

\begin{figure}[htb]
\begin{center}
\fcolorbox{white}{white}{
  \begin{picture}(165,91) (105,-209)
    \SetWidth{0.5}
    \SetColor{Black}
    \ArrowLine(105,-179)(150,-179)
    \Vertex(150,-179){2.83}
    \ArrowLine(210,-179)(255,-179)
    \Vertex(210,-179){2.83}
    \ArrowLine(210,-179)(150,-179)
    \DashArrowArc(180,-179)(30,-0,180){10}
    \Text(230,-209)[lb]{\Large{\Black{$\lambda$}}}
    \Text(120,-209)[lb]{\Large{\Black{$\lambda$}}}
    \Text(175,-134)[lb]{\Large{\Black{$\Phi$}}}
    \Text(175,-209)[lb]{\Large{\Black{$\Phi$}}}
  \end{picture}
}
\caption{Gaugino mass: 1 loop.}
\label{fig:gauge1}
\end{center}
\end{figure}
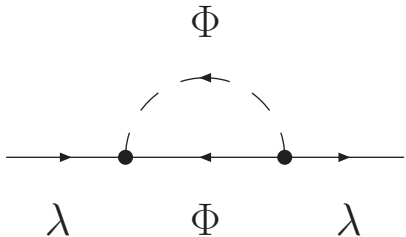

Fortunately, gauge mediated models naturally provide such a suppression
as long as the flavons have no direct coupling to the SUSY breaking sector.
This follows because the gauginos do not couple directly to the flavons
(the flavons are not charged under the \SM \ gauge group) and so the
contributions to the flavon masses occur at one loop order higher than the
contributions to the sfermion masses. To see this explicitly, note that the
gaugino masses are generated as a one loop effect, with the heavy
messenger(s) of SUSY breaking, $\Phi$, coupling directly to the gaugino $\lambda$, as seen in figure \ref{fig:gauge1}.

\begin{figure}[htb]
\begin{center}
\fcolorbox{white}{white}{
  \begin{picture}(270,106) (60,-193)
    \SetWidth{0.5}
    \SetColor{Black}
    \ArrowLine(105,-164)(150,-164)
    \Vertex(150,-164){2.83}
    \ArrowLine(210,-164)(255,-164)
    \Vertex(210,-164){2.83}
    \ArrowLine(210,-164)(150,-164)
    \DashArrowArc(180,-164)(30,-0,180){10}
    \Text(230,-194)[lb]{\Large{\Black{$\lambda$}}}
    \Text(120,-194)[lb]{\Large{\Black{$\lambda$}}}
    \Text(175,-119)[lb]{\Large{\Black{$\Phi$}}}
    \Text(175,-194)[lb]{\Large{\Black{$\Phi$}}}
    \Vertex(105,-164){2.83}
    \Vertex(255,-164){2.83}
    \ArrowArc(180,-164)(75,-0,180)
    \DashArrowLine(105,-164)(60,-164){10}
    \DashArrowLine(300,-164)(255,-164){10}
    \Text(75,-194)[lb]{\Large{\Black{$\tilde{f}$}}}
    \Text(275,-194)[lb]{\Large{\Black{$\tilde{f}$}}}
  \end{picture}
}
\caption{Sfermion mass: 2 loops.}
\label{fig:gauge2}
\end{center}
\end{figure}
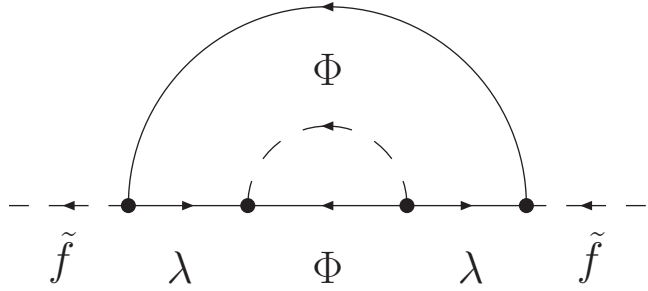

In turn, the (generation blind) contributions to sfermion ($\tilde{f}$) masses are two loop effects 
\cite{Alvarez-Gaume:1981wy,Giudice:1997ni,Giudice:1998bp} through
their coupling to gauginos, as seen in figure \ref{fig:gauge2}. 

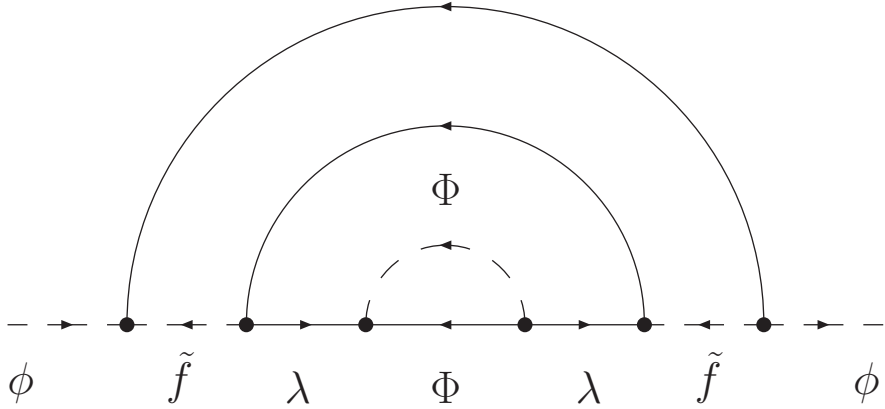
\begin{figure}[htb]
\begin{center}
\fcolorbox{white}{white}{
  \begin{picture}(360,151) (15,-148)
    \SetWidth{0.5}
    \SetColor{Black}
    \ArrowLine(105,-119)(150,-119)
    \Vertex(150,-119){2.83}
    \ArrowLine(210,-119)(255,-119)
    \Vertex(210,-119){2.83}
    \ArrowLine(210,-119)(150,-119)
    \DashArrowArc(180,-119)(30,-0,180){10}
    \Text(230,-149)[lb]{\Large{\Black{$\lambda$}}}
    \Text(120,-149)[lb]{\Large{\Black{$\lambda$}}}
    \Text(175,-74)[lb]{\Large{\Black{$\Phi$}}}
    \Text(175,-149)[lb]{\Large{\Black{$\Phi$}}}
    \Vertex(105,-119){2.83}
    \Vertex(255,-119){2.83}
    \ArrowArc(180,-119)(75,-0,180)
    \DashArrowLine(105,-119)(60,-119){10}
    \DashArrowLine(300,-119)(255,-119){10}
    \Text(75,-149)[lb]{\Large{\Black{$\tilde{f}$}}}
    \Text(275,-149)[lb]{\Large{\Black{$\tilde{f}$}}}
    \Vertex(60,-119){2.83}
    \Vertex(300,-119){2.83}
    \DashArrowLine(15,-119)(60,-119){10}
    \DashArrowLine(300,-119)(345,-119){10}
    \Text(15,-149)[lb]{\Large{\Black{$\phi$}}}
    \Text(335,-149)[lb]{\Large{\Black{$\phi$}}}
    \ArrowArc(180,-119)(120,-0,180)
  \end{picture}
}
\caption{Flavon mass: 3 loops.}
\label{fig:gauge3}
\end{center}
\end{figure}

The only way for the gauginos to
communicate the SUSY breaking to the flavon sector is through 
the flavon coupling to the sfermions - so it is a three loop effect with an
additional loop suppression which depends on the \FS \ gauge
coupling strength - as seen in the diagram of figure \ref{fig:gauge3}.
For low gauge mediation scale, $(m^{2}+\bar{m}^{2}/c)/\langle
m_{\tilde{q}}^{2}\rangle $ needs to take values as small as $6
\times 10^{-4}$. This is possible if the \FS \ gauge coupling is very small, $\alpha_{f}/4\pi <10^{-3}$. In practise one
might expect a combination of the loop factor and a mixing angle
suppression below the maximum used in deriving the upper bounds will
allow for a solution with a larger gauge coupling. Alternatively it
may be that the gauge mediation scale is higher than $200$ TeV leading
to a further suppression of the bound compared to those shown in table
\ref{ta:compare}. Further, we have once again the special case $c=-1$, where the \FS \ gauge contribution is completely suppressed: the contribution to the flavon mass is proportional to the square
of the family charge, and thus generates $m^{2}=\bar{m}^{2}$. In this
case the bounds are satisfied for any value of the \FS \ gauge coupling, as
the non-gauge interactions are very heavily suppressed.

\section{Summary and conclusions \label{sec:Sumfp}}

We have re-examined the constraints on continuous \FSs \
coming from the need to suppress the associated $D$-term contributions to
sfermion masses below the experimental bounds coming from \FCNC \ processes.
The $D$-term contributions depend on unknown mixing angles, so we first derived upper bounds which are independent of
the mixing angles. We then compared these upper bounds with the experimental
bounds in each sector, accounting for the weakening of the constraints at
higher energy scales. 

Albeit the analysis was performed in a simple model, the results are valid
in more general cases (including non-Abelian \FSs \ and models with
more than two dominant flavons).

For the case of gravity mediation the constraints are
only significant for the mixing of the first two generations. We identified
several ways in which these constraints are automatically satisfied without
appealing to a suppression involving alignment between the fermions and
sfermions. In the \SUGRA \ and anomaly mediated cases, if the flavon fields
spontaneously breaking a \FS \ relating the first two
generations have the same magnitude of family charge, the $D$-terms vanish
up to radiative corrections which may readily be within the constraints.
Even if the radiative corrections are large, the $D$-terms may still be
within the limits if there is an underlying symmetry relating the couplings
of the flavon fields and this may happen quite readily in family schemes
relying on the \FN \ mechanism to generate the \FMM \ (such as those discussed in chapter \ref{ch:fs} and chapter \ref{ch:fsd}). Yet
another possibility, motivated by orbifold string 
compactified models, is that the flavons have modular weights such that
they are anomalously light. This mechanism works irrespective of the family
charge carried by the flavons.

For the case of gauge mediated models, the
lower mediation scale leads to stronger constraints which are non-trivial
for all three generations. For arbitrary
flavon charges these bounds can be satisfied by having a small family gauge
coupling if, as is generally the case, the flavons couple to the
messenger sector only via the quark and lepton sector. For special
cases in which the magnitude of flavon charges are equal, the bounds are
satisfied for arbitrary gauge coupling.

In conclusion, the $D$-terms associated with continuous \FSs \
may be consistent with the experimental bounds on \FCNC \ for a large class of
family models and SUSY breaking schemes. It is relevant to note that
models including just a single flavon have less possibilities of
avoiding the bounds.

Finally, in almost all cases the present bounds on the mixing between
the first two families are quite close to the expected signals in
these models, demonstrating yet again the importance of improving the
experimental searches for \FCNC \ effects.

\chapter{Conclusion \label{ch:conc}}

In this chapter we present the conclusions by means of a very brief global summary. Detailed summaries of chapter \ref{ch:fs}, chapter \ref{ch:fsd} and chapter \ref{ch:fcnc} are presented in section \ref{sec:Confmm1}, section \ref{sec:Confmm2} and section \ref{sec:Sumfp} respectively.

We have constructed theories of \FMM \ based on
\FSs . The model presented in section \ref{sec:Symmetries}
is based on $SU(3)_{f}$, and a relatively simpler model based on its
discrete subgroup $\Delta(27)$ is presented in section
\ref{sec:D27}. The two models produce phenomenologically successful
patterns of \FMM \ and particularly, the leptonic mixing
angles are predicted to be close to the \TBM \ structure that is
observed.

Being based on a continuous symmetry, the $SU(3)_{f}$ model predicts
contributions to \FCNC s through the $D$-term associated with the
\FS . The sort of constraints this produces in the context
of the SUSY \FP \ is analysed for general continuous \FSs \ in chapter \ref{ch:fcnc}.

\chapter*{Final comments}

This thesis was supported by FCT under the grant SFRH/BD/12218/2003.

All Feynman diagrams were created with JaxoDraw \cite{Jaxo}.

Figures \ref{fig:susy_mass}, \ref{fig:susy_gauge} and \ref{fig:TBM} used with authors' permission - namely, Steve Martin and Alexei Smirnov - duly referenced in the respective captions.

\appendix

\chapter{Vacuum alignment: $SU(3)_{f}$ model \label{app_A}}


As discussed in section \ref{sec:Symmetries}, the alignment of \VEV s
is crucial to 
the generation of \TBM. In this appendix we consider
in detail how the
minimisation of the scalar potential proceeds, and show that the soft
masses and the
superpotential terms allowed by the symmetries of the $SU(3)_{f}$ model
lead to the desired \VEV \
alignment. The field content and their symmetry properties are given in table
\ref{Ta:Table 1}.

We start by considering the $\theta _{i}$ and $\bar{\theta}^{i}$ fields. Their
soft masses $m_{\theta}$ and $m_{\bar{\theta}}$ have radiative
corrections coming from gauge and Yukawa related couplings. The gauge
couplings contribute positively to their squared soft masses while the
Yukawa couplings contribute negatively. If the latter dominates, the squared masses
can be driven negative, triggering radiative breaking and
driving a
non-zero \VEV \ for the field. Since $\theta$ and $\bar{\theta }$ are in
conjugate representations there is a $D$-flat direction with 
equal \VEV s for  $\theta$ and $\bar{\theta}$. Assuming this
direction is also $F-$flat (this depends on the structure of the massive
sector of the theory which is not specified here) the scale of the
\VEV s is
close to the scale $\Lambda$, at which $m_{\theta}^{2}(\Lambda
)+m_{\bar{\theta}}^{2}(\Lambda )=0$. 
Without loss of generality the
basis is chosen such that this initial breaking of the $SU(3)_{f}$ is
aligned along the $3$rd family direction.

Turning now to the $\phi _{3}$ and $\bar{\phi}_{3}$ fields we consider the
superpotential of the form: 

\begin{equation}
P_{3} = X_{3} \left(Tr[\bar{\phi}_{3}^{i}] \phi_{3_{i}} - M S_{3} \right)
\end{equation}%
$P_3$ is allowed by the symmetries of the theory with the trace $Tr$ taken to yield the $SU(3)_{f}\times
SU(2)_{R}$ invariant component of the first term.
$M$ is a mass scale (it can arise from a singlet like $S_{3}$, as long as the charges are suitable). We assume that the field $S_{3}$ undergoes radiative breaking with \VEV \ $\mu _{3}$. Then
the $\phi _{3}$ and $\bar{\phi}_{3}$ \VEV s  are triggered by the $F$-term $\left\vert F_{X_{3}}\right\vert ^{2}$.
These \VEV s  develop
along the $D$-flat direction:
\begin{equation}
\left\langle \bar{\phi}_{3}\right\rangle =\left( 
\begin{array}{ccc}
0 & 0 & 1%
\end{array}%
\right) \otimes \left( 
\begin{array}{cc}
a_{u} & 0 \\ 
0 & a_{d}%
\end{array}%
\right)
\label{eq:AP3} 
\end{equation}%
With the $SU(3)_{f}\times SU(2)_{R}$ structure exhibited in eq.(\ref{eq:AP3}), and: 
\begin{equation}
\left\langle \phi _{3}\right\rangle =\sqrt{a_{u}^{2}+a_{d}^{2}}\left( 
\begin{array}{c}
0 \\ 
0 \\ 
1%
\end{array}%
\right) 
\end{equation}%
We note that these \VEV s  naturally align along the same direction as
$\langle \theta \rangle$ and $\langle \bar{\theta} \rangle$. The reason is that
$\theta$ and $\bar{\theta}$ break $SU(3)_{f}$ to $SU(2)_{f}$ so that
the stabilising gauge radiative corrections (positive) to the squared soft masses of
$\bar{\phi}_{3}$ and $\phi_{3}$ act more on the 1st and 2nd 
family elements than on the third family elements,
$\bar{\phi}_{3}^{3}$ and $\phi_{3_{3}}$.
As a result these third components are the ones that have the smallest
mass squared and so it is energetically favourable for the \VEV s  to
develop along 
them. We have thus obtained the structure of eq.(\ref{eq:P3B vev}),
eq.(\ref{eq:P3 vev}), eq.(\ref{eq:T3B vev}) and
eq.(\ref{eq:T3 vev}). In fact it is not relevant whether
$\theta$, $\bar{\theta}$ acquire their \VEV s  after or before
$\bar{\phi}_{3}$, just that they all share the same direction, which is guaranteed by the residual $SU(2)_{f}$ as described.

Consider now the adjoint field $\Sigma _{i}^{j}$, which we introduce for
alignment reasons. The symmetries of the theory allow the following
renormalisable terms in the superpotential:
\begin{equation}
P_{\Sigma }=\frac{\beta _{3}}{3}Tr\left( \Sigma ^{3}\right) +M' \frac{\beta
_{2}}{2}Tr\left( \Sigma ^{2}\right) 
\end{equation}%
$M'$ is a mass scale in the effective theory associated with the
UV completion of the theory. It can arise from the \VEV \ of a field with correct $R$ charge ($R=2/3$,
same $R$ charge as that of $\Sigma $ so that the term is allowed in
the superpotential). This superpotential induces a \VEV \ of the form:
\begin{equation}
\left\langle \Sigma \right\rangle =\left( 
\begin{array}{ccc}
a & 0 & 0 \\ 
0 & a & 0 \\ 
0 & 0 & -2a%
\end{array}%
\right) 
\label{eq:Sigma vev}
\end{equation}%
The relative alignment of $\left\langle \Sigma \right\rangle $ and $%
\left\langle \bar{\phi}_{3}\right\rangle $ (i.e. which direction has
the \VEV \ with the largest magnitude, $2 a$) again follows from the residual $SU(2)_{f}$, for the same reasons leading to the relative alignment of $\langle \bar{\phi}_{3}
\rangle$ with the $\theta$ \VEV s: the \VEV \ in eq.(\ref{eq:Sigma vev})
preserves the yet unbroken $SU(2)_{f}$, so that the stabilising
(positive) gauge radiative corrections act more on the 1st and 2nd
direction, favouring the 3rd direction to acquire the largest
magnitude, in order to minimise the residual vacuum energy.

The second stage of symmetry breaking will break the remaining $SU(2)_{f}$. Consider the fields $\phi _{2}$, $\bar{\phi }_{23}$. In discussing the
cancellation of the $D$-terms we allow for the presence of additional fields 
$\bar{\phi}_{2}$, $\phi_{23}$ transforming in conjugate representations to $%
\phi_{2}$ and $\bar{\phi}_{23}$ respectively. Their alignment is driven by the following superpotential terms:

\begin{equation}
P_{23}=Y_{2}Tr[\bar{\phi}_{3}^{i}]\phi _{2_{i}}+X_{23}\left( \bar{\phi}%
_{23}^{i}\phi _{3_{i}}\bar{\phi}_{23}^{j}\phi _{2_{j}}-\mu _{23}^{4}\right)
+Y_{23}\left( \bar{\phi}_{23}^{i}\phi _{23_{i}}\right) 
\end{equation}%
The trace is taken as before over the $SU(2)_{R}$ indices, and we only represent explicitly the $SU(3)_{f}$
family indices. The quantity $\mu _{23}$ is a mass scale, and
similarly to $\mu _{3}$ it can arise from an appropriate singlet $S_{23}$.
Following from this superpotential are $F$-term contributions to the scalar
potential which force the \VEV s  of the form of eq.(\ref{eq:P2 vev}),
eq.(\ref{eq:P23B vev}). The term
$\left\vert F_{X_{23}}\right\vert^{2}=
\left\vert\bar{\phi}_{23}^{i}\phi_{3_{i}}\bar{\phi}_{23}^{j}\phi_{2_{j}}
-\mu_{23}^{4}\right\vert^{2}$ forces the \VEV \ of both $\phi _{2}$
and $\bar{\phi}_{23}$ to be nonzero. The term $\left\vert
F_{Y_{2}}\right\vert ^{2}=$$\left\vert \bar{\phi}_{3}^{i}\phi
_{2_{i}}\right\vert ^{2}$ forces the $\phi _{2}$ \VEV \ to be orthogonal to
that of $\bar{\phi}_{3}$. Without loss of generality we may choose a basis
in which it is aligned along the $2$nd direction as in eq.(\ref{eq:P2 vev}):

\begin{equation}
\left\langle \phi _{2}\right\rangle =\left( 
\begin{array}{c}
0 \\ 
y \\ 
0%
\end{array}%
\right)
\end{equation}%
$\left\vert F_{X_{23}}\right\vert ^{2}$ forces $\bar{\phi}_{23}$ 
to have non-vanishing \VEV s in both the $2$nd and $3$rd directions:

\begin{equation}
\left\langle \bar{\phi}_{23}\right\rangle =\left( 
\begin{array}{ccc}
0 & b & b_{3}%
\end{array}%
\right)  \label{eq:P23B vev2}
\end{equation}

Consider now the remaining pair of fields $\phi_{123}$, $\bar{\phi}_{123}$. The alignment of
these will complete the alignment discussion. The relevant terms are:

\begin{equation}
P_{123} = X_{123}\left( \bar{\phi}_{123}^{i}\phi _{123_{i}}-\mu
_{123}^{2}\right)  \label{eq:A_P123_X}
\end{equation}%
\begin{equation}
+Y_{123}\left( \bar{\phi}_{23}^{i}\phi _{123_{i}}\right)  \label{eq:A_P123_Y}
\end{equation}
\begin{equation}
+Z_{123}\left( \bar{\phi}_{123}^{i}\Sigma _{i}^{j}\phi _{123_{j}}\right)
\label{eq:A_P123_Z}
\end{equation}
The operators in eq.(\ref{eq:A_P123_X}), eq.(\ref{eq:A_P123_Y}) and
eq.(\ref{eq:A_P123_Z}) trigger and align the \VEV s  of
$\bar{\phi}_{123}$ (as in eq.(\ref{eq:P123B vev})) and $\phi _{123}$
(as in eq.(\ref{eq:P123 vev})) through the \VEV s  of $\bar{\phi}_{23}$,
$\Sigma $ and the mass scale $\mu _{123}$ (like with $\mu_{23}$, this mass
scale can be obtained through the \VEV \ of a singlet $S_{123}$).

Throughout the analysis, the important effect of $D$-terms and soft
SUSY breaking mass terms $m_{i}$ must be taken into account. We will analyse these conditions perturbatively in an
expansion involving small mass ratios, assuming the ordering $m_{i}\ll
c,\bar{c}\ll b,b_{3},y\ll a_{u},$ $a_{d}$ which is the
phenomenologically viable range and which is readily obtained by a choice of
the free parameters of the theory. In this case minimisation of the
potential in leading order proceeds by setting the $D$-terms and $F$-terms to zero
and minimising the contribution to the potential coming from the soft terms.
Of course the true minimum corresponds to the case that the $D$-terms and $F$-terms
terms are not zero, but instead have a magnitude comparable to the contribution of
the soft terms. This only involves a non-leading change in the \VEV s found
by setting the $D$-terms and $F$-terms to zero, and so can be dropped in leading
order.

The desired \VEV \ structure has the ($D$-flat and $F$-flat) form:

\begin{equation}
\left\langle \bar{\phi}_{3}\right\rangle =\left( 
\begin{array}{ccc}
0 & 0 & 1%
\end{array}%
\right) \otimes \left( 
\begin{array}{cc}
a_{u} & 0 \\ 
0 & a_{d}%
\end{array}%
\right)
\label{eq:Pvev1}
\end{equation}

\begin{equation}
\left\langle \phi_{3}\right\rangle =\left( 
\begin{array}{c}
0 \\ 
0 \\ 
\sqrt{a_{u}^{2}+a_{d}^{2}}%
\end{array}%
\right)
\end{equation}

\begin{equation}
\left\langle \bar{\phi}_{2}\right\rangle = \left( 
\begin{array}{ccc}
0 & y & 0%
\end{array}%
\right)
\end{equation}

\begin{equation}
\left\langle {\phi}_{2}\right\rangle = \left( 
\begin{array}{c}
0 \\ 
y \\ 
0%
\end{array}%
\right)
\end{equation}

\begin{equation}
\left\langle \bar{\phi}_{23}\right\rangle =\left( 
\begin{array}{ccc}
0 & b & -b%
\end{array}%
\right)
\end{equation}

\begin{equation}
\left\langle {\phi }_{23}\right\rangle = \left( 
\begin{array}{c}
0 \\ 
b \\ 
b%
\end{array}%
\right) e^{i \beta}
\end{equation}

\begin{equation}
\left\langle \bar{\phi}_{123}\right\rangle =\left( 
\begin{array}{ccc}
\bar{c} & \bar{c} & \bar{c}%
\end{array}%
\right)
\end{equation}

\begin{equation}
\left\langle \phi_{123}\right\rangle =\left(%
\begin{array}{c}
\bar{c} \\ 
\bar{c} \\ 
\bar{c}%
\end{array}%
\right) e^{i \gamma}
\label{eq:Pvev2}
\end{equation}
The overall phases are factored into the definitions of $b$ and $\bar{c%
}$ leaving only relative phases which are explicitly shown. These
phases uniquely preserve the $F$-flatness in this configuration. The
relative phase of $\phi _{23}$ is connected with the relative 
phase of $\bar{\phi}_{23}$ due to the $\left\vert 
F_{Y_{23}}\right\vert ^{2}=\left\vert \bar{\phi}_{23}^{i}\phi
_{23_{i}}\right\vert ^{2}$ orthogonality condition. The $D$-term flatness
conditions constrain the relative phases of the $\phi _{123}$ and $\bar{\phi}%
_{123}$ components to be zero (i.e. each of these \VEV s  has solely an overall
undetermined phase). The relative phase of the $\bar{\phi}_{23}$
field components is $\pi$. This follows because the $F-$term $\left\vert
F_{Y_{123}}\right\vert ^{2}=$$\left\vert \bar{\phi}_{23}^{i}\phi
_{123_{i}}\right\vert ^{2}$ forces the respective \VEV s  to be orthogonal.

This is not the only $D$-flat configuration possible, and among the
possible $D$-flat configurations, the vacuum is going to be
established by the soft mass terms. In practise we only need to
ensure that $\langle \bar{\phi}_{23} \rangle$ and $\langle
\bar{\phi}_{123} \rangle$ acquire their specific \VEV s , and to demonstrate that this is indeed the case, we now study
the effect of the soft mass terms on their alignment. We assume that all the fields relevant
for the discussion have positive mass squared. The \VEV s  of
$\bar{\phi}_{2},$ $\phi _{23},$ $\phi _{2}$ and $\bar{\phi}_{23}$ are
triggered by minimising $\left\vert 
F_{X_{23}}\right\vert ^{2}$, which requires $<ybb_{3}>=\mu
_{23}^{4}/\sqrt{a_{u}^{2}+a_{d}^{2}}$. The mass term
$\bar{m}_{23}^{2}\left\vert \bar{\phi}_{23}\right\vert
^{2}+m_{2}^{2}\left\vert \phi _{2}\right\vert 
^{2}+m_{23}^{2}\left\vert \phi _{23}\right\vert
^{2}+\bar{m}_{2}^{2}\left\vert \bar{\phi}_{2}\right\vert ^{2}$ is then
minimised by:
\begin{eqnarray}
\left\vert b\right\vert  &=&\left\vert b_{3}\right\vert  \\
y^{2} &=&\frac{m_{23}^{2}+\bar{m}_{23}^{2}}{m_{2}^{2}+\bar{m}%
_{2}^{2}}b^{2}.
\end{eqnarray}
The correct alignment follows as long as $\bar{m}_{23}^{2}$ is the largest soft mass and its
contribution dominates over the other soft mass terms. The equality of
\VEV s  in the $2$nd and $3$rd directions is then a direct 
result of the soft mass degeneracy of $\bar{\phi}_{23,2}^{2}$ and
$\bar{\phi}_{23,3}^{2}$.
This follows from the underlying \FS \ ensuring that $\bar{m}_{23}^{2}\left\vert 
\bar{\phi}_{23}\right\vert ^{2}=\bar{m}_{23}^{2} \left( \left\vert \bar{\phi}%
_{23,1}\right\vert ^{2}+\left\vert \bar{\phi}_{23,2}\right\vert
^{2}+\left\vert \bar{\phi}_{23,3}\right\vert ^{2} \right)$ : the
required \VEV \ alignment arises due to the \FS \ even
though at this stage it has been spontaneously broken.

The vacuum structure applies in a given region of soft mass
parameter space: the correct structure is obtained if $\left\vert \bar{m%
}_{23}\right\vert >\left\vert m_{23}\right\vert$, $\left\vert \bar{m}%
_{123}\right\vert $, $\left\vert {m}_{123}\right\vert > \left\vert
\bar{m}_{_{2}}\right\vert$. The domination of $\bar{m}_{23}^{2}$
ensures soft mass minimisation won't allow $\bar{\phi}_{23}$ to be
perturbed away from the required \VEV. Then, as long as the $\bar{\phi}_{2}$
soft mass is light, its respective \VEV \ ensures $D$-flatness, and
minimisation of the soft mass terms together with $F$-flatness conditions force each entry of $\phi
_{123}$ and $\bar{\phi}_{123}$ to be of equal magnitude. It is also due to the $D$-term that many of the phases are highly
constrained, allowing only overall phase ambiguities shown in the \VEV \ equations from eq.(\ref{eq:Pvev1}) to (\ref{eq:Pvev2}).

\addcontentsline{toc}{chapter}{Bibliography}
\bibliography{refs}
\bibliographystyle{hepv}

\end{document}